\documentclass[12pt,letterpaper]{article}
\usepackage[usenames, dvipsnames]{xcolor}
\usepackage{jcapmod}

\usepackage{tocloft}
\usepackage[]{todonotes}
\usepackage{verbatim}
\usepackage{mathrsfs}
\usepackage{cleveref}
\usepackage[shortlabels]{enumitem}
\usepackage{pgf}
\usepackage{tikz-cd}
\usetikzlibrary{shapes,arrows}
\usetikzlibrary{calc}
\usepackage{pgfplots}
\pgfplotsset{compat=1.12}
\usepackage{tikz-3dplot}
\usepackage{slashed}
\usetikzlibrary{decorations.markings}

\usepackage{amsthm}

\usepackage{tikz}
\usepackage{setspace,caption}
\usepackage{lipsum}
\usetikzlibrary{matrix,arrows,calc}
\makeatletter
\newsavebox\myboxA
\newsavebox\myboxB
\newlength\mylenA

\definecolor{cornellRed}{HTML}{B31B1B}
\definecolor{cornellBlue}{HTML}{0068AC}
\definecolor{cornellGreen}{HTML}{6EB43F}

\usepackage{bbm} 					
\usepackage{slashed} 				
\usepackage{graphicx}				
\usepackage{subcaption}			
\usepackage{psfrag}				
\usepackage{tensor}				
\usepackage{fouridx}				
\usepackage{bm}					
\usepackage{mdframed}				
\usepackage{multirow}				
\usepackage{soul}					
\usepackage{bbold}				
\usepackage{multicol}				
\usepackage{tikz-cd}
\usepackage{rotating}

\usetikzlibrary{arrows}

\tikzset{
commutative diagrams/.cd,
arrow style=tikz,
diagrams={>=latex}}

\usepackage{amsmath}
\usepackage{amssymb}
\usepackage{amsfonts}
\usepackage{mathtools}

\usepackage{feynmf}
\usepackage{marvosym}

\usepackage{import}

\newcommand*\xoverline[2][0.75]{%
    \sbox{\myboxA}{$\m@th#2$}%
    \setbox\myboxB\null
    \ht\myboxB=\ht\myboxA%
    \dp\myboxB=\dp\myboxA%
    \wd\myboxB=#1\wd\myboxA
    \sbox\myboxB{$\m@th\overline{\copy\myboxB}$}
    \setlength\mylenA{\the\wd\myboxA}
    \addtolength\mylenA{-\the\wd\myboxB}%
    \ifdim\wd\myboxB<\wd\myboxA%
       \rlap{\hskip 0.5\mylenA\usebox\myboxB}{\usebox\myboxA}%
    \else
        \hskip -0.5\mylenA\rlap{\usebox\myboxA}{\hskip 0.5\mylenA\usebox\myboxB}%
    \fi}
\makeatother

\newcommand{\im}{\,\mathrm{Im}\,}
\newcommand{\re}{\,\mathrm{Re}\,}

\definecolor{cobalt}{RGB}{44, 98, 120}
\definecolor{celadon}{rgb}{0.67, 0.88, 0.69}
\definecolor{dm}{cmyk}{.20, 0, .30, 0}
\definecolor{burgundy}{rgb}{0.5, 0.0, 0.13}
\definecolor{plotBlue}{RGB}{94, 130, 181}
\definecolor{bisque}{rgb}{1.0, 0.89, 0.77}

\hypersetup{
  colorlinks,
  citecolor=Violet,
  linkcolor=cobalt,
  urlcolor=Blue}

\DeclareSymbolFontAlphabet{\mathbb}{AMSb}
\NewDocumentCommand{\xrightarrows}{ O{}O{} }{%
\mathrel{%
\vcenter{\hbox{%
\begin{tikzpicture}
  \node[minimum width=1cm,minimum height=1ex,anchor=south,align=center] (a){\text{\vphantom{hg}#1}\\[0.5ex] \vphantom{hg}#2};
  \draw[<-] ([yshift=0.35ex]a.west) -- ([yshift=0.35ex]a.east);
  \draw[->] ([yshift=-0.35ex]a.west) -- ([yshift=-0.35ex]a.east);
\end{tikzpicture}
}}%
}%
}

\newif\iffastcompile

\fastcompilefalse

\iffastcompile
\newcommand{\mk}[1]{}
\else
\newcommand{\mk}[1]{\todo[color=burgundy!30, size=\scriptsize, bordercolor=burgundy!30]{MK: #1}}
\fi

\iffastcompile
\newcommand{\sht}[1]{}
\else
\newcommand{\sht}[1]{\todo[color=bisque!30, size=\scriptsize, bordercolor=bisque!30]{ST: #1}}
\fi

\iffastcompile
\newcommand{\sjs}[1]{}
\else
\newcommand{\sjs}[1]{\todo[color=green!30, size=\scriptsize, bordercolor=green!30]{SJS: #1}}
\fi

\iffastcompile
\newcommand{\fc}[1]{}
\else
\newcommand{\fc}[1]{\todo[color=orange!30, size=\scriptsize, bordercolor=orange!30]{FC: #1}}
\fi

\ProvideTextCommandDefault{\Dbar}{%
\leavevmode\lower.5ex\rlap{\hskip-.07em\accent"16}D%
}

\usepackage{environ}
\usepackage{changepage}

\newcommand{\p}{\partial}

\usepackage{dsfont}

\begin{document}
	\newcommand{\main}{.}
\begin{titlepage}

\setcounter{page}{1} \baselineskip=15.5pt \thispagestyle{empty}
\setcounter{tocdepth}{2}
\bigskip\

\vspace{1cm}
\begin{center}
{\large \bfseries On string loops in Calabi-Yau orientifolds in large volume}
\end{center}

\vspace{0.55cm}

\begin{center}
\scalebox{0.95}[0.95]{{\fontsize{14}{30}\selectfont  Manki Kim$^{a}$\vspace{0.25cm}}}

\end{center}

\begin{center}

\vspace{0.15 cm}
{\fontsize{11}{30}
\textsl{$^{a}$Leinweber Institute for Theoretical Physics,
Stanford University, Stanford, CA 94305}}\\
\vspace{0.25cm}

\vskip .5cm
\end{center}

\vspace{0.8cm}
\noindent

\vspace{1.1cm}
We explain and illustrate how to compute string-loop amplitudes in Calabi-Yau orientifold compactification in the large volume limit with the help of the patch-by-patch description of string field theory. We compute the one-loop partition function of the D1-instanton in type IIB string theory compactified on an O9 orientifold of a Calabi-Yau threefold to the first order in the large volume expansion. We show that the unphysical divergence arising from a naive choice of PCOs is canceled by vertical integration. The corollaries of this result, including the universal part of the normalization of the D-instanton superpotential and the one-loop renormalization of K\"ahler moduli, will be presented elsewhere.

\vspace{3.1cm}

\noindent\today

\end{titlepage}
\tableofcontents\newpage

\section{Introduction}
String theory provides a self-consistent framework to study quantum gravity in idealized backgrounds. An important question of whether string theory can be useful for understanding quantum gravity in realistic cosmologies remains elusive.\footnote{For review on the recent progress on the construction of de Sitter vacua in string theory, see \cite{McAllister:2025qwq}.}

The attempt to construct realistic cosmologies in string theory is hampered by the famous Dine-Seiberg problem \cite{Dine:1985he}. To obtain realistic late-time cosmology, such as de Sitter solutions, the vacuum expectation values of moduli need to be fixed. Therefore, to retain a theoretical control of a vacuum solution with moduli stabilization, not only do we need to control string loop expansions but also non-perturbative corrections both in $\alpha'$ and $g_s.$

However, computing string loop amplitudes and non-perturbative effects in cosmological solutions in string theory has proved extremely challenging. First, the details of worldsheet CFTs for generic Calabi-Yau orientifold compactifications, even in the large-volume limit, remain largely unknown. Hence, only quantities that are protected by supersymmetry \cite{Bershadsky:1993cx,Antoniadis:1993ze,Antoniadis:1996qg,Antoniadis:1996vw,Antoniadis:1997eg} or genuine amplitudes in toroidal backgrounds were computed \cite{Berg:2004ek,Berg:2005ja,Haack:2015pbv}. Second, moduli stabilization schemes in type II string theories make heavy use of Ramond-Ramond fluxes \cite{Kachru:2003aw,Balasubramanian:2005zx,DeWolfe:2005uu}, for which the conventional RNS formulation fails to be useful. Third, D-instanton amplitudes suffer from IR ambiguities that are difficult to resolve within the first-quantization approach to string theory. Lastly, generic perturbative string amplitudes suffer from unphysical divergences that stem from the wrong choice of PCOs, which is hard to handle systematically in the RNS formalism.

In light of this difficulty, we would like to study the computation of string loop amplitudes in the formalism of superstring field theory, as string field theory was recently used to systematize conformal perturbation theory \cite{Mazel:2024alu,Scheinpflug:2023osi}, study curved backgrounds \cite{Mazel:2025fxj,Frenkel:2025wko,Mamade:2025htb}, study Ramond-Ramond backgrounds \cite{Cho:2018nfn,Cho:2023mhw,Kim:2024dnw,Cho:2025coy}, and compute the D-instanton amplitudes without ambiguities \cite{Sen:2020cef,Sen:2020eck,Sen:2021qdk,Sen:2021tpp,Eniceicu:2022nay,Eniceicu:2022xvk, Chakravarty:2022cgj,Agmon:2022vdj, Alexandrov:2021shf,Alexandrov:2021dyl, Alexandrov:2022mmy}.\footnote{Recently, the recoil problem of D0-brane scattering amplitudes was understood in \cite{Sen:2025xaj} within string field theory.} We shall show explicitly in this work that with the proper treatment of PCO prescribed by string field theory, we can cancel the unphysical divergence of string loop amplitudes to arrive at a manifestly finite answer.

In search of a one-loop amplitude that is relatively simple yet interesting, with many potential applications, we chose to study the partition function of the D1-instanton to first order in the large-volume expansion. To be concrete, we shall work with type IIB string theory compactified on an O9 orientifold of a Calabi-Yau threefold with the trivial gauge bundle on the D9-branes in the large volume limit. The Ramond-Ramond tadpole induced by $\alpha'^2$ terms on the D9-branes and O9-plane worldvolume is canceled by placing an appropriate number of D5-branes in the compactification. Note that O5-planes are not included in the compactification. We shall assume that a D1-instanton is wrapped on a curve $\Sigma$ in the Calabi-Yau orientifold, while being far away from all D5-branes. Successful evaluation of the D1-instanton partition function will teach us about the new supersymmetric index of the D1-brane \cite{Kim:2023cbh}, thereby allowing the determination of the overall normalization of the D1-instanton amplitude \cite{Alexandrov:2022mmy, Kim:2023cbh}, the one-loop renormalization of K\"ahler moduli \cite{Andreas}, and the renormalization of the Einstein-Hilbert action and the dilaton kinetic terms \cite{Kim:2023sfs, Kim:2023eut}.

As we previously stated, details of the worldsheet CFTs for Calabi-Yau orientifold compactifications, even in the large-volume limit, remain unknown. To overcome this difficulty, we shall use the recently developed non-linear sigma model approach to string field theory \cite{Frenkel:2025wko}. 

The main idea behind the NLSM approach to SFT is pleasantly familiar. Provided that the target space is locally flat, at any point in the target space, we can expand the spacetime metric in the normal coordinate expansion 
\begin{equation}
    g_{ab}(x)=\eta_{ab}-\frac{1}{3}R_{acbd}x^cx^d+\dots\,.
\end{equation}
Note that $\mathcal{O}(x)=\mathcal{O}(\alpha'^{1/2}).$
From the point of view of string perturbation theory, one can treat 
\begin{equation}
    -\frac{1}{3}R_{acbd}x^cx^d\,,
\end{equation}
and the higher order terms in $\alpha'$ as a marginal deformation to the free field CFT. Correspondingly, in string field theory, we construct a string field theory based on the free field worldsheet CFT. Then, the marginal deformation of the worldsheet CFT shall be represented by the perturbative background solution of string field theory that solves the background equation of motion
\begin{equation}
    Q_B\Psi_0+\sum_n\frac{1}{n!}[\Psi_0^n]=0\,.\label{eqn:SFT eom}
\end{equation}
Provided that a proper background solution $\Psi_0$ is found, we can now use string field theory to compute string amplitudes in a local patch of the Calabi-Yau orientifold compactification.

We want to stress the necessity of the off-shell formulation in the context of the non-linear sigma model backgrounds. In general, even in perturbation theory, higher-order background solutions are not annihilated by the BRST charge constructed from the original CFT, as can be seen from the structure of equations of motion \eqref{eqn:SFT eom}. Therefore, to even compute on-shell quantities in the shifted background, in the intermediate steps, one must compute off-shell amplitudes from the point of view of the original string background. This necessitates a self-consistent formulation of off-shell amplitudes, which is beyond the capabilities of the conventional on-shell formulation of string perturbation theory. String field theory, on the other hand, provides a self-consistent off-shell formulation that is well suited for the question at hand. 

There is one more important point to stress. The locations of PCOs are off-shell data. The common lore, therefore, is that the locations of PCOs can be chosen arbitrarily. Based on this common lore, one often chooses a PCO configuration that simplifies the computation of the worldsheet CFT correlation function. For example, a popular choice for n-point insertions of the NS-NS states in the one-loop diagrams is to use the NS-NS vertex operators in the $(0,0)$ picture. However, using such a simple choice leads to a problem. An important requirement of the string perturbation theory is that PCOs will distribute correctly in the degeneration limit, such that each Riemann surface with Euler characteristic $\chi$ has the picture number $-\chi.$ The choice to use $(0,0)$ picture for all the NS-NS states does not lead to the correct distribution of the PCOs in the degeneration limit. Failure of this leads to the inconsistencies. For example, the n-point function at hand is divergent in moduli integral, and even worse, a naive attempt to fix the divergence will force amplitudes to depend on the gauge choice. 

A common approach to address this problem is to perform analytic continuation, see, for example, \cite{Eberhardt:2023xck}. Although finding analytic continuation contour may be a pragmatic way to make progress, it is important to understand how to fix the inconsistencies of the naive PCO choices. And most importantly, it may not be possible to perform analytic continuation to fix the inconsistencies of the PCO treatments especially in the context of D-instanton amplitudes, or in the presence of zero-momentum vertex operators.

String field theory, on the other hand, provides a systematic approach to handle PCOs through the main identity \cite{deLacroix:2017lif}. The BV structure of string field theory imposes boundary conditions on the locations of PCOs in a constructive way. Once the boundary condition on the PCO locations is found correctly, one may attempt a hybrid approach where one uses a simple PCO choice in the interior of the moduli space and perform vertical integration on the boundary of the moduli space \cite{Sen:2015hia} as was explained in \cite{Sen:2019jpm}. In this draft, we shall adopt this hybrid approach.

In addition to the proper treatment of the off-shell amplitudes and PCOs in the intermediate steps, one notable utility of string field theory comes about as we are trying to consistently glue local patches to construct the CFT data for a compact Calabi-Yau orientifold. The idea is to build a suitable partition of unity on the target space and average over the string field theories constructed on local patches that cover it. Provided that we restrict ourselves to perturbative treatments in $\alpha'$ expansion, one can show that such an averaging constructs a global version of string field theory action that solves the BV master equation, which, therefore, provides a consistent framework to study string amplitudes in a compact Calabi-Yau orientifold background. Although we do not prove it in this paper, as the partition of unity is off-shell data, on-shell quantities should not depend on the details of the partition of unity. 

We shall conclude the introduction with a remark on the structure of $\alpha'$ expansion in the D1-instanton partition function. In total, we shall consider four diagrams. 
\begin{enumerate}
    \item Annulus with both ends on the D1-instanton,
    \item Möbius strip,
    \item Annulus with one end on the D9-branes,
    \item Annulus with one end on D5-branes, or equivalently, the D1-instanton disk amplitude with insertions of backreaction induced by D5-branes.
\end{enumerate}
The first diagram vanishes due to the spacetime supersymmetry. The fourth diagram is subleading to the first three by $\mathcal{O}(\alpha'^2) $. It is relatively intuitive to see the origin of the suppression by the $\alpha'^2$ factor in the closed string channel. D5-brane tadpole is created by the $\alpha'^2$ terms in the D9-brane and O9-plane worldvolume action. Therefore, the closed string tadpole cancellation requires that the D5-brane amplitude is paired up with the $\alpha'^2$ terms in the D9-brane and O9-plane amplitudes. As only the first-order perturbation in the large volume expansion is concerned in the draft, we shall therefore ignore the annulus with one end on D5-branes. As we will show in the draft, the closed string tadpole is canceled without the need to introduce D5-branes to the first order in the expansion.

This paper is organized as follows. In \S\ref{sec:NLSM}, we review the non-linear sigma model approach to string field theory. In \S\ref{sec:structure}, we study the structure of the one-loop amplitudes. In \S\ref{sec:corr zero modes}, we explain how to compute correlation functions if vertex operators contain polynomials in X. In \S\ref{sec:zero-th}, we compute the partition function at the zero-th order in the $\alpha'$ expansion. In \S\ref{sec:first d1-d1}, as a warm-up, we compute the annulus with both ends on the D1-instanton at the first order in the expansion. We show that explicit calculation indeed agrees with the expectation from the spacetime supersymmetry. In \S\ref{sec:first Mob}, we compute the Möbius at the first order in the expansion. In \S\ref{sec:d1-d9}, we compute the annulus with one end on the D9-branes at the first order in the expansion. In \S\ref{sec:first order full}, we combine the first order results to compute the full one-loop partition function. In \S\ref{sec:conclusions}, we conclude. In \S\ref{sec:conventions}, we set our conventions. In \S\ref{app:theta functions}, we collect useful theta function identities. In \S\ref{app:Green}, we collect Green's functions. In \S\ref{app:bc ghost}, we write the bc ghost partition function. In \S\ref{app:xi vertical}, we collect useful results for the vertical integral. In \S\ref{app:one-loop reps}, we collect different worldsheet representations of the one-loop diagrams. In \S\ref{app:vertices}, we collect relevant string vertices.

\subsection{Summary of the draft}
For the reader’s convenience, we summarize here the main points and results of this draft.

We develop a constructive prescription for computing one-loop string amplitudes in compact Calabi–Yau orientifolds in the large-volume regime using the patch-by-patch formulation of superstring field theory together with the $\alpha'$ expansion \cite{Frenkel:2025wko}. The central technical issue is a careful treatment of   PCO location. The widely used “convenient” prescriptions (for example, taking all NS–NS vertices in the $(0,0)$ picture for one-loop amplitudes) generally fail to reproduce the required picture-number distribution in degeneration limits. This mismatch can lead to spurious divergences in moduli integrals and can introduce an unphysical dependence on gauge choices. String field theory provides a systematic resolution: the BV structure yields consistent boundary conditions for PCO insertions \cite{deLacroix:2017lif}, and we implement a hybrid prescription in which one uses a simple PCO choice in the interior of moduli space and restores correct factorization by adding the corresponding vertical-integration contributions on its boundaries \cite{Sen:2019jpm}.

As an explicit application, we compute the D1-instanton one-loop partition function to first order in the large-volume expansion for type IIB string theory compactified on an O9 orientifold of a Calabi–Yau threefold with a trivial gauge bundle on the D9-branes. The computation is organized by one-loop diagrams: the D1–D1 annulus, the Möbius strip, and the D1–D9 annulus. The D1–D1 annulus vanishes at first order, in agreement with the expectation from spacetime supersymmetry. We explain why contributions associated with the D5 sector enter only at subleading order in the $\alpha'$ expansion and can be neglected at the first order. Combining the remaining first-order contributions, we obtain the full D1-instanton one-loop partition function at this order in a way that makes the cancelation between the PCO-mismatch “exact pieces” and the corresponding vertical-integration explicit.

\paragraph{What this enables} The D1-instanton partition function computed here is intended as an input for several quantities of direct interest in four-dimensional effective theories arising from string compactifications, including the D1 “new supersymmetric index” \cite{Kim:2023cbh}, the overall normalization of D1-instanton amplitudes \cite{Alexandrov:2022mmy,Kim:2023cbh}, and one-loop renormalization of couplings such as the Einstein–Hilbert term \cite{Kim:2023sfs}, the dilaton kinetic terms \cite{Kim:2023eut}, and the metric of K\"ahler moduli, all within the regime of validity of the large-volume expansion.

Although our explicit computation is carried out only to the first order in the large-volume expansion, the patch-by-patch SFT formulation is designed to be systematic: in principle, one can extend the analysis to higher orders by constructing higher-order background solutions $\Psi_0,$ and including the corresponding higher-order interaction vertices and vertical-integration contributions required by the BV consistency conditions. This provides a path to computing more general loop amplitudes and D-instanton normalizations beyond the D1-instanton studied here, including other Euclidean D-branes and, more broadly, amplitudes in backgrounds with additional ingredients. In particular, since superstring field theory offers a framework that can incorporate Ramond–Ramond backgrounds, the same strategy is expected to be applicable—after the appropriate background solution is specified—to loop and non-perturbative computations in flux compactifications with Ramond–Ramond fluxes.

\section{Non-linear sigma model in string field theory}\label{sec:NLSM}
\subsection{Review of string field theory}
In this section, we shall collect relevant aspects of string field theory. For more in-depth reviews, see, for example, \cite{deLacroix:2017lif,FarooghMoosavian:2019yke,Erler:2019loq,Erler:2019vhl,Sen:2024nfd,Erbin:2021smf,Maccaferri:2023vns}. 

String field theory is an attempt to formulate string theory off-shell by constructing a self-consistent path integral
\begin{equation}
    \int\mathcal{D}\Psi e^{iS(\Psi)}\,,
\end{equation}
with the action $S(\Psi)$ for string field $\Psi$, which is a collection of string excitations. 

To precisely define what we mean by $\Psi,$ we need first to introduce state spaces. Let us denote the space of GSO even closed string states in the small Hilbert space of CFT with the worldsheet picture number $(p,q)$ by $\mathcal{H}_{p,q}^c,$ and the space of GSO even open string states with picture number $p$ by $\mathcal{H}_{p}^o.$ We shall further require that every state $|\psi\rangle\in \mathcal{H}_{p,q}^c$ satisfies
\begin{equation}
    b_0^-|\psi\rangle=L_0^-|\psi\rangle=0\,,
\end{equation}
where $b_0^\pm:=b_0\pm \bar{b}_0,$ and $L_0^\pm:=L_0\pm\bar{L}_0.$ We now prepare two subspaces in the closed string state space and two subspaces in the open string state space
\begin{align}
    &\mathcal{H}^c:=\mathcal{H}_{-1,-1}^c\oplus\mathcal{H}_{-1/2,-1}^c\oplus\mathcal{H}^c_{-1,-1/2}\oplus\mathcal{H}_{-1/2,-1/2}^c\,,\\
    &\widetilde{\mathcal{H}}^c:=\mathcal{H}_{-1,-1}^c\oplus\mathcal{H}_{-3/2,-1}^c\oplus\mathcal{H}_{-1,-3/2}^c\oplus\mathcal{H}_{-3/2,-3/2}^c\,,\\
    &\mathcal{H}^o:=\mathcal{H}_{-1}^o\oplus\mathcal{H}_{-1/2}^o\,,\\
    &\widetilde{\mathcal{H}}^o:=\mathcal{H}_{-1}^o\oplus\mathcal{H}_{-3/2}^o\,.
\end{align}
We shall define string fields as states living in the above string spaces
\begin{equation}
    |\Psi^c\rangle\in\mathcal{H}^c\,,\quad |\widetilde{\Psi}^c\rangle\in\widetilde{\mathcal{H}}^c\,,\quad |\Psi^o\rangle\in\mathcal{H}^o\,,\quad |\widetilde{\Psi}^o\rangle\in\widetilde{\mathcal{H}}^o\,.
\end{equation}

To construct the off-shell action of string field theory, we shall construct off-shell string vertices 
\begin{equation}
    \{\Psi_1^c\otimes\Psi_2^c\otimes\cdots\otimes \Psi_n^c;\Psi_1^o\otimes\cdots\otimes\Psi_m^o\}
\end{equation}
for generic $m$ and $n,$ and
\begin{equation}
    \{\widetilde{\Psi}^c\}_{D^2}\,,\quad \{\widetilde{\Psi}^c\}_{\Bbb{RP}^2}\,.
\end{equation}
An important consistency requirement is that the moduli space of punctured Riemann surfaces is covered exactly once by string vertices. For the details on string vertices, see \cite{Sen:2019jpm,Sen:2020eck}. 

Important data that goes into the construction of the string vertices is the locations of the PCOs in the vertex region of the moduli space. For genuine off-shell amplitudes, the results explicitly depend on the location of the PCOs, even in the interior of the moduli space.  Since we will compute only the partition function of the D1-instanton to first order in perturbation theory, the vertex operator we insert is closed under the BRST charge. So, one may naively disregard the need for string field theory altogether. 

However, there are subtleties regarding the PCO locations in the degeneration limits. For string amplitudes to be well defined, in the degeneration limits, PCOs must be correctly distributed over the worldsheet such that each Riemann surface with genus $g$ appearing in the degeneration limit comes with the correct picture numbers. For example, a Riemann surface with genus $g$ must have picture number $2(g-1)$. The construction of string vertices in string field theory automatically imposes boundary conditions for the PCO locations in the vertex regions through main identities \cite{FarooghMoosavian:2019yke}. In appendix \S\ref{app:vertices}, we shall explicitly construct relevant string vertices following the method of \cite{Sen:2020eck}.

Once the correct boundary conditions of PCO locations are found, one can proceed as follows, provided that the inserted vertex operators are on-shell. In the interior of the vertex region, one can use the PCO configuration, which is convenient for computing CFT correlators. On the boundary of the vertex region, the convenient choice of PCO configuration will not, in general, agree with the boundary conditions imposed by the main identity. Such a disagreement will manifest itself as an unphysical divergence in the amplitudes. To rectify such a problem, one can then perform the vertical integral \cite{Sen:2015hia} to fill the gaps between the wrong and correct locations of PCOs. Once the vertical integration is added to the amplitude, the divergence is canceled, yielding a consistent answer. 

It is important to note that for generic string vertices, only the states from untilded string spaces participate, while the Disk and Cross-cap one-point functions depend only on the tilded string states. This is related to the fact that the Ramond-Ramond states in $(-1/2,-1/2)$ picture are field strengths, whereas the fields in $(-3/2,-3/2)$ represent the gauge fields. We also define string brackets 
\begin{equation}
    [\Psi_1^c\otimes\cdots\otimes \Psi_n^c;\Psi_1^o\otimes\cdots\otimes\Psi_m^o ]^c\in\widetilde{\mathcal{H}}^c\,,\quad [\Psi_1^c\otimes\cdots\otimes \Psi_n^c;\Psi_1^o\otimes\cdots\otimes\Psi_m^o ]^o\in\widetilde{\mathcal{H}}^o
\end{equation}
\begin{equation}
    []_{D^2}\in\mathcal{H}^c\,,\quad []_{\Bbb{RP}^2}\in\mathcal{H}^c\,,
\end{equation}
via
\begin{align}
    &\langle \Psi_0^c|c_0^-|[\Psi_1^c\otimes\cdots\otimes \Psi_n^c;\Psi_1^o\otimes\cdots\otimes\Psi_m^o ]^c\rangle =\{\Psi_0^c\otimes\Psi_1^c\otimes\cdots\otimes \Psi_n^c;\Psi_1^o\otimes\cdots\otimes\Psi_m^o \}\,,\\
    &\langle\Psi_0^o|[\Psi_1^c\otimes\cdots\otimes \Psi_n^c;\Psi_1^o\otimes\cdots\otimes\Psi_m^o]^o\rangle=\{ \Psi_1^c\otimes\cdots\otimes \Psi_n^c;\Psi_0^o\otimes\Psi_1^o\otimes\cdots\otimes\Psi_m^o\}\,,\\
    &\langle \widetilde{\Psi}_0^c|c_0^-|[]_{D^2,\Bbb{RP}^2}\rangle= \{\widetilde{\Psi}_0^c\}_{D^2,\Bbb{RP}^2}\,,
\end{align}
for all states $|\Psi_0^c\rangle\in\mathcal{H}^c,$ and $|\Psi_0^o\rangle\in\mathcal{H}^o\,.$ Note that $\langle \Psi|$ is the BPZ conjugate of $|\Psi\rangle$ obtained by the map $z\mapsto 1/z$ for closed strings and $z\mapsto -1/z$ for open strings. We shall define the maps $\mathcal{G}^c$ and $\mathcal{G}^o$ 
\begin{equation}
    \mathcal{G}^c:=\widetilde{\mathcal{H}}^c\rightarrow\mathcal{H}^c\,,
\end{equation}
and
\begin{equation}
    \mathcal{G}^o:=\widetilde{\mathcal{H}}^o\rightarrow\mathcal{H}^o\,,
\end{equation}
that act as the zero modes of the PCOs to Ramond states. Note that we will omit the superscript of $\mathcal{G}^c$ and $\mathcal{G}^o$ from now on.

We are finally ready to write the string field theory action
\begin{align}
    S=&-\frac{1}{2g_s^2} \langle\widetilde{\Psi}^c|c_0^-Q_B\mathcal{G}|\widetilde{\Psi}^c\rangle+\frac{1}{g_s^2}\langle\widetilde{\Psi}^c|c_0^-Q_B|\Psi^c\rangle-\frac{1}{2g_s}\langle\widetilde{\Psi}^o|Q_B\mathcal{G}|\widetilde{\Psi}^o\rangle+\frac{1}{g_s} \langle\widetilde{\Psi}^o|Q_B|\Psi_o\rangle\nonumber\\
    &+\{\widetilde{\Psi}^c\}_{D^2+\Bbb{RP}^2}+\sum_{n,m}\frac{1}{n!m!}\{ (\Psi^c)^n;(\Psi^o)^m\}\,.
\end{align}
The action above is invariant under the gauge transformation \cite{FarooghMoosavian:2019yke}
\begin{align}
    &|\delta\Psi^c\rangle=Q_B|\Lambda^c\rangle +g_s^2\sum_{n,m}\frac{1}{n!m!}\mathcal{G}[\Lambda^c(\Psi^c)^n;(\Psi^o)^m]^c+g_s^2\sum_{n,m}\frac{1}{n!m!}\mathcal{G}[(\Psi^c)^n;\Lambda^o(\Psi^o)^m]^c\,,\\
    &|\delta\Psi^o\rangle=Q_B|\Lambda^o\rangle-g_s \sum_{n,m}\frac{1}{n!m!}\mathcal{G}[\Lambda^c(\Psi^c)^n;(\Psi^o)^m]^o-g_s\sum_{n,m}\frac{1}{n!m!}\mathcal{G}[(\Psi^c)^n;\Lambda^o(\Psi^o)^m]^o\,,\\
    &|\delta\widetilde{\Psi}^c\rangle=Q_B|\widetilde{\Lambda}^c\rangle +g_s^2\sum_{n,m}\frac{1}{n!m!}[\Lambda^c(\Psi^c)^n;(\Psi^o)^m]^c+g_s^2\sum_{n,m}\frac{1}{n!m!}[(\Psi^c)^n;\Lambda^o(\Psi^o)^m]^c\,,\\
    &|\delta\widetilde{\Psi}^o\rangle=Q_B|\widetilde{\Lambda}^o\rangle +g_s^2\sum_{n,m}\frac{1}{n!m!}[\Lambda^c(\Psi^c)^n;(\Psi^o)^m]^o+g_s^2\sum_{n,m}\frac{1}{n!m!}[(\Psi^c)^n;\Lambda^o(\Psi^o)^m]^o\,,
\end{align}
and solves the BV master equation. From the SFT action, we can find the equations of motion
\begin{align}
    &Q_B(|\Psi^c\rangle-\mathcal{G}|\widetilde{\Psi}^c\rangle)+g_s^2|[]_{D^2+\Bbb{RP}^2}\rangle=0\,,\\
    &Q_B|\widetilde{\Psi}^c\rangle+g_s^2 \sum_{n=1}^\infty \sum_{m=0}^\infty \frac{1}{(n-1)!m!}[(\Psi^c)^{n-1};(\Psi^o)^m]^c=0\,,\\
    &Q_B (|\Psi^o\rangle-\mathcal{G}|\widetilde{\Psi}^o\rangle)=0\,,\\
    &Q_B|\widetilde{\Psi}^o\rangle+g_s \sum_{n=0}^\infty\sum_{m=1}^\infty \frac{1}{n!(m-1)!}[(\Psi^c)^n;(\Psi^o)^{m-1}]^o=0\,.
\end{align}
We can combine the eom for the tilded fields and the untilded fields to write 
\begin{align}
    &Q_B|\Psi^c\rangle+g_s^2\sum_{n=1}^\infty \sum_{m=0}^\infty \frac{1}{(n-1)!m!}\mathcal{G}[(\Psi^c)^{n-1};(\Psi^o)^m]^c+g_s^2|[]_{D^2+\Bbb{RP}^2}\rangle=0\,,\\
    & Q_B|\Psi^o\rangle +g_s \sum_{n=0}^\infty\sum_{m=1}^\infty \frac{1}{n!(m-1)!}\mathcal{G}[(\Psi^c)^n;(\Psi^o)^{m-1}]^o=0\,.
\end{align}

Now, suppose that one found the background solution $\Psi_0^c$ and $\Psi_0^o$ by solving the string equations of motion. To find the scattering states $\delta \Psi^c$ and $\delta\Psi^o$, we can solve the linearized equation of motion
\begin{equation}
    Q_B|\delta\Psi^c\rangle+g_s^2\sum_{n,m}\frac{1}{n!m!}\mathcal{G}[\delta\Psi^c (\Psi_0^c)^n;(\Psi_0^o)^m]^c+g_s^2 \sum_{n,m}\frac{1}{n!m!}\mathcal{G}[(\Psi_0^c)^n;\delta\Psi^o(\Psi_0^o)^m]^c=0\,,
\end{equation}
\begin{equation}
    Q_B|\delta\Psi^o\rangle+g_s\sum_{n,m}\frac{1}{n!m!}\mathcal{G}[\delta\Psi^c (\Psi_0^c)^n;(\Psi_0^o)^m]^o+g_s \sum_{n,m}\frac{1}{n!m!}\mathcal{G}[(\Psi_0^c)^n;\delta\Psi^o(\Psi_0^o)^m]^o=0\,.
\end{equation}
One can then use the scattering states $\delta\Psi$ and their associated Feynman rules to compute string amplitudes in the new background represented by $\Psi_0.$
\subsection{Patch-by-patch description}
In this section, we will briefly outline the main ideas of \cite{Frenkel:2025wko}. For more details, we refer the readers to \cite{Frenkel:2025wko}.

For the formulation of string perturbation theory in the RNS formalism, one must start with a worldsheet CFT for a fixed target space. However, it is uncommon for the target space in question to admit a known worldsheet CFT description. In a favorable situation, one might study a spacetime by deforming a known CFT. If the deformation is exactly marginal and in the NSNS sector, one can employ the standard technique of conformal perturbation theory. 

However, not every background is built for our convenience, and we often need to study either a deformation that is not exactly marginal or a deformation induced by Ramond-Ramond states. In such cases, conventional string perturbation theory is hard to make sense of. 

If the deformation is not exactly marginal, the higher-order deformations are not represented by BRST-closed states. Correspondingly, scattering states found at high orders of perturbation theory are, in general, written as off-shell states from the point of view of the original CFT, even if such scattering states are supposed to be on-shell in the new background. As even the calculations of on-shell amplitudes in the new background now involve the insertion of the off-shell states, \emph{from the point of view of the original CFT we began with}, one must understand how to formulate off-shell amplitudes consistently. This is beyond the reach of the conventional string perturbation theory.

The problem caused by the Ramond-Ramond deformation is perhaps more concerning. Not only is Ramond-Ramond deformation, in general, not exactly marginal, but it is also potentially non-local on the worldsheet \cite{Berenstein:1999ip,Berenstein:1999jq}. In the RNS formalism based on the picture-changing operators, Ramond-Ramond vertex operators live in the CFT Hilbert space with half-integral picture numbers. On the other hand, the worldsheet action has an integral picture number. Therefore, it is not possible to directly perturb the worldsheet action with the Ramond-Ramond vertex operator. A way to remedy this problem was suggested in \cite{Berenstein:1999ip,Berenstein:1999jq}, which, however, involves an operation that is not well-defined and additionally induces a non-locality on the worldsheet. 

However, as was recently shown, string field theory provides a powerful toolkit for circumventing the aforementioned problems. Therefore, it is worthwhile to explore the possibility of studying a non-trivial target space that admits some notion of perturbation theories. With this in mind, we shall study a curved background in string theory in $g_s$ and $\alpha'$ expansion.

The target spaces we are interested in are smooth and curved manifolds that admit a well-defined perturbative $\alpha'$ expansion. An important example we shall study in detail is a Calabi-Yau manifold without a singularity. Any good smooth manifold $Y$ is by definition locally Euclidean, meaning that we can always choose coordinates around any point $p\in U_i$ in a coordinate chart $U_i\subset Y$ such that the metric admits the Riemann normal coordinate expansion
\begin{equation}
    g_{ab}^{(i)}(x)=\eta_{ab}-\frac{1}{3} R_{acbd}^{(i)}x^cx^d+\dots\,.\label{eqn:curved metric}
\end{equation}
The target space with a constant metric can be studied using the well-known free-field worldsheet CFT. Now, then, the question is whether we can describe the higher-order terms in the metric as a background shift.

In the context of string field theory, we can formulate the question as follows. Let us suppose that we constructed string field theory by perturbing around the flat ten-dimensional spacetime described by the free field CFT. The question is, then, can we find the background solution that represents the curved metric \eqref{eqn:curved metric}? 

Let us expand the (closed-string) background field in $\alpha'$ expansion
\begin{equation}
    \Psi_0^{(i)}=\sum_n \Psi_{0,n}^{(i)}\,,
\end{equation}
where the superscript $i$ denotes the index of the local coordinate chart $U_i,$ and $\mathcal{O}(\Psi_{0,n}^{(i)})=\mathcal{O}(\alpha'^n).$ The equation of motion governing the background field is 
\begin{equation}
    Q_B|\Psi_{0,1}^{(i)}\rangle=0\,,
\end{equation}
\begin{equation}
    Q_B|\Psi_{0,2}^{(i)}\rangle+\frac{1}{2}[(\Psi_{0,1}^{(i)})^2]=0\,,
\end{equation}
and so on. As it turns out, the first-order equation of motion states that the first-order deformation should be on-shell. This should not be surprising, as any perturbative background shift starts out its life as a marginal deformation. The second- and higher-order background equations of motion also have a simple interpretation: $\alpha'$-corrected Einstein's equation and $\alpha'$-corrected equations for string excitations. To be more concrete, let us write down an ansatz for the first-order background solution
\begin{equation}
    \Psi_{0,1}^{(i)}=\frac{1}{12\pi}R_{acbd}X^cX^dc\bar{c}e^{-\phi}\psi^ae^{-\bar{\phi}}\bar{\psi}^b\,.
\end{equation}
The BRST condition requires
\begin{equation}
    R_{ab}=0\,,
\end{equation}
which is Einstein's equation. Provided that one finds the background solution in a local coordinate patch $ U_i$, one can then study the local properties of the target space.

So far, we have only found the local description. To study a compact Calabi-Yau manifold, for example, we must learn how to glue up the local patches in a consistent manner. Two consistency conditions the global version of the string field theory must satisfy are: the gauge invariance of the action, and the solvability of the BV master equation. 

Then readers may worry that constructing the global version of string field theory on curved backgrounds may not be easy. As the background fields are polynomial in the spacetime coordinates, the background fields do not die off quickly on the boundary of the closure of the local coordinate chart. When the field configuration does not die off quickly, the gauge invariance of the action is broken on the boundary. A common way to rectify such a breakdown of gauge invariance is to introduce a boundary action that will restore the consistency of the formulation. However, introducing a boundary action will introduce new degrees of freedom that we do not encounter in string compactifications. Therefore, we are forced to restore the gauge invariance \emph{without} introducing boundary action. 

However, it is important to note that the same worry equally applies to general relativity and gauge theory on a curved background. Upon performing the Taylor expansion, metric and gauge fields do extend beyond the local coordinates, and one may equally worry that gauge invariance is broken. This worry is unwarranted, as the addition of local contributions with a partition of unity removes any inconsistency, provided that fields in different coordinate charts on their overlap are related via a gauge transformation. 

Essentially, we can use the same idea of partition of unity to prove that the gauge invariance of the global version of string field theory is restored. Furthermore, it is relatively straightforward to show that the action of SFT solves the BV master equation, hence providing a consistent field theory. 

The general proof of gauge invariance and the solvability of the BV master equation for the patch-by-patch description of string field theory requires rather laborious index bookkeeping. To avoid overloading the indices, we shall illustrate the main idea with a target space $X$ that can be covered by two coordinate charts $U_1$ and $U_2.$ We shall also assume that $U_1$ and $U_2$ have a non-trivial overlap $U_{12}:=U_1\cap U_2.$ We will then construct string field theories constructed from the free field CFT for $U_1,$ $U_2,$ and $U_{12}.$ For the general proof, see \cite{Frenkel:2025wko}.

Because string fields $\Psi^{(1)}$ and $\Psi^{(2)}$ are formally constructed from the same state space, it makes sense to compare $\Psi^{(1)}$ and $\Psi^{(2)}$ in the overlap $U_{12}.$ In general, string fields $\Psi^{(1)}$ and $\Psi^{(2)}$ do not agree. This is not a fatal problem, as long as $\Psi^{(1)}$ and $\Psi^{(2)}$ are related by gauge transformation. This is precisely what we shall require. Now, we need one more ingredient. We shall define an interpolating string field $\Psi^{(12)}$ that satisfies the following conditions
\begin{equation}
    \Psi^{(12)}|_{\partial U_1\cap U_{12}}=\Psi^{(1)}|_{\partial U_1\cap U_{12}}\,,\quad  \Psi^{(12)}|_{\partial U_2\cap U_{12}}=\Psi^{(2)}|_{\partial U_2\cap U_{12}}\,.\label{eqn:interpolating sf}
\end{equation}
The detail of how the string field $\Psi^{(12)}$ interpolates $\Psi^{(1)}$ and $\Psi^{(2)}$ does not matter. 

With the ingredients at hand, we can construct the global form of the string field theory action
\begin{equation}
    S(\Psi)=S^{(1)}(\Psi^{(1)})+S^{(2)}(\Psi^{(2)})-S^{(12)}(\Psi^{(12)})\,,
\end{equation}
where the action $S^{(i)}$ is defined over the coordinate chart $U_i,$ and the global form of string field $\Psi$ is defined via
\begin{equation}
    \Psi|_{U_1}=\Psi^{(1)}\,,\quad \Psi|_{U_2}=\Psi^{(2)}\,,\quad \Psi|_{U_{12}}=\Psi^{(12)}\,.
\end{equation}
The motivation behind the global form of the action is rather simple. As we add up $S^{(1)}$ and $S^{(2)},$ we double counted the contributions from $U_{12}.$ Subtracting $S^{(12)}$ then achieves the partition of unity. Under the gauge transformation
\begin{equation}
    |\delta\Psi\rangle=Q_B|\Lambda\rangle+ \sum_n \frac{1}{n!}[\Psi^n\Lambda]\,,
\end{equation}
the local action is no longer invariant. However, the breakdown of the gauge invariance of the local action is localized on the boundary
\begin{equation}
    \delta_\Lambda S^{(i)}= \delta_\Lambda S^{(i)}|_{\partial U_i}\,.
\end{equation}
Due to the interpolating condition, \eqref{eqn:interpolating sf}, the breakdown of gauge invariance of the local actions cancels out. Therefore, the global form of string field theory action is gauge invariant. One can also similarly show that the global form of the string field theory action solves the BV master equation \cite{Frenkel:2025wko}.

A remark is in order. As is evident from the discussion above, for consistency in explicit calculations, it is crucial to correctly identify the boundary terms that must be canceled when summing the contributions from all local coordinate patches.\footnote{For recent discussions on boundary action in string field theory, see \cite{Stettinger:2024uus,Firat:2024kxq,Maccaferri:2025orz,Maccaferri:2025onc}.} Once the boundary terms are identified, one can equivalently drop them altogether. The prescription of \cite{Belopolsky:1995vi} will be used heavily to identify the boundary terms correctly.

\section{Structure of the one-loop partition function}\label{sec:structure}
In this section, we shall discuss the structure of the partition function of a D1-instanton to the first order in the large volume expansion. Although it is strictly not necessary, we will assume that the D1-instanton is wrapped on a $\Bbb{P}^1$ in the Calabi-Yau orientifold. Hence, the D1-instanton itself is also rigid. Furthermore, the D1-instanton has a gauge group $O(1).$ Also, to perform the moduli integral over $ t$, we shall ignore the open-string zero-mode contribution, as light modes of solitons cannot be treated perturbatively and should be handled with string field theory. The main idea of such an analysis was already presented in \cite{Alexandrov:2022mmy}. However, as we are performing the large volume expansion, despite the fact that the D-instanton is rigid, there are more than four bosonic and two fermionic light states. That is because some of the light modes have weights that go to zero in the strict infinite volume limit. A careful treatment of such light open-string states will be presented in \cite{D-instanton}, to determine the overall normalization of the D-instanton amplitude in CY orientifolds. Note also that determining the overall normalization of the D-instanton amplitude automatically determines the one-loop renormalization of the D-instanton tension, which is K\"ahler moduli in the case of the D1-instanton \cite{Alexandrov:2022mmy}.

We shall explicitly lay out the strategy for the computation of the contribution from the Möbius strip diagram $Z_M$. The strategy for computing the annulus diagrams is similar, so we won't repeat it. We shall divide $Z_M$ into the vertex contribution and the Feynman contributions
\begin{align}
    Z_M=&\sum_{(a)}Z_{M}^{(a)}\,,\\
    =& \{ \Psi_{0,1};\}_{M}+\sum_o \{\Psi_{0,1};\varphi_o\otimes \varphi_o\} _{D^2}\left\langle\varphi_o^\dagger\left|\Omega\frac{b_0}{L_0}\right|\varphi_o^\dagger\right\rangle+\sum_{o,o'}\{\Psi_{0,1};\varphi_{o'}\}_{D^2} \{;\varphi_{o'}\otimes\varphi_o\otimes \varphi_o\}_{D^2} \nonumber\\&\times\left\langle \varphi_{o'}^\dagger \left|\frac{b_0}{L_0}\right|\varphi_{o'}^\dagger \right\rangle \left\langle\varphi_o^\dagger\left|\Omega\frac{b_0}{L_0}\right|\varphi_o^\dagger\right\rangle+\sum_o \{\Psi_{0,1}; \varphi_o\}_{D^2} \{;\varphi_o\}_M\left\langle\varphi_o^\dagger\left|\frac{b_0}{L_0}\right|\varphi_o^\dagger\right\rangle\nonumber\\
    &+\sum_{o,c}\{\Psi_{0,1};\varphi_o\}_{D^2} \{V_c;\varphi_o\}_{D^2}\{V_c\}_{\Bbb{RP}^2}\left\langle\varphi_o^\dagger\left|\frac{b_0}{L_0}\right|\varphi_o^\dagger\right\rangle \left\langle V_c^\dagger\left|c_0^-\frac{b_0^+}{L_0^+}\right|V_c^\dagger\right\rangle\nonumber\\
    &+\left(\sum_c \{\Psi_{0,1} \otimes V_c;\}_{D^2}\{V_c;\}_{
    \Bbb{RP}^2} +\sum_c \{\Psi_{0,1} \otimes V_c\}_{\Bbb{RP}^2}\{V_c\}_{D^2}\right) \left\langle V_c^\dagger\left|c_0^-\frac{b_0^+}{L_0^+}\right|V_c^\dagger\right\rangle\nonumber\\
    &+\sum_{c,c'} \{ V_c;\}_{D^2}\{\Psi_{0,1} \otimes V_c\otimes V_{c'}\}_{S^2}\otimes \{V_{c'};\}_{\Bbb{RP}^2}\left\langle V_c^\dagger\left|c_0^-\frac{b_0^+}{L_0^+}\right|V_c^\dagger\right\rangle\left\langle V_{c'}^\dagger\left|c_0^-\frac{b_0^+}{L_0^+}\right|V_{c'}^\dagger\right\rangle\,,
\end{align}
where
\begin{equation}
    \Psi_{0,1}=-\frac{1}{4\pi}\delta g_{ab}c\bar{c}e^{-\phi}\psi^a e^{-\bar{\phi}}\bar{\psi}^b\,,
\end{equation}
with 
\begin{equation}
    \delta g_{ab}=-\frac{1}{3}R_{acbd}X^cX^d\,,
\end{equation}
is the solution to the first-order background equation of motion, and $\dagger$ denotes the conjugate state. Note that $\Omega$ denotes the orientifolding. As we shall take the large stub limit, insertion of off-shell states into string vertices with positive weights gives rise to a vanishing contribution. Therefore, the Feynman region contributions only receive non-trivial contributions from the exchange of massless states. 

First, we note that we are placing the Euclidean D1-brane with an $O(1)$ gauge group; therefore, the gauge fields are projected out. Second, the open string zero modes describing the deformation of the ED1-brane are lifted, and hence 
\begin{equation}
    \Bbb{P}[\Psi_{0,1}^c;]^o_{D^2}=0\,,
\end{equation}
where $\Bbb{P}$ is the projection operator to $L_0^+$-nilpotent states of string theory. Third, the exchanges of the closed string zero modes will be canceled upon summing over the annuli and Möbius strip diagrams. This will be explicitly shown later in the draft. Hence, the diagrams we need to evaluate explicitly are
\begin{align}
    \{\Psi_{0,1};\}_{M}\,,\quad \sum_o \{\Psi_{0,1};\varphi_o\otimes \varphi_o\} _{D^2}\left\langle\varphi_o^\dagger\left|\Omega\frac{b_0}{L_0}\right|\varphi_o^\dagger\right\rangle\,.\label{eqn:terms to compute}
\end{align}

To evaluate the Möbius one-point string vertex, we will divide the moduli space into four regions. First, the open string channel with $t_o>t_o^*.$ Second, the open string channel with $1/4<t_o\leq t_o^*.$ Third, the closed string channel with $1/(4t_c^*)<t_o\leq1/4.$ Fourth, the closed string channel with $t_o\leq 1/(4t_c^*).$  In the large stub limit, the first region only contains the open string zero mode contributions. Similarly, the fourth region contains only the closed-string zero-mode contributions, which we can effectively omit, as the tadpole cancellation is guaranteed.  By taking the limit $t\rightarrow \infty,$ we can read off the open-string zero-mode contributions from evaluating the contribution from the second region. We can similarly read off the closed string zero mode contributions by evaluating the contribution from the third region and taking the $t\rightarrow0$ limit. Therefore, we only need to evaluate the second and the third regions explicitly. 

Note that in the large stub limit, the second term in \eqref{eqn:terms to compute} receives contributions only from the zero modes. As such contributions can be read off from the Mobius one-point, and we will defer the proper treatment of the solitonic zero modes to \cite{D-instanton}, we shall not explicitly evaluate this term in this draft.

To compute $\{\Psi_{0,1};\}_{M},$ we must insert two PCOs. To simplify the calculation, in the even spin structures, in the interior of the moduli space, we shall use the $(0,0)$ picture form of the background field. The odd-spin structure contribution can be shown to vanish \cite{Alexandrov:2022mmy}. We shall justify the use of such a picture later in the draft. However, placing two PCOs on top of the vertex operator leads to inconsistencies as the PCOs are wrongly distributed in the degeneration limit, in disagreement with the Feynman regions. The incorrect placement of the PCOs will manifest as an unphysical divergence that must be regularized. To fix the problem, on the boundary of the vertex region of the moduli space, we shall perform the vertical integration \cite{Sen:2015hia} by filling the gaps in the PCOs. We will find that the vertical integral precisely cancels the unphysical divergence.

A few comments are in order. First, we study the D1-instanton amplitudes for which there is no continuous internal momentum, since the D1-instanton is assumed to wrap a compact two-dimensional manifold. On the other hand, in the one-loop amplitude, in the perturbative expansion in the large volume, we are summing over the continuous momentum $k,$ as if the D1-brane is extended along a non-compact spacetime. This can be understood as follows. If we were capable of computing the spectrum at finite volume, we would've written the one-loop partition function as
\begin{equation}
    \int\frac{dt}{2t}\text{Tr}(e^{-2\pi (-\nabla^2+h)t})=\int \frac{dt}{2t} \text{Tr}(\sum_l d_l e^{-2\pi (\lambda_l+h)t})\,,
\end{equation}
where the sum over $l$ is over the appropriate spherical harmonics, where $\lambda_l$ is the eigenvalue of $-\nabla^2.$ At finite volume, the degeneracies and the eigenvalues under $-\nabla^2$ are discrete. On the other hand, in the large volume expansion, the gaps between the eigenvalues approach zero, which forces the eigenvalue spectrum to be continuous. This continuous spectrum is the source of the integral over the internal momentum
\begin{equation}
    \int\frac{d^2k}{(2\pi)^2}\,.
\end{equation}

\subsection{Comments on general loop amplitudes}
Before closing this section, we shall explain a general strategy one can follow to compute string one-loop amplitudes in Calabi-Yau orientifolds in the large volume expansion. Two and higher loops will have similar structures, but with a little more complications. We shall assume that a perturbative background solution in the closed string theory is obtained to an arbitrary precision, solving
\begin{equation}
    Q_B|\Psi_0^c\rangle+g_s^2\sum_{n=1}^\infty \frac{1}{n!}\mathcal{G}[(\Psi^c_0)^n]^c_{S^2}=0\,.
\end{equation}
For an actual calculation, one will have to truncate the background solution to a finite order. Now, we shall add D-branes and O-planes to the closed string backgrounds, assuming that orientifolding does not induce curvature singularities in the spacetime. For simplicity, we shall further assume that D-branes and O-planes break only half of the spacetime supersymmetry, and that the tadpole is canceled. We shall also, for simplicity, the gauge bundles on D-branes are trivial. If the gauge bundles are non-trivial, we should find a coupled background solution both in the closed string and open string sectors. The tadpole cancellation condition is written as
\begin{equation}
    \Bbb{P} \sum_n\left(\frac{1}{n!}\mathcal{G}\left[(\Psi_0^c)^n;\right]^c_{D^2+\Bbb{RP}^2}\right)=Q_B(\mathfrak{T})\,,\label{eqn:tadpole}
\end{equation}
where $\Bbb{P}$ denotes a projection to the massless states. The tadpole cancellation condition \eqref{eqn:tadpole}, however, does not, in general, imply that the one-point function of massive states vanish 
\begin{equation}
    (1-\Bbb{P}) \left(\sum_n\frac{1}{n!}\mathcal{G}\left[(\Psi_0^c)^n;\right]^c_{D^2+\Bbb{RP}^2}\right)\neq Q_B(\mathfrak{T})\,.
\end{equation}
We shall denote the backreaction caused by the disk and crosscap on the massive states by $\mathfrak{B}$
\begin{equation}
    \mathfrak{B}:=-g_s^2\frac{b_0^+}{L_0^+}(1-\Bbb{P})\left(\sum_n \frac{1}{n!}\left[(\Psi_0^c)^n;\right]^c_{D^2+\Bbb{RP}^2} \right)\,,
\end{equation}
and we shall redefine
\begin{equation}
    \Psi_0^c\mapsto\Psi_0^c+\mathfrak{B}\,.
\end{equation}
For the study of string loops in flux compactifications, one must modify the above tadpole cancellation condition to allow the tadpole dissolved in the flux to cancel the mismatch of the RR tadpole. For example, see \cite{Cho:2023mhw}. 

The next step is to quantize scattering states. Here, we shall assume that there is no D-brane state that is light enough to recoil. If such light D-branes exist, one shall need to modify the discussion that will be presented slightly to take into account the recoils properly \cite{Sen:2025xaj}. In the presence of D-branes, in general, closed and open string scattering states mix at high order in perturbation theory. So, one should solve the mixed linearized equations of motion to quantize scattering states. 
\begin{align}
    &Q_B|\delta\Psi^c\rangle+g_s^2 \sum_{n} \frac{1}{n!}\mathcal{G} \left([\delta\Psi^c(\Psi_0^c)^n;]^c+[(\Psi_0^c)^n;\delta\Psi^o]^c\right)=0\,,\\
    &Q_B|\delta\Psi^o\rangle+g_s\sum_{n}\frac{1}{n!}\mathcal{G} \left([\delta\Psi^c(\Psi_0)^c;]^o+[(\Psi_0^c)^n;\delta\Psi^o]^o\right)=0\,.
\end{align}
Note that to compute scattering amplitudes to order $\mathcal{O}(g_s^2),$ we must find $\delta\Psi^c$ up to $\mathcal{O}(g_s^2)$ terms and $\delta\Psi^o$ up to $\mathcal{O}(g_s)$ terms
\begin{equation}
    \delta\Psi^c=\delta\Psi^{c,(0)}+g_s\delta\Psi^{c,(1)}+g_s^2\delta\Psi^{c,(2)}+\dots\,,
\end{equation}
\begin{equation}
    \delta\Psi^o=\delta\Psi^{o,(0)}+g_s\delta\Psi^{o,(1)}+\dots\,.
\end{equation}

We have now collected all the essential tools to compute the one-loop amplitudes. Suppose that we would like to compute a four-point amplitude. Let us first define a few symbols
\begin{align}
    [\delta\Psi_1\otimes\dots\otimes \delta\Psi_n]':=& \biggr(\sum_{n,\mathcal{P}} \frac{1}{n!}[\delta\Psi^c_{\mathcal{P}_{1,1}}\otimes \dots \otimes \delta\Psi_{\mathcal{P}_{1,l_1}}^c (\Psi_0^c)^n;\delta\Psi^o_{\mathcal{P}_{2,1}}\otimes \dots \otimes \delta\Psi_{\mathcal{P}_{2,l_2}}^o]^c\nonumber\\
    &\,,\sum_n \frac{1}{n!}[\delta\Psi^c_{\mathcal{P}_{1,1}}\otimes \dots \otimes \delta\Psi_{\mathcal{P}_{1,l_1}}^c (\Psi_0^c)^n;\delta\Psi^o_{\mathcal{P}_{2,1}}\otimes \dots \otimes \delta\Psi_{\mathcal{P}_{2,l_2}}^o]^o\biggr)\in (\mathcal{H}^c,\mathcal{H}^o)\,, 
\end{align}
\begin{align}
    \langle [\delta\Psi_1\dots\delta\Psi_n]'| |[\delta\Psi_1\dots \delta\Psi_m]'\rangle:=&  \langle [\delta\Psi_1\dots\delta\Psi_n]'_1|c_0^- |[\delta\Psi_1\dots \delta\Psi_m]'_1\rangle\nonumber\\
    &+\langle [\delta\Psi_1\dots\delta\Psi_n]'_2| |[\delta\Psi_1\dots \delta\Psi_m]'_2\rangle\,,
\end{align}
\begin{equation}
    \{\delta\Psi_1\dots\delta\Psi_n\}':=\{\delta\Psi^c_{\mathcal{P}_{1,1}}\otimes \dots \otimes \delta\Psi_{\mathcal{P}_{1,l_1}}^c (\Psi_0^c)^n;\delta\Psi^o_{\mathcal{P}_{2,1}}\otimes \dots \otimes \delta\Psi_{\mathcal{P}_{2,l_2}}^o\}
\end{equation}
where $\mathcal{P}$ is a partition of a list $[1,\dots,n]$ into two lists $\mathcal{P}_1$ and $\mathcal{P}_2$ with lengths $l_1$ and $l_2,$ respectively. Note that the sum over partitions also sums over $ l_1$. Then, one can compute the four-point amplitude by evaluating
\begin{align}
    \mathcal{A}_4= \{\delta\Psi_1\otimes\delta\Psi_2\otimes\delta\Psi_3\otimes\delta\Psi_4\}'+\sum_{\mathcal{P}^2} \{ \delta\Psi_{\mathcal{P}^2_{1,1}}\delta\Psi_{\mathcal{P}^2_{1,2}}[\delta\Psi_{\mathcal{P}^2_{2,1}}\delta\Psi_{\mathcal{P}^2_{2,2}} ]'\}'\,,
\end{align}
where $\mathcal{P}^2$ is a partition of $[1,2,3,4]$ into the lists of equal length. We have assumed that a diagonal basis of scattering states was found. Note that by construction, every term in the above expression is finite except at physical poles. Although the above expression is simple, it is important to note that, actually, the string vertices and brackets with an apostrophe are highly complicated objects. For general cases, it would not be feasible to evaluate them with pen and paper, and developing an efficient computer algorithm will be necessary.

\section{Correlation functions involving zero momentum operators}\label{sec:corr zero modes}
A crucial subtlety we encounter in the perturbative background of string field theory is that the background solutions come with non-trivial zero-momentum operators. To evaluate the correlation functions with the zero momentum vertex operators, we shall adopt the prescription proposed in \cite{Belopolsky:1995vi}. 

The prescription is rather simple to state. Suppose that a vertex operator in a correlation function takes the following form
\begin{equation}
    V_0=\sum_i c_i \prod_l (X^{i_l})^{p_l}  W_i\,,
\end{equation}
where $c_i$ is a numerical constant, $W_i$ is a regular local operator of the matter CFT. We shall assume for simplicity that all the other vertex operators take the usual form 
\begin{equation}
    V_i e^{ik_i\cdot X}\,.
\end{equation} 
To calculate
\begin{equation}
    \langle V_0 \otimes V_1\otimes V_2\otimes\dots V_n\rangle\,,
\end{equation}
we shall first replace $V_0$ with
\begin{equation}
    \sum_i c_i W_i e^{ik_0\cdot X}\,,
\end{equation}
and compute the CFT correlation function as if $V_0$ is a regular vertex operator. The resulting CFT correlation function will take the following form
\begin{equation}
    \mathcal{C}=\sum_i c_i  f_i(k)\delta^{(d)}(\sum_{j}k_j)\,,\label{eqn:corre1}
\end{equation}
where $d$ is the dimension of the spacetime of interest. Crucially, to arrive at the expression \eqref{eqn:corre1}, we shall not impose the BRST closed conditions even if some of the vertex operators are BRST-closed. At the final step of the calculation, we replace $c_i$ with
\begin{equation}
    c_i \prod_l \left(-i\frac{\p}{\p k_{0,i_l}}\right)^{p_l}\,,
\end{equation}
and set $k_0$ to be zero and impose the BRST conditions if there were any BRST closed vertex operators at the end of the calculation.

A few comments are in order. It is crucial to impose the BRST condition at the end of the calculation. If the BRST condition is imposed before applying the replacement rules, one may find that the result is ambiguous. But when the result is unaffected by imposing the BRST condition early, one can judiciously decide to do so. One main goal of this draft is to calculate loop amplitudes in the large-volume expansion in the non-linear sigma-model background. For the consistency of the BV master action, it is absolutely crucial to identify the ``boundary terms" in each coordinate patch \cite{Frenkel:2025wko}, as the cancelation of such terms guarantee the BV solvability of the action of string field theory. We shall interpret any term with the following structure 
\begin{equation}
    (\sum_i k_i)^l \delta^{(d)}(\sum_i k_i)\,,
\end{equation}
with $l>1,$ and the $n-$th derivatives of the above expressions with respect to a momentum with $n<l$ as the boundary term. It is intuitive to understand such an interpretation, as we can replace the sum of momenta with the derivative with respect to the center of mass $\sum_i k_i\equiv i\p/\p x_0.$ Once the total derivative terms are identified in each coordinate patch, we can drop them in the calculation, as such total derivative terms will cancel upon adding up contributions from different coordinate patches \cite{Frenkel:2025wko}. Finally, we emphasize that it is important to correctly take into account the effect of the local coordinates in the intermediate calculations. Otherwise, one may arrive at an inconsistent result.\footnote{We thank Atakan Hilmi F\i rat for discussions.} 
\subsection{Sample calculations}
We shall perform some sample calculations to better understand the rules of the calculations.

\subsubsection{Disk one point function}
The disk one-point function, where the closed string insertion is the first-order background solution, must represent the first-order term of the volume of the D-brane in the normal coordinate expansion. 

To compute this, we shall proceed as follows
\begin{align}
    \{\Psi_{0,1}^c\}=&\frac{1}{2\pi}\langle c_0^- \Psi_{0,1}^c\rangle_{D^2}\,,\\
    =&\frac{1}{2\pi} \times \left(-\frac{1}{4\pi}\right) \delta g_{ab}\langle c_0^- c\bar{c} e^{-\phi}\psi^ae^{-\bar{\phi}}\bar{\psi}^be^{ik\cdot X}(0)\rangle\,,\\
    =& \frac{C_{D^2}}{8\pi^2} \delta^{(p+1)}(k_{||})\delta g_{ab} D^{ab} e^{ik_\perp\cdot x_{\perp}} \,,
\end{align}
where
\begin{equation}
    \delta^{(p+1)}(k_{||}):=\int d^{p+1}x_0 e^{ik_{||} \cdot x_0}\,. 
\end{equation}
Note that, as explained in \cite{Belopolsky:1995vi}, the on-shell condition for the momentum square shall be imposed at the end of the calculation. 

We shall replace
\begin{equation}
    \delta g_{ab} \mapsto \frac{1}{3}R_{acbd} \frac{\partial}{\partial k^c}\frac{\partial}{\partial k^d}\,.
\end{equation}
After imposing the BRST condition, we find the correct result for the first-order term of the volume in the normal coordinate expansion
\begin{equation}
    -\frac{1}{2}T_p \int d^{p+1}X  \left( \delta g_{\mu\nu}\eta^{\mu\nu}_{||}\right)\,.
\end{equation}

\subsubsection{Disk with C-O}
We shall now insert one closed-string vertex operator and one open-string operator. The closed string insertion will again be provided by the background solution. For the open string, we shall insert the normal bundle deformation represented by
\begin{equation}
    ce^{-\phi}\psi^c e^{ik_{||}\cdot X}\,.
\end{equation}
Note that $c$ ranges only over the normal directions. We shall insert the PCO at the location $p.$ 

Then we compute
\begin{align}
    \mathcal{A}=&    i \delta g_{ab}\left\langle c\bar{c} e^{-\phi}\psi^ae^{-\bar{\phi}}\bar{\psi}^be^{ik\cdot X}(i) \otimes c e^{-\phi}\psi^c e^{ip_{||}\cdot X}(0)\otimes \mathcal{X}(y)\right\rangle\,,\\
    \propto&\delta g_{ab} \langle \partial X_d(y) \otimes e^{ik\cdot X}(i) \otimes e^{ip_{||}\cdot X}(0)\rangle\times D^b_{b'} \langle \psi^a(i)\psi^{b'}(-i)\psi^c(0)\psi^d(y)\rangle\nonumber\\
    &\times \langle e^{-\phi}(i)\otimes e^{-\phi}(-i)\otimes e^{-\phi}(0)\otimes e^{\phi}(y)\rangle\,,\\
    \propto& \delta g_{ab} D^b_{b'}\left( \frac{\eta^{ab'}\eta^{cd}}{2i(-y)}-\frac{\eta^{ac}\eta^{b'd}}{i\times (-i-y)}+\frac{\eta^{ad}\eta^{b'c}}{(i-y)(-i)}\right)\times \frac{(i-y)(-i-y)(-y)}{(2i)\times i\times (-i)}\nonumber\\
    &\times \left(\frac{ik_d}{y-i}+\frac{i(k\cdot D)_d}{y+i}+\frac{2ip_{||}}{y}\right)\times \delta^{(p+1)}\left(k_{||}+p_{||}\right)\,.
\end{align}
As one can check, naively, the two point function depends on the PCO location $y.$ Upon imposing the on-shell conditions, the PCO dependence drops out as expected for the usual amplitudes. However, we shall now show that for the zero-momentum vertex operators with polynomial dependence on $ X$, the PCO dependence is non-trivial.

There are three types of derivatives we shall take. First, two derivatives along the normal directions. Second, one derivative along normal directions and one derivative along parallel directions. Third, two derivatives along the parallel directions. Since there are no two insertions of $k,$ we conclude that the first type of derivatives yields a trivial result. The second type of derivatives yield
\begin{align}
    \mathcal{A}_{(2)}\propto &R_{aibj}\frac{\partial}{\partial k^i_{||}}\frac{\partial}{\partial k^j_\perp} D^b_{b'} \left( \frac{\eta^{ab'}\eta^{cd}}{2i(-y)}-\frac{\eta^{ac}\eta^{b'd}}{i\times (-i-y)}+\frac{\eta^{ad}\eta^{b'c}}{(i-y)(-i)}\right)\times \frac{(i-y)(-i-y)(-y)}{(2i)\times i\times (-i)}\nonumber\\
    &\times \left(\frac{ik_d}{y-i}+\frac{i(k\cdot D)_d}{y+i}+\frac{2ip_{||}}{y}\right)\times \delta^{(p+1)}\left(k_{||}+p_{||}\right)+(i \leftrightarrow j)\,,\\
    =& R_{aibj}D^b_{b'}\left(\frac{i\eta^{ab'}\eta^{cd}}{2y}-\frac{\eta^{ac}\eta^{b'd}}{1-iy}+\frac{\eta^{ad}\eta^{b'c}}{1+iy}\right)\frac{iy(1+y^2)}{2}\times  \frac{-2\eta^j_{d,\perp}}{1+y^2}\times \delta^{(p+1)}_{,i}(k_{||}+p_{||})+(i \leftrightarrow j)\,.
\end{align}
The first term in the parentheses is independent of the PCO location $ y$  and represents the variation of the volume with respect to the embedding coordinate. Let us inspect the rest of the terms more closely
\begin{align}
    \mathcal{A}_{(2)}'=&R_{aibj} \left(-\frac{\eta^{ac}_\perp D^{bd}}{1-iy}-\frac{\eta^{ad}\eta^{bc}}{1+iy}\right) (-iy)\times \frac{1}{2}(\eta-D)^j_d \delta_{,i}^{(p+1)}(k_{||}+p_{||})+(i\leftrightarrow j)\,,\\
    =&\frac{1}{2}R_{aibj} \frac{iy}{1+y^2}\eta^{ac}_{\perp}D^{bj}\delta_{,i}^{(p+1)}(k_{||}+p_{||})+(i\leftrightarrow j)\,.
\end{align}    
Note that we used the symmetry $(a\leftrightarrow b).$ We can check that one way to remove the PCO-dependent term is to average over PCO locations by placing the PCO at $\pm 3/\sqrt{3} $.

Now, finally, let us study the third type of the contractions. We find
\begin{align}
    \mathcal{A}_{(3)}\propto& R_{aibj} D^b_{b'}\left(\frac{i\eta^{ab'}\eta^{cd}}{2y}-\frac{\eta^{ac}\eta^{b'd}}{1-iy}+\frac{\eta^{ad}\eta^{b'c}}{1+iy}\right) \times y(1+y^2)\nonumber\\
    &\times\frac{\partial}{\partial k^i_{||}}\frac{\partial}{\partial k_{||}^j}\left[\left( \frac{ik_d}{y-i}+\frac{i(k\cdot D)_d}{i+y}+\frac{2ip_{d}}{y} \right)\delta^{(p+1)}(k_{||}+p_{||})\right]\,,\\
    =&-R_{aibj}D^{bi} \frac{2y
    ^2}{(1+y^2)^2}\eta^{ac}\delta_{,j}^{(p+1)}+(i\leftrightarrow j)\nonumber\\
    &+R_{aibj}D^b_{b'} \left(-\frac{\eta^{ab'}k^c}{1+y^2}-i\frac{\eta^{ac}}{(1-iy)}\left\{\frac{k^{b'}}{y-i}+\frac{2p^{b'}1}{y}\right\}+i\frac{\eta^{b'c}}{1+iy}\left\{\frac{(k\cdot D)^a}{i+y}+\frac{2p^a}{y}\right\}\right)\delta_{,ij}^{(p+1)}\,,\\
    =&-R_{aibj}D^{bi} \frac{2y
    ^2}{(1+y^2)^2}\eta^{ac}\delta_{,j}^{(p+1)}+(i\leftrightarrow j)\nonumber\\
    &+R_{aibj}D^b_{b'} \left(-\frac{\eta^{ab'}k^c}{1+y^2}+\frac{2\eta^{ac}k^{b'}}{1+y^2}-\frac{4p^{b'}\eta^{ac}}{1+y^2}\right)\delta_{,ij}^{(p+1)}\,.
\end{align}
Note that we used the symmetry $(a\leftrightarrow b).$ Again, we find that the answer depends on the PCO location. Note that we used
\begin{align}
    \mathcal{C}^{ij}_d:=&\frac{\partial}{\partial k^i_{||}}\frac{\partial}{\partial k^j_{||}} \left[\left( \frac{ik_d}{y-i}+\frac{i(k\cdot D)_d}{i+y}+\frac{2ip_{||}}{y} \right)\delta^{(p+1)}(k_{||}+p_{||})\right]\,,\\
    =&\frac{i (\eta+D)^i_d  y}{1+y^2}\delta^{(p+1)}_{,j}+(i\leftrightarrow j)+\left( \frac{ik_d}{y-i}+\frac{i(k\cdot D)_d}{i+y}+\frac{2ip_{d,||}}{y} \right)\delta^{(p+1)}_{,ij}(k_{||}+p_{||})
\end{align}
and
\begin{align}
    (k_{||}+p_{||})^a\delta_{,i}^{(p+1)}(k_{||}+p_{||})+\frac{1}{2}(\eta+D)^a_i\delta^{(p+1)}(k_{||}+p_{||})=0\,,
\end{align}
\begin{align}
    (k_{||}+p_{||})^a\delta_{,ij}^{(p+1)}(k_{||}+p_{||})+\left(\frac{1}{2}(\eta+D)^a_i\delta_{,j}^{(p+1)}(k_{||}+p_{||})+(i\leftrightarrow j) \right)=0\,.
\end{align}

We now encounter a puzzle. As is commonly said, on-shell ``amplitudes" are independent of the PCO locations in the interior of the moduli space. However, we find the explicit dependence on the PCO location. We should understand why this happens for the zero-momentum operators. Physically, we are not computing amplitudes. Instead, as the closed string insertion is a background field, it should be understood as the open-string tadpole in the new background described by $\Psi_{0,1}^c.$ The tadpole itself is an off-shell quantity that is not well defined. What is well defined is the on-shell value of the action in the saddle point. Therefore, there is no contradiction. Nevertheless, we shall understand why, in the case of the open string tadpole, there is an explicit PCO dependence.

Since the disk ``amplitude" with one closed-string and one open-string insertion has no moduli, the PCO location is treated as a parameter. The effect of the change in the PCO location is then given as
\begin{equation}
    \left\langle (\mathcal{X}(y_1)-\mathcal{X}(y_2)) \otimes  \Psi_{0,1}^c(i)\otimes  \delta \Psi^o(0)\right\rangle\,.
\end{equation}
Although $\xi$ is not an allowed operator in the small Hilbert space, a difference of $\xi$ in the insertions is. Therefore, we can write the disk amplitude as
\begin{equation}
    \mathcal{A}=\left\langle \{Q_B,\xi(y_1)-\xi(y_2)\} \otimes \Psi_{0,1}^c(i)\otimes \delta\Psi^o(0)\right\rangle\,.
\end{equation}
By deforming the BRST contour, we obtain
\begin{equation}
    \mathcal{A}= -\left\langle (\xi(y_1)-\xi(y_2)) \otimes Q_B\Psi_{0,1}^c(i)\otimes \delta\Psi^o(0)\right\rangle-\left\langle(\xi(y_1)-\xi(y_2)) \otimes\Psi_{0,1}^c(i)\otimes Q_B\delta\Psi^o\right\rangle\,.\label{eqn:A puzzle}
\end{equation}
Now, we are arriving at a puzzle. The vertex operators we inserted are BRST-closed. Therefore, \eqref{eqn:A puzzle} must vanish as well. On the other hand, direct calculation shows that the amplitude explicitly depends on the PCO location. How can we understand this apparent contradiction?

As was explained in \cite{Belopolsky:1995vi}, in the context of bosonic string field theory, when vertex operators with the polynomial dependence in $X$ are inserted, much care is needed. The crucial subtlety originates from the fact that the correlators involving polynomials in $X$ are distributions rather than a function. In practice, this means the BRST conditions must be imposed at the end of the calculations, after properly treating the Dirac delta in momentum space. Without this careful treatment, one would find a contradiction as was carefully shown in \cite{Belopolsky:1995vi}.

Then, let us compute $\mathcal{A}$ using the representation of \eqref{eqn:A puzzle} to understand the correct treatment of the distribution. Due to the ghost number conservation, the only non-vanishing contributions from the $Q_B$ action are the action of $\eta e^\phi T_F.$ Effectively, what this does is as follows
\begin{align}
     Q_B\Psi_{0,1}^c\equiv&\left(i\sqrt{\frac{2}{\alpha'}}\oint dz \times \eta e^\phi \partial X\cdot \psi +c.c.\right)\left(-\frac{1}{4\pi}\delta g_{ab}c\bar{c}e^{-\phi}\psi^ae^{-\bar{\phi}}\bar{\psi}^b e^{ik\cdot X}\right)\,,\\
     =&\sqrt{\frac{\alpha'}{2}} \left(-\frac{1}{4\pi}\right) k^a\delta g_{ab} \eta c\bar{c}e^{-\bar{\phi}}\bar{\psi}^be^{ik\cdot X} +c.c.\,,\label{eqn:QB c-background}
\end{align}
\begin{align}
    Q_B\delta\Psi^o\equiv& i\sqrt{\frac{2}{\alpha'}}\oint dz \times \eta e^\phi \partial X\cdot \psi  \times  ce^{-\phi}\psi^c e^{ip\cdot X}\,,\\
    =&\sqrt{2\alpha'} p^c \eta c e^{ip\cdot X}=0\,.
\end{align}
Note that we have chosen $c$ to vary only along the transverse directions.

We then compute
\begin{align}
    \mathcal{A}=&\frac{1}{4\pi}\sqrt{\frac{\alpha'}{2}} \left\langle(\xi(y_1)-\xi(y_2)) \otimes k^a \delta g_{ab}\eta c\bar{c}e^{-\bar{\phi}}\bar{\psi}^be^{ik\cdot X}(i)\otimes ce^{-\phi}\psi^c e^{ip\cdot X}(0) \right\rangle+h.c\,,\\
    =&\frac{1}{4\pi}\sqrt{\frac{\alpha'}{2}}\left( \frac{y_2-y_1}{(y_1-i)(y_2-i)} k^a\delta g_{ab}D^{bc} + \frac{y_2-y_1}{(y_2+i)(y_1+i)}k^a\delta g_{ab}\eta^{bc}\right)\delta^{(p+1)}(k_{||}+p)\,.
\end{align}
We shall now, as explained in \cite{Belopolsky:1995vi}, replace $\delta g_{ab}$ with
\begin{equation}
    \frac{1}{3}R_{aibj}\frac{\p}{\p k_i}\frac{\p}{\p k_j}\,,
\end{equation}
to obtain
\begin{align}
    \mathcal{A}=&\frac{1}{4\pi}\sqrt{\frac{\alpha'}{2}}k^a R_{aibj}\left( \frac{y_2-y_1}{(y_1-i)(y_2-i)}  D^{bc} + \frac{y_2-y_1}{(y_2+i)(y_1+i)}\eta^{bc}\right)\delta^{(p+1)}_{,ij}(k_{||}+p)\,.
\end{align}
By using
\begin{equation}
    0=(x\delta (x))''=x\delta(x)''+\delta (x)'\,,
\end{equation}
we find
\begin{align}
    \mathcal{A}=&-\frac{1}{2\pi}\sqrt{\frac{\alpha'}{2}}D^{aj} R_{aibj}\left( \frac{y_2-y_1}{(y_1-i)(y_2-i)}  D^{bc} + \frac{y_2-y_1}{(y_2+i)(y_1+i)}\eta^{bc}\right)\delta^{(p+1)}_{,i}(k_{||}+p)\,,
\end{align}
which does not vanish as claimed. However, again, the above expression denotes the open-string tadpole in a different field basis. Provided that the open-string tadpole vanishes, the above expression vanishes altogether as well.

\section{One-loop amplitudes at zero-th order}\label{sec:zero-th}
In this section, as a warm-up, we shall compute the partition function of the D1-instanton to zeroth order in the curvature expansion.

\subsection{Annulus: D1-D1}
Let us start with the Annulus with both ends on the D1-instanton. We compute
\begin{align}
    f_{00}=& q^{-\frac{1}{2}}\prod_{n} (1-q^n)^{-10}\prod_n (1+q^{n-\frac{1}{2}})^{10} \prod (1-q^n)^2\prod (1+q^{n-\frac{1}{2}})^{-2}=\frac{\vartheta_{00}^4(\tau)}{\eta^{12}(\tau)}\,, 
\end{align}
\begin{equation}
    f_{01}=-q^{-\frac{1}{2}}\prod_{n} (1-q^n)^{-10}\prod_n (1-q^{n-\frac{1}{2}})^{10} \prod (1-q^n)^2\prod (1-q^{n-\frac{1}{2}})^{-2}=-\frac{\vartheta_{01}^4(\tau)}{\eta^{12}(\tau)}\,, 
\end{equation}
\begin{equation}
    f_{10}=-\frac{1}{2}(16+16)\prod_{n} (1-q^n)^{-10}\prod_n (1+q^{n})^{10} \prod (1-q^n)^2\prod (1+q^{n})^{-2}=-\frac{\vartheta_{10}^4(\tau)}{\eta^{12}(\tau)}\,, 
\end{equation}
\begin{equation}
    f_{11}=-\frac{1}{2}(16-16)\prod_{n} (1-q^n)^{-10}\prod_n (1-q^{n})^{10} \prod (1-q^n)^2\prod (1-q^{n})^{-2}=-\frac{\vartheta_{11}^4(\tau)}{\eta^{12}(\tau)}=0\,. 
\end{equation}
As expected, after the spin sum, we find a trivial result. This can be shown by using
\begin{equation}
    \vartheta_{00}^4(\tau)-\vartheta_{01}^4(\tau)-\vartheta_{10}^4(\tau)=0\,.
\end{equation}

\subsection{Möbius}
Let us first compute the Möbius strip $Z_M^{(0)}.$ There are in total 8 Dirichlet directions and 2 Neumann directions. The partition function is then given as
\begin{align}
    f_{00}=&-\hat{q}^{-\frac{1}{2}} \prod (1+\hat{q}^n)^{-8}(1-\hat{q}^n)^{-2}\prod (1-\hat{q}^{n-\frac{1}{2}})^8(1+\hat{q}^{n-\frac{1}{2}})^2\prod (1-\hat{q}^n)^2\prod(1+\hat{q}^{n-\frac{1}{2}})^{-2}\,,\\
    =&-2^4\frac{\vartheta_{0,1}(\hat{\tau})^4}{\vartheta_{1,0}(\hat{\tau})^4}\,,\\
    f_{01}=&\hat{q}^{-\frac{1}{2}}\prod (1+\hat{q}^n)^{-8}(1-\hat{q}^n)^{-2}\prod (1+\hat{q}^{n-\frac{1}{2}})^8(1-\hat{q}^{n-\frac{1}{2}})^2\prod (1-\hat{q}^n)^2\prod(1-\hat{q}^{n-\frac{1}{2}})^{-2}\,,\\
    =&2^4\frac{\vartheta_{0,0}(\hat{\tau})^4}{\vartheta_{1,0}(\hat{\tau})^4}\,,\\
    f_{10}=&\frac{1}{2}(16-16)\prod (1+\hat{q}^n)^{-8}(1-\hat{q}^n)^{-2} (1-\hat{q}^n)^8(1+\hat{q}^n)^2(1-\hat{q}^n)^2(1+\hat{q}^n)^{-2}\,,\\
    =&0\,,\\
    f_{11}=&-\frac{1}{2}(16-16) \prod (1+\hat{q}^n)^{-8}(1-\hat{q}^n)^{-2}(1+\hat{q}^n)^8(1-\hat{q}^n)^{2}(1-\hat{q}^n)^2(1-\hat{q}^n)^{-2}\,,\\
    =&0\,.
\end{align}
The zero mode contribution in the Ramond sector can be understood as follows. There are two kinds of Ramond zero modes
\begin{equation}
    V_{S,1}=\lambda^\alpha ce^{-\phi/2}\Sigma_\alpha\,,\quad V_{S,2}=\lambda_\alpha ce^{-\phi/2}\Sigma^\alpha\,.
\end{equation}
The D1-instanton boundary condition implies that the internal spinor satisfies the following boundary condition
\begin{equation}
    e^{\frac{1}{2}(\pm H_1+v_2H_2+v_3H_3) }=\mp e^{\frac{1}{2}(\pm \bar{H}_1+v_2\bar{H}_2+v_3\bar{H}_3)}\,.
\end{equation}
Similarly, the orientifolding maps the spin fields as
\begin{equation}
    c e^{-\phi/2} \Sigma_\alpha\mapsto \bar{c} e^{-\bar{\phi}/2}\overline{\Sigma}_\alpha\,,\quad ce^{-\phi/2}\Sigma^\alpha \mapsto \bar{c}e^{-\bar{\phi}/2}\overline{\Sigma}^\alpha\,.
\end{equation}
We shall normalize $F$ such that
\begin{equation}
    (-1)^F ce^{-\phi/2}\Sigma_\alpha=ce^{-\phi/2}\Sigma_\alpha\,,\quad (-1)^F ce^{-\phi/2}\Sigma^\alpha =-ce^{-\phi/2}\Sigma^\alpha\,.
\end{equation}
Then, in the Ramond sector, among 16 GSO even spacetime fermions, 8 of them are even under the orientifolding, and the rest are odd under the orientifolding. Similarly, among 16 GSO odd spacetime fermions, 8 of them are even under the orientifolding, and the rest are again odd. Therefore, the zero mode contributions to $f_{10}$ and $f_{00}$ both cancel. After summing over the spin structures, we find
\begin{equation}
    Z_M=\frac{1}{(8\pi^2\alpha't)}4\,.
\end{equation}
Note
\begin{equation}
    4=2^4\times \frac{1}{2}\times \frac{1}{2}\,,
\end{equation}
where two factors of $1/2$ are included due to the GSO projection and the orientifold projection. Note that we included the integral of the momentum modes along the Neumann directions. The final result agrees with the expected answer. As there are eight bosonic zero modes and eight fermionic zero modes that are projected in both under the GSO and the orientifolding, the expected zero mode contribution to $Z_M$ is $8-8/2,$ which is exactly what we found.

Now we shall perform the same calculation, but by computing the new supersymmetric index of the internal CFT. The internal part of the CFT at a generic spin structure is determined as
\begin{equation}
   Z_{\alpha,\beta}:=  2^2 (-1)^{\alpha+\beta}\frac{\vartheta_{\alpha,\beta+1}(\hat{\tau})^2\vartheta_{\alpha,\beta}(\hat{\tau})}{\vartheta_{1,0}(\hat{\tau})^2\eta(\hat{\tau})^3}\,.
\end{equation}
We can rewrite the theta function at a generic spin structure as
\begin{equation}
    Z_{\alpha,\beta}=2^2 (-1)^{\alpha+\beta} \frac{\vartheta_{0,1}(\alpha/2+\beta\hat{\tau}/2;\hat{\tau})^2\vartheta_{0,0}(\alpha/2+\beta\hat{\tau}/2;\hat{\tau})}{\vartheta_{1,0}(\hat{\tau})^2\eta(\hat{\tau})^3}\,.
\end{equation}
The new supersymmetric index is computed by 
\begin{equation}
    \frac{1}{2\pi } \frac{d }{d\beta} Z_{\alpha,\beta}|_{(\alpha,\beta)=(1,1)}= 4\,.
\end{equation}
As it should, we reproduce $Z_M=4.$

\subsection{Annulus: D1-D9}

We shall now compute the annulus $Z_A^{(0)}.$ On the one end of the boundary we impose the D1-instanton boundary condition, and on the other boundary we impose the D9-brane boundary condition. Therefore, there are in total 8 Neumann-Dirichlet boundary conditions and 2 Neumann-Neumann boundary conditions. We compute
\begin{align}
    f_{00}=&2^4q^{\frac{1}{2}}\prod (1-q^{n-\frac{1}{2}})^{-8}(1-q^n)^{-2}\prod(1+q^n)^8 (1+q^{n-\frac{1}{2}})^2\prod (1-q^n)^2\prod (1+q^{n-\frac{1}{2}})^{-2}\,,\\
    =&\frac{\vartheta_{1,0}(\tau)^4}{\vartheta_{0,1}(\tau)^4}\,,\\
    f_{01}=&-(8-8) q^{\frac{1}{2}}\prod (1-q^{n-\frac{1}{2}})^{-8}(1-q^n)^{-2}(1-q^n)^{-8}(1-q^{n-\frac{1}{2}})^2(1-q^n)^2(1-q^{n-\frac{1}{2}})^{-2}\,\\
    =&0\,,\\
    f_{10}=&-\frac{1}{2}(1+1) \prod (1-q^{n-\frac{1}{2}})^{-8}(1-q^n)^{-2} (1+q^{n-\frac{1}{2}})^{8}(1+q^n)^2(1-q^n)^2(1+q^n)^{-2}\,,\\
    =&-\frac{\vartheta_{0,0}(\tau)^4}{\vartheta_{0,1}(\tau)^4}\,,\\
    f_{11}=&-\frac{1}{2}(1-1)\prod(1-q^{n-\frac{1}{2}})^{-8}(1-q^n)^{-2}(1-q^{n-\frac{1}{2}})^8(1-q^n)^2(1-q^n)^2(1-q^n)^{-2}\,,\\
    =&0\,.
\end{align}
By summing over the Chan-Paton factors, we find 
\begin{equation}
    Z_A=16\frac{1}{(8\pi^2\alpha't)}\frac{\vartheta_{1,0}(\tau)^4-\vartheta_{0,0}(\tau)^4}{\vartheta_{0,1}(\tau)^4}=-\frac{16}{8\pi^2\alpha't}\,.
\end{equation}
Note
\begin{equation}
    16=32\times \frac{1}{2}\times\frac{1}{2}\times 2\,,
\end{equation}
where $32$ is due to the 32 branes in the upstairs picture, two factors of $1/2$ are due to the GSO projection and the orientifolding, and the factor of $2$ is due to two different orientations. Note that we included the integral of the momentum modes along the Neumann-Dirichlet directions. As one can check, the closed string tadpole is appropriately cancelled.

\section{First order: Annulus D1-D1}\label{sec:first d1-d1}
As a warm-up, we shall compute the annulus with both ends on the D1-instanton.
\subsection{Normalization of the one-point function}
We shall treat the annulus as a square 
\begin{equation}
    [0,\pi]\times[0,2\pi t_o]\,,
\end{equation}
in the open string channel. We shall construct the annulus by quotienting a torus
\begin{equation}
    [0,2\pi]\times[0,2\pi t_o]\,,
\end{equation}
by the following involution
\begin{equation}
    \mathcal{I}(z)=2\pi -\bar{z}\,.
\end{equation}
The boundaries are located at $\text{Re}(z)=0$ and $\text{Re}(z)=1/2.$ Due to the $\Bbb{Z}_2$ symmetry
\begin{equation}
    z\mapsto \frac{1}{2}-\bar{z}\,,
\end{equation}
for $\text{Re}(z)$ we shall only integrate over $[0,1/4].$

We write the annulus one-point function in the interior of the moduli space as
\begin{equation}
    \mathcal{A}^{D1-D1}=\frac{1}{4}\sum_s \int dt_o \int_0^{1/4}dx \frac{1}{8\pi^2\alpha' t_o}\biggr\langle\mathcal{B}_t \mathcal{B}_x c\bar{c} V^{-1,-1}(2\pi x)\mathcal{X}(y_1)\mathcal{X}(y_2) \biggr\rangle_s\,,\label{eqn:A int d1d1}
\end{equation}
where the Beltrami differentials are given as
\begin{equation}
    \mathcal{B}_x=-2\pi \oint_{C_1}dzb(z)+2\pi\oint_{C_1}d\bar{z}\bar{b}(\bar{z})\,,
\end{equation}
\begin{equation}
    \mathcal{B}_t=2\pi i\oint_{C_3} dzb(z)+2\pi i\oint_{C_3}d\bar{z}\bar{b}(\bar{z})\,.
\end{equation}
As in the Möbius one-point function, we shall average over the PCO locations
\begin{equation}
    \oint_{C_1}\frac{dp_1}{p_1}\mathcal{X}(p_1)\,,\quad -\oint_{C_1}\frac{d\bar{p}_2}{\bar{p}_2}\overline{\mathcal{X}}(\bar{p}_2)\,.
\end{equation}

The choice of PCOs we make for the interior of the moduli space is not compatible with the choice of PCOs we make for the Feynman regions. Therefore, on the boundary of the vertex regions, we shall perform vertical integration. We shall normalize the vertical integration as
\begin{align}
    \mathcal{V}^{D1-D1}=&\frac{1}{4}\sum_s \int dm_b \frac{1}{8\pi^2\alpha't}\biggr\langle \mathcal{B}_{m_b} c\bar{c} V^{-1,-1}(2\pi x)\left[\left(\xi(y_1)-\xi(W_1) \right) \mathcal{X}(y_2)+\left(\xi(y_2)-\xi(W_2)\right)\mathcal{X}(W_1)\right] \biggr\rangle_s\,.
\end{align}
Note, however, upon the spin sum, \eqref{eqn:A int d1d1} vanishes, and so does the vertical integral. We shall not, therefore, explicitly evaluate the vertical integral for the annulus diagram with both ends on the D1-instanton.

\subsection{Even spin structure}

\begin{align}
    \mathcal{A}_{NS}^{(1)}=&\frac{1}{8\pi\alpha'}\sum_{s=(0,0),(0,1)}\int dt_o\int_0^{\frac{1}{4}}dx \frac{1}{8\pi^2\alpha't_o} \times 2(2\pi)^2e^{i\vartheta}\eta^2(\tau)\nonumber\\
    &\times \left\langle \delta g_{ab} \left(\partial X^a-i\frac{\alpha'}{2}k\cdot \psi\psi^a\right)\left(\bar{\partial}X^b-i\frac{\alpha'}{2}k\cdot\bar{\psi}\bar{\psi}^b\right)e^{ik\cdot X}(2\pi x)\right\rangle_s\,,\\
    =&\sum_{s=(0,0),(0,1)}\int dt_o\int_0^{\frac{1}{4}}dx \frac{(-1)^{\alpha+\beta}e^{i\vartheta}}{8\pi\alpha'^2t_o}\delta g_{ab} e^{-(k\cdot D\cdot k)\mathcal{G}_{T^{1,1}}(4\pi x)}\frac{\vartheta_{\alpha\beta}^4}{\eta^{12}}\biggr[ 2\eta_{||}^{ab}\mathcal{G}''_{T^{1,1}}(4\pi x)\nonumber\\
    &-(k\cdot D)^a(k\cdot D)^b(\mathcal{G}'_{T^{1,1}}(4\pi x))^2-\frac{\alpha'^2}{4}((k\cdot D\cdot k)D^{ab}-(k\cdot D)^a(k\cdot D)^b)(\mathcal{G}_{F,T^{1,1}}^{\alpha\beta}(4\pi x)^2)\biggr]\nonumber\\&\times \delta^{(2)}(k_{||})\,.
\end{align}
We shall now replace
\begin{equation}
    \delta g_{ab}=\frac{1}{3}R_{aibj}\frac{\partial}{\p k^i}\frac{\p}{\p k^j}\,,
\end{equation}
or, equivalently, 
\begin{equation}\label{eqn:metric pert rule}
    \delta g_{ab}= -R_{a\bar{b}i\bar{j}}\frac{\partial}{\partial k^i}\frac{\partial}{\partial k^{\bar{j}}}\,.
\end{equation}

Before explicitly evaluating the CFT correlators, a few comments are in order. The momenta in the CFT correlation function represent the contractions of $X$'s appearing in the CFT correlator. For example, $k$ in the fermionic correlator is induced from contracting $\partial X$ in the PCO against $e^{ik\cdot X}$ of the background field. Derivative of the Dirac delta with respect to the parallel momenta, on the other hand, represents the insertion of the zero modes of $X.$ With this in mind, let's inspect one of the possible terms that can arise by naively applying the replacement rule \eqref{eqn:replace rule}
\begin{equation}
    \eta_{\perp}^2k_{||}^2 (\mathcal{G}_{F,T}^{\alpha,\beta})^2 \delta_{,ij}^{(2)}(k_{||})\,,\label{eqn:term to be dropped}
\end{equation}
which arises from the contraction of the four-fermi term with two derivatives acting on the Dirac delta. We argue that such a term must be absent. For the term \eqref{eqn:term to be dropped} to be generated, there must be free bosons along the parallel directions that are not yet contracted so that they can be used for the zero mode insertions. However, the four-fermi term of the background field already saturates the bosons, and therefore, there are not enough bosonic zero modes. We therefore conclude that a term like \eqref{eqn:term to be dropped} must not be included for the calculation of interest. Dropping terms like \eqref{eqn:term to be dropped} leads to the result consistent with the results one can find by directly contracting all the bosons manually. However, we shall continue to use the prescription of \cite{Belopolsky:1995vi} with care, as it makes manifest the boundary contributions that must be dropped in the formulation of string field theory for non-linear sigma model backgrounds \cite{Frenkel:2025wko}.

Collecting all the non-trivial terms, we find
\begin{align}
    \mathcal{A}_{NS}^{(1)}=&-\sum_s\int dt_o \int_0^{\frac{1}{4}}\frac{(-1)^{\alpha+\beta}e^{i\vartheta}}{8\pi\alpha'^2 t_o}( 2R_{a\bar{b}i\bar{j}}) \frac{\vartheta_{\alpha\beta}^4(\tau)}{\eta^{12}(\tau)} \biggr[ 2\eta_{||}^{a\bar{b}} \mathcal{G}''_{T^{1,1}}(4\pi x)\delta_{,i\bar{j}}^{(2)}\nonumber\\
    &-\biggr\{8\eta^{a\bar{b}}_{||}\eta^{i\bar{j}}_{||}\mathcal{G}''_{T^{1,1}}(4\pi x)\mathcal{G}_{T^{1,1}}(4\pi x)+4\eta_{||}^{a\bar{j}}\eta_{||}^{i\bar{b}}(\mathcal{G}'_{T^{1,1}}(4\pi x))^2\nonumber\\
    &\quad+\frac{\alpha'^2}{4}\left(8\eta_{||}^{i\bar{j}}\eta^{a\bar{b}}_{||}-4\eta_{||}^{a\bar{j}}\eta_{||}^{i\bar{b}}\right)(\mathcal{G}_{F,T^{1,1}}^{\alpha\beta}(4\pi x))^2\biggr\}\delta^{(2)}\biggr]\,,\\
\end{align}
We shall show that the NS sector contribution completely cancels the R sector contribution, as it should. Due to the abstruse identity
\begin{equation}
    \sum_{s} (-1)^{\alpha+\beta}\vartheta_{\alpha\beta}^4(\tau)=0\,,
\end{equation}
the bosonic contractions can be shown to vanish. The Green's function for the worldsheet fermions satisfies the following identity
\begin{equation}
    (\mathcal{G}_{F,T^{1,1}}^{\alpha\beta}(4\pi x))^2\propto \frac{\vartheta_{\alpha\beta}''(0|\tau)}{\vartheta_{\alpha\beta}(0|\tau)}-\partial_z^2 \log\vartheta_{11}(z|\tau)|_{z=2x}\,.\label{eqn:G2}
\end{equation}
The second term in \eqref{eqn:G2} vanishes upon the spin sum, for the same reason as in the bosonic contractions. To show that the first term in \eqref{eqn:G2} vanishes, we use the following relation
\begin{equation}
    \vartheta_{\alpha\beta}''(z|\tau)=c \partial_\tau \vartheta_{\alpha\beta}(z|\tau)\,,
\end{equation}
where $c$ is a constant. Therefore, we find
\begin{equation}
    \sum_s (-1)^{\alpha+\beta}\vartheta_{\alpha\beta}^3\vartheta_{\alpha\beta}''(0|\tau)=0\,,
\end{equation}
as well. Hence, we conclude that the annulus with both ends on the D1-instanton vanishes as it should.

\section{First order: Möbius}\label{sec:first Mob}
In this section, we shall evaluate the one-loop amplitudes at the first order in $\alpha'$ expansion.

\subsection{Möbius one-point function}
\paragraph{Choice of PCOs}
As we explained in \S\ref{sec:corr zero modes} and, to the best of the author's knowledge, first observed in \cite{Belopolsky:1995vi}, the inclusion of polynomials in $x$ in the vertex operators leads to the breakdown of the decoupling of BRST-exact states. Although string field theory is still consistently formulated, as the main identities remain true and hence the string field theory action solves the BV master equation \cite{Frenkel:2025wko}, this poses a complication in the treatment of the picture-changing operators. 

Provided that the boundary conditions on the location of the PCOs are not changed, one can normally argue that the PCO locations in the interior of the moduli space do not affect the result for on-shell amplitudes as follows. The change in the location of PCOs can be represented by inserting the following operator instead of the PCO
\begin{equation}
    \{Q_B,\xi(W)-\xi(p)\}\,.
\end{equation}
One can then deform the BRST contour so that the BRST operator acts on the BRST-closed vertex operators and the Beltrami differentials. Since we are assuming that the PCO locations are not deformed on the boundary of the moduli space, $Q_B$ acts trivially on the Beltrami differentials. Therefore, one concludes that the PCO location in the interior of the moduli space does not affect the string vertices with on-shell string fields.

The subtlety that arises when computing string vertices with the string fields in the polynomials in $X$ is as follows. The correct prescription is to impose the BRST conditions for string fields at the end of the CFT calculation \cite{Belopolsky:1995vi}, as incorrectly imposing the BRST condition early on in the calculation induces ambiguities that render the string vertex not well-defined. Once the calculation is complete, the string vertex takes the following form
\begin{equation}
    \mathcal{V}= \sum_j F_{i_1\dots i_j}\delta^{(d)}_{,i_1\dots i_j}(\sum_l k_l)\,.
\end{equation}
One can show that the BRST decoupling argument leads to the vanishing of $ F$ upon imposing momentum conservation. However, although the coefficients that multiply the Dirac delta vanish provided that the momentum conservation condition is imposed, their derivatives in the conserved momentum may not. If the proper derivative of the coefficients does not vanish on-shell, that can lead to the failure of the BRST decoupling argument as we reviewed and illustrated in \cite{Belopolsky:1995vi}.

What this subtlety due to the failure of the BRST decoupling argument entails is that the string vertices may actually depend on the locations of PCOs in the interior of the moduli space, even if all the string fields are on-shell. Therefore, we need to digress on the proper choice of PCOs. 

First, on the physical ground, we must emphasize that the location of the PCOs is an off-shell data that must not affect the proper on-shell observables had we formulated the problem correctly. Also, we note that in the formulation of the supersymmetric non-linear sigma model, the deformation of the worldsheet action is given by the $(0,0)$ picture of the background field in the NS-NS sector. Therefore, a natural choice for the PCO locations in the interior of the moduli space is to average over PCOs around the vertices such that all vertex operators are in the $(0,0)$ picture. 

Therefore, we shall use $(0,0)$ picture vertex operators in the interior of the moduli space, while fixing the mismatch of the PCO locations on the boundary of the moduli space with the vertical integrations \cite{Sen:2015hia}.

\paragraph{Normalization of the one-point function}
We shall determine the normalization of the Möbius one-point function. Let us denote that the vertex operator we shall insert by
\begin{equation}
    c\bar{c} V^{-1,-1}(x)\,.
\end{equation}

We shall treat the Möbius strip as a square 
\begin{equation}
    [0,\pi]\times[0,2\pi t]\,,
\end{equation}
with the identification
\begin{equation}
    (x,y)\equiv (\pi-x,y+2\pi t)\,.
\end{equation}
Under the complexified coordinate $z:= x+iy,$ the identification is given as
\begin{equation}
    z\equiv -\bar{z}+2\pi \tau\,,
\end{equation}
with $\tau=1/2+it.$ The boundaries are located at $x=0$ and $x=\pi,$ whereas the crosscap is located at $x=\pi/2.$

First, there are a number of overall factors we need to take into account. The integral of the momentum modes along the Neumann directions introduces a factor of
\begin{equation}
    \frac{1}{8\pi^2\alpha' t}
\end{equation}
\cite{Polchinski}. We shall also introduce an overall factor $1/2$ to take into account the GSO projection, and an additional factor of $1/2$ to take into account the $Z_2$ symmetry $x\equiv 1/2-x$ of the Möbius strip diagram.

To compute the Möbius one-point function, we should therefore evaluate the following amplitude
\begin{equation}
   \frac{1}{4}\sum_s \int dt \int dx \frac{1}{8\pi^2\alpha't} \biggr\langle \mathcal{B}_t\mathcal{B}_x c\bar{c} V^{-1,-1}(2\pi x) \mathcal{X} (y_1)\mathcal{X}(y_2)\biggr\rangle_s\,,
\end{equation}
where $\mathcal{B}_m$ is the Beltrami differential defined as
\begin{equation}
    \mathcal{B}_m:=  \sum_{C_{nm}}\left( \oint dz_m \frac{\partial z_m}{\partial m}\biggr|_{z_n} b - \oint d\bar{z}_m \frac{\partial \bar{z}_m}{\partial m}\biggr|_{\bar{z}_n} \bar{b} \right)\,,\label{eqn:belt}
\end{equation}
and $y_1$ and $y_2$ are the locations of the PCO insertions. Note that $\oint$ contains a factor of $1/(2\pi i),$ and the contour is taken counter-clockwise. 

The Beltrami differential \eqref{eqn:belt} is defined in terms of the transition functions. First, we shall treat the Möbius strip as a quotient of a torus $z\equiv z+2\pi \equiv z+2\pi\tau$ by
\begin{equation}
    I(z)=2\pi-\bar{z}\,.
\end{equation}
Equivalently, we can treat the Möbius strip as a strip
\begin{equation}
    [\pi,2\pi]\times [0,2\pi t]\,,
\end{equation}
where the following conditions are imposed
\begin{equation}
    \bar{z}=2\pi \tau-z\,.
\end{equation}
We shall divide the punctured Möbius into three regions: a disk $D_1$ with the local coordinate $w$ around the closed string puncture, a sphere $S_1$ with three holes $C_1,~C_2,~C_3,$ and a sphere $S_2$ with two holes $C_2,~C_3.$ We shall denote the coordinates of $S_1$ and $S_2$ by $z_1$ and $z_2.$ Also, we shall identify the boundary of $D_1$ with $C_1.$ $C_2$ and $C_3$ are given as lines $[\pi,2\pi ]\times y^C_i.$ The transition functions are then given as
\begin{align}
  C_1&: \qquad  w= f(t) (z_1-2\pi x)\,,\\
  C_2&: \qquad z_1=\bar{z}_2\,,\\
  C_3&: \qquad z_2= \bar{z}_1-2\pi \tau \,,
\end{align}
with $z_1=z.$ 

We find
\begin{align}
    B_x=& -2\pi f(t) \oint_{C_1} dw b(w) +2\pi \bar{f}(\bar{t})\oint_{C_1} d\bar{w} \bar{b}(\bar{w})\,,\\
    =&-2\pi \oint_{C_1} dz b(z) +2\pi \oint_{C_1} d\bar{z}\bar{b}(\bar{z})\,,
\end{align}
\begin{align}
    B_t=&2\pi i \oint_{C_3} dz b(z)+2\pi i \oint_{C_3} d\bar{z} \bar{b}(\bar{z})+\dots\,,
\end{align}
where $\dots$ contain the derivatives of $f(t)$ that would not affect our calculations. 

We shall also make a convenient choice of PCOs. As we previously mentioned, we will raise the picture of $V^{-1,-1}$ to $V^{0,0}.$ We shall impose this picture-raising operation by taking the zero modes of the PCOs. This is equivalent to averaging over the PCO locations as 
\begin{equation}
    \oint_{C_1}\frac{dy_1}{y_1} \mathcal{X}(y_1)\,,\quad -\oint_{C_1}\frac{d\bar{y}_2}{\bar{y}_2} \overline{\mathcal{X}}(\bar{y_2})\,.
\end{equation}
As it is manifest, this choice of PCO locations is incompatible with the PCO locations of the Feynman regions. 

To address the PCO mismatch, we shall perform vertical integration \cite{Sen:2015hia}. We shall parametrize the boundary of the vertex region of the moduli space with $m_b.$ Then, the corresponding Beltrami differential is given as
\begin{align}
    B_{m_b}=&  \oint_{C_1} dw b(w) \partial_{m_b}\left(f(t)(z_1 -2\pi x)\right)+2\pi \oint_{C_3} dz b(z) \partial_{m_b}\tau+c.c.\,,\\
    =&-2\pi \partial_{m_b}x \oint_{C_1} dz b(z)+\frac{\partial_{m_b}f(t)}{f(t)}\oint_{C_1} dz (z-2\pi x) b(z)+2\pi\partial_{m_b} \tau\oint_{C_3}dzb(z) +c.c.\,.
\end{align}

To perform the vertical integration, we shall move the PCO at $y_1$ to $W_1,$ and sequentially move the PCO at $y_2$ to $W_2.$ Then the vertical integration is given as
\begin{align}
    \mathcal{V}=&\frac{1}{4}\sum_s \int dm_b \frac{1}{8\pi^2\alpha't}\biggr\langle \mathcal{B}_{m_b} c\bar{c} V^{-1,-1}(2\pi x)\left[\left(\xi(y_1)-\xi(W_1) \right) \mathcal{X}(y_2)+\left(\xi(y_2)-\xi(W_2)\right)\mathcal{X}(W_1)\right] \biggr\rangle_s\,.
\end{align}

Using the relation
\begin{equation}
    V^{-1,-1}= e^{-\phi} e^{-\bar{\phi}} \mathcal{W}\,,
\end{equation}
we can decompose the integrand of the moduli integral into the ghost and matter contributions. 

Let us first compute the ghost contribution to the integrand. We shall decompose the integrand into the interior contribution $\mathcal{I}$ and the vertical integration contribution $\mathcal{V}.$ We write
\begin{align}
    \mathcal{I}_{gh}:= \langle \mathcal{B}_t \mathcal{B}_x c\bar{c}(2\pi x)\rangle_s\,, 
\end{align}
\begin{equation}
    \mathcal{I}_{m}:=- \oint_{C_1} \frac{dy_1}{y_1} \oint_{C_1}\frac{d\bar{y}_2}{\bar{y}_2} \langle V^{-1,-1}(2\pi x) \mathcal{X}(y_1) \mathcal{X}(y_2)\rangle_s\,,
\end{equation}
\begin{equation}
    \mathcal{V}_{gh}:=-\oint_{C_1} \frac{dy_1}{y_1}\oint\frac{d\bar{y}_2}{\bar{y}_2} \langle \mathcal{B}_{m_b} c\bar{c} e^{-\phi}e^{-\bar{\phi}}(2\pi x) \left[\left(\xi(y_1)-\xi(W_1) \right) \mathcal{X}(y_2)+\left(\xi(y_2)-\xi(W_2)\right)\mathcal{X}(W_1)\right] \rangle_s\,,
\end{equation}
\begin{equation}
    \mathcal{V}_{m}:=\langle \mathcal{W}(2\pi x)\rangle_s\,.
\end{equation}

Let us first compute $\mathcal{I}_{gh}$ and $\mathcal{I}_m.$ For the ghost correlator, we find
\begin{align}
    \mathcal{I}_{gh}=&2\pi\langle \mathcal{B}_t (c-\bar{c})(2\pi x)\rangle_{s}\,,\\
    =&2(2\pi)^2 e^{i\vartheta}\eta (\tau)^2\,,
\end{align}
modulo the phase. Note that we normalized the b-c ghost partition function as
\begin{equation}
    \langle b_0 c_0\rangle= e^{i\vartheta} \eta(\tau)^2\,.
\end{equation}
As one can check, modulo the overall phase, our ghost correlator agrees with the convention of \cite{Polchinski}.\footnote{In bosonic string, following \cite{Polchinski}, we can start with the measure $-\int\frac{dt}{2 t} \int d^2z V^{0,0}(z),$ with an appropriate insertion of ghosts. By integrating over the vertical direction, we arrive at $-\int \frac{dt}{2}\int dx 2(2\pi)^2 V^{0,0}(2\pi x). $} 

\subsection{Interior contribution}
\subsubsection{CFT correlator: Even spin structure}

We shall compute $\mathcal{I}_m$
\begin{align}
    \mathcal{I}_{m}=&-\oint_{C_1}\frac{dy_1}{y_1}\oint_{C_2}\frac{d\bar{y}_2}{\bar{y}_2}\langle V^{-1,-1}(2\pi x) \mathcal{X}(2\pi y_1)\mathcal{X} (2\pi y_2)\rangle_s\,,\\
    =&\frac{1}{2\pi\alpha'}\biggr\langle \delta g_{ab} \left(\partial X^a-i\frac{\alpha'}{2}k\cdot \psi \psi^a\right)\left(\bar{\p}X^b-i\frac{\alpha'}{2}k\cdot\bar{\psi}\bar{\psi}^b\right)e^{ik\cdot X}\biggr\rangle_s\,,\\
    =&\frac{1}{2\pi\alpha'} \exp\left(-k_{||}^2\mathcal{G}_{NN}(4\pi x)-k_{\perp}^2\mathcal{G}_{DN}(4\pi x)\right)\delta g_{ab}\biggr[  \biggr( 2\eta_{||}^{ab}\mathcal{G}''_{T^{1,2},a}(4\pi x)\nonumber\\
    &-(k^a_{||}\mathcal{G}_{NN}'(4\pi x)+k_{\perp}^a\mathcal{G}_{DN}'(4\pi x))(k^b_{||}\mathcal{G}_{NN}'(4\pi x)+k_{\perp}^b\mathcal{G}_{DN}'(4\pi x))\biggr)\nonumber\\
    &-\frac{\alpha'^2}{4} \left( (\eta_{||}^{ab}\mathcal{G}_{F,T}^{\alpha\beta}-\eta^{ab}_{\perp}\mathcal{G}_{F,T}^{\alpha,\beta+1}) (k_{||}^2\mathcal{G}_{F,T}^{\alpha\beta}-k_{\perp}^2 \mathcal{G}_{F,T}^{\alpha,\beta+1}\right)(4\pi x -2\pi)\nonumber\\
    &+\frac{\alpha'^2}{4}\left(k_{||}^a\mathcal{G}_{F,T}^{\alpha,\beta}(4\pi x)-k_{\perp}^a\mathcal{G}_{F,T}^{\alpha,\beta+1}(4\pi x)\right)\left(k_{||}^b\mathcal{G}_{F,T}^{\alpha,\beta}(4\pi x)-k_{\perp}^b\mathcal{G}_{F,T}^{\alpha,\beta+1}(4\pi x)\right)\biggr]\nonumber\\
    &\times (-1)^{\alpha+\beta+1}\frac{2^4\vartheta_{\alpha,\beta+1}(\hat{\tau})^4}{\vartheta_{1,0}(\hat{\tau})^4\eta(\hat{\tau})^2}\times\delta^{(2)}(k_{||})\,,
\end{align}
where we used
\begin{align}
    \langle \psi(2\pi z_1)\bar{\psi}(2\pi \bar{z}_2)\rangle_{M,DN}^{\alpha\beta}=&-i\mathcal{G}_{F,T}^{\alpha,\beta+1}(2\pi z_1;2\pi-2\pi\bar{z}_2)\,,\\
    =&-i(\mathcal{G}_{F,T^{1,2}}^{\alpha,1}(2\pi z_1;2\pi-2\pi\bar{z}_2)+(-1)^\beta\mathcal{G}_{F,T^{1,2}}^{\alpha,1}(2\pi z_1;2\pi-2\pi\bar{z}_2+2\pi \hat{\tau}))\,,
\end{align}
and
\begin{equation}
    2\mathcal{G}_{F,T^{1,2}}^{\alpha,1}(2\pi z_1;2\pi-2\pi\bar{z}_2)=\mathcal{G}_{F,T^{1,1}}^{\alpha,\beta}(2\pi z_1;2\pi-2\pi\bar{z}_2)+\mathcal{G}_{F,T^{1,1}}^{\alpha,\beta+1}(2\pi z_1;2\pi-2\pi\bar{z}_2)\,.
\end{equation}
Note that the b-c ghost correlator gives
\begin{equation}
    2(2\pi)^2e^{i\vartheta}\eta(\tau)^2\,.
\end{equation}

Collecting every term, we find
\begin{equation}
    \mathcal{I}=\mathcal{I}_b+\mathcal{I}_f\,,
\end{equation}
where $\mathcal{I}_b$ is a sum over bosonic contractions, and $\mathcal{I}_f$ is a sum over fermionic contractions
\begin{align}
    \mathcal{I}_b=&\frac{1}{4}\sum_s\int dt_o\int dx \frac{2(2\pi)^2}{16\pi^3\alpha'^2t_o}(-1)^{\alpha+\beta+1} \delta g_{ab} \frac{2^4\vartheta_{\alpha,\beta+1}^4(\hat{\tau})}{\vartheta_{1,0}^4(\hat{\tau})}\exp\left(-k_{||}^2\mathcal{G}_{NN}(4\pi x)-k_{\perp}^2\mathcal{G}_{DN}(4\pi x)\right)\nonumber\\
    &\times \biggr[ 2\eta_{||}^{ab}\mathcal{G}_{T^{1,2}}''(4\pi x)-(k^a_{||}\mathcal{G}_{NN}'(4\pi x)+k_{\perp}^a\mathcal{G}_{DN}'(4\pi x))(k^b_{||}\mathcal{G}_{NN}'(4\pi x)+k_{\perp}^b\mathcal{G}_{DN}'(4\pi x))\biggr)\biggr]\delta^{(2)}(k_{||})\,,
\end{align}
To evaluate $\mathcal{I}_b,$ we shall replace $\delta g_{ab}$ with
\begin{equation}
    \delta g_{ab}\mapsto -R_{a\bar{b}i\bar{j}}\frac{\p}{\p k_{i}}\frac{\p}{\p k_{\bar{j}}}\,,\label{eqn:replace rule}
\end{equation}
and set $k=0$ and impose BRST conditions at the end of the calculation.

We, therefore, find
\begin{align}
    \mathcal{I}_b=&-\int dt_o\int dx \frac{2e^{i\vartheta}}{\pi\alpha'^2t_o} R_{a\bar{b}i\bar{j}}\biggr[ 4\eta_{||}^{a\bar{b}} \mathcal{G}_{T^{1,2}}''(4\pi x)\delta_{,i\bar{j}}^{(2)}-16\eta_{||}^{a\bar{b}}\eta^{i\bar{j}}_{||} (\mathcal{G}'_{T^{1,2}}(4\pi x))^2\delta^{(2)}\nonumber\\
    &-8\eta_{||}^{a\bar{j}}\eta_{||}^{i\bar{b}}(\mathcal{G}_{T^{1,2}}'(4\pi x))^2 \delta^{(2)}\biggr]\,,
    \end{align}

Similarly, we compute
\begin{align}
    \mathcal{I}_f=&\frac{1}{4}\sum_{even~s}\int dt_o\int dx \frac{2(2\pi)^2}{16\pi^3\alpha'^2t_o} (-1)^{\alpha+\beta+1}\delta g_{ab} \frac{2^4\vartheta_{\alpha,\beta+1}^4(\hat{\tau})}{\vartheta_{1,0}^4(\hat{\tau)}} \times\left(\frac{\alpha'^2}{4}\right)\nonumber\\
    &\times\left(4\eta_{||}^{ab}k_{||}^2 -4k_{||}^ak_{||}^b\right)(\mathcal{G}_{F,T^{1,2}}^{0,1})^2\delta^{(2)}(k_{||})\,.
\end{align}
Note that we dropped the terms that do not contribute after replacing $\delta g_{ab}$ with $-R_{a\bar{b}i\bar{j}}\p/\p k_i\p/\p k_{\bar{j}}.$ To simplify the spin sum, we shall use the following identity
\begin{equation}
    \vartheta_{00}^4(0|\hat{\tau})-\vartheta_{01}^4(0|\hat{\tau})-\vartheta_{10}^4(0|\hat{\tau})=0\,.
\end{equation}
Therefore, after performing the spin sum, we find
\begin{align}
    \mathcal{I}_f=&\int dt_o\int dx \frac{2}{\pi t_o}\delta g_{ab} \left(\eta_{||}^{ab}k_{||}^2-k_{||}^ak_{\perp}^b\right)(\mathcal{G}_{F,T^{1,2}}^{0,1})^2\delta^{(2)}(k_{||})\,.
\end{align}
Applying the replacement rule \eqref{eqn:replace rule}, we find
\begin{align}
    \mathcal{I}_f=&-\int dt_o \int dx \frac{4}{\pi t_o} \hat{R}(\mathcal{G}_{F,T^{1,2}}^{0,1})^2\delta^{(2)}(k_{||})\,.
\end{align}

\subsubsection{Moduli integral: open string channel}
In this region, we shall integrate the moduli over
\begin{equation}
    \int_{\frac{1}{4}}^{t_o^*}dt_o\int^{\frac{1}{4}}_{x^*}dx\,,
\end{equation}
where
\begin{equation}
    t_o^*=\frac{1}{2\pi}\log\mu^2+\frac{1}{4\pi\mu^2}+\dots\,,
\end{equation}
\begin{equation}
    x^*=\frac{1}{2\pi\mu\nu}+\frac{1}{2\pi\mu^3\nu}+\dots\,.
\end{equation}

We compute
\begin{align}
    \mathcal{I}=&\mathcal{I}_{ex}+\mathcal{I}_{rest}\,,
\end{align}
where $\mathcal{I}_{ex}$ is the total derivative in x
\begin{align}
    \mathcal{I}_{ex}=&-\int_{\frac{1}{4}}^{t_o^*} dt_o\int_{x^*}^{\frac{1}{4}} dx \frac{2e^{i\vartheta}}{\pi\alpha'^2 t_o}R_{a\bar{b}i\bar{j}} \left(4\eta_{||}^{a\bar{b}}\mathcal{G}''_{T^{1,2}}(4\pi x)\delta_{,i\bar{j}}^{(2)}-16\eta_{||}^{a\bar{b}}\eta_{||}^{i\bar{j}}(\mathcal{G}'_{T^{1,2}}\mathcal{G}_{T^{1,2}})'(4\pi x)\delta^{(2)}\right)\,,\\
    =&\int_{\frac{1}{4}}^{t_o^*}dt_o \frac{e^{i\vartheta}}{2\pi^2\alpha'^2t_o}R_{a\bar{b}i\bar{j}} \left(4\eta_{||}^{a\bar{b}}\mathcal{G}'_{T^{1,2}}(4\pi x^*)\delta_{,i\bar{j}}^{(2)}-16\eta_{||}^{a\bar{b}}\eta_{||}^{i\bar{j}}(\mathcal{G}'_{T^{1,2}}\mathcal{G}_{T^{1,2}})(4\pi x^*)\delta^{(2)}\right)\,.\label{eqn:M ex}
\end{align}
and $\mathcal{I}_{rest}$ is the rest of the terms
\begin{align}
    \mathcal{I}_{rest}=&\int_{\frac{1}{4}}^{t_o^*} dt_o\int_{x^*}^{\frac{1}{4}} dx \frac{4e^{i\vartheta}}{\pi\alpha'^2t_o}\hat{R} \biggr[ -4(\mathcal{G}'_{T^{1,2}}(4\pi x))^2+\alpha'^2(\mathcal{G}_{F,T^{1,2}}^{0,1}(4\pi x))^2\biggr]\delta^{(2)}(k_{||})\,,\label{eqn:M rest}\\
    =&\int_{\frac{1}{4}}^{t_o^*}dt_o\int_{x^*}^{\frac{1}{4}}dx \frac{e^{i\vartheta}}{\pi^3t_o}\hat{R} \left[-\left(\frac{\vartheta_{11}'(2x|2\hat{\tau})}{\vartheta_{11}(2x|2\hat{\tau})}\right)^2+\frac{\vartheta_{01}''(0|2\hat{\tau})}{\vartheta_{01}(0|2\hat{\tau})} -\partial_z^2\log\vartheta_{11}(z|2\hat{\tau})|_{z=2x}\right]\delta^{(2)}(k_{||})\,,\label{eqn:IM rest}
\end{align}
and we define
\begin{equation}
    \hat{R}=R_{a\bar{b}i\bar{j}}\eta_{||}^{a\bar{b}}\eta_{||}^{i\bar{j}}\,.
\end{equation}

The first term in $\mathcal{I}_{rest}$ is computed as
\begin{align}
    \mathcal{I}_{rest,1}=&-\int_{\frac{1}{4}}^{t_o^*}dt_o\int_{x^*}^{\frac{1}{4}} dx \frac{e^{i\vartheta}}{\pi t_o}\hat{R} \biggr[ \cot(2\pi x)+4\sum_{n=1}^\infty \frac{q^{2n}\sin(4\pi x)}{q^{4n}-2q^{2n}\cos(4\pi x)+1}\biggr]^2\delta^{(2)}(k_{||})\,,\\
    =&-\int_{\frac{1}{4}}^{t_o^*}dt_o \frac{e^{i\vartheta}}{\pi t_o}\hat{R}\biggr[-\frac{1}{4}+\frac{1}{2\pi}\cot(2\pi x^*)+\sum_{n=1}^\infty \left(\frac{2q^{2n}}{1-q^{2n}}+\sum_{m=1}^{\infty} \frac{2q^{2m+2n}}{1-q^{2m+2n}} \right)\biggr]\delta^{(2)}\,,
\end{align}
the second term is
\begin{equation}
    \mathcal{I}_{rest,2}=-\int_{\frac{1}{4}}^{t_o^*}dt_o \frac{e^{i\vartheta}}{\pi t_o}\hat{R}  \sum_{n=1}^{\infty}\frac{2q^{2n-1}}{(1+q^{2n-1})^2}\delta^{(2)}\,,
\end{equation}
and the last term is
\begin{equation}
    \mathcal{I}_{rest,3}=\int_{\frac{1}{4}}^{t_o^*}dt_o\frac{e^{i\vartheta}}{2\pi^2 t_o}\hat{R}\cot(2\pi x^*)\delta^{(2)}\,,
\end{equation}
in the large stub limit. We, therefore, find
\begin{align}
    \mathcal{I}_{rest}=&-\int_{\frac{1}{4}}^{t_o^*} dt_o \frac{e^{i\vartheta}}{\pi t_o}\hat{R}\biggr[ -\frac{1}{4} +\sum_{n=1}^\infty \left(\frac{2q^{2n}}{1-q^{2n}}+\frac{2q^{2n-1}}{(1+q^{2n-1})^2}+\sum_{m=1}^{\infty} \frac{2q^{2n+2m}}{1-q^{2m+2n}}\right)\biggr]\delta^{(2)}\,.
\end{align}
Note that, as we will show, $\mathcal{I}_{ex}$ is cancelled by the vertical integral.
\subsubsection{Moduli integral: closed string channel}
We shall rewrite the one-loop amplitude in terms of the closed string channel variables. As in the open string channel, $\mathcal{I}_{ex}$ will be canceled against the vertical integration. So, we shall only rewrite $\mathcal{I}_{rest}$ explicitly.

We shall perform a chain of modular transformations $TST^2S$ to transform $\hat{\tau}$ and $x$ into 
\begin{equation}
    \check{\tau}:=\frac{1}{2}+il= TST^2S(\hat{\tau}) =\frac{\hat{\tau}-1}{2\hat{\tau}-1}=\frac{1}{2}+i\frac{1}{4t_o}\,,\quad i\check{x}=TST^2S(x)=-\frac{x}{2\hat{\tau}-1}=\frac{i}{2t_o}x=2ilx\,.
\end{equation}
We shall use a shorthand notation
\begin{equation}
    \check{q}:=\exp(-2\pi l)\,.
\end{equation}
We also note the modular transformation of the theta functions
\begin{align}
    &\vartheta_{00}\left(\frac{z}{\tau}\biggr|-\frac{1}{\tau}\right)=(-i\tau)^{1/2}e^{\frac{\pi iz^2}{\tau}}\vartheta_{00}(z|\tau)\,,\quad \vartheta_{00}(z|\tau+1)=\vartheta_{01}(z|\tau)\,,\\
    &\vartheta_{01}\left(\frac{z}{\tau}\biggr|-\frac{1}{\tau}\right)=(-i\tau)^{1/2}e^{\frac{\pi iz^2}{\tau}}\vartheta_{10}(z|\tau)\,,\quad \vartheta_{01}(z|\tau+1)=\vartheta_{00}(z|\tau)\,,\\
    &\vartheta_{10}\left(\frac{z}{\tau}\biggr|-\frac{1}{\tau}\right)=(-i\tau)^{1/2}e^{\frac{\pi iz^2}{\tau}}\vartheta_{01}(z|\tau)\,,\quad \vartheta_{10}(z|\tau+1)=e^{\frac{\pi i}{4}}\vartheta_{10}(z|\tau)\,,\\
    &\vartheta_{11}\left(\frac{z}{\tau}\biggr|-\frac{1}{\tau}\right)=-i(-i\tau)^{1/2}e^{\frac{\pi iz^2}{\tau}}\vartheta_{11}(z|\tau)\,,\quad \vartheta_{11}(z|\tau+1)=e^{\frac{\pi i}{4}}\vartheta_{11}(z|\tau)\,.
\end{align}
We therefore find
\begin{align}
    \frac{\vartheta_{11}'(2x|2\tau)}{\vartheta_{11}(2x|2\tau)}=&-2il \frac{\vartheta_{11}'(2i\check{x}|2il)}{\vartheta_{11}(2i\check{x}|2il)}+4\pi \check{x}\,,
\end{align}
\begin{align}
    \frac{\vartheta_{00}''(0|2\tau)}{\vartheta_{00}(0|2\tau)}=&-4l^2\frac{\vartheta_{00}''(0|2il)}{\vartheta_{00}(0|2il)}-4\pi l\,.
\end{align}
Note that the domain of the integral for $\check{x}$ is given as
\begin{equation}
   \hat{x}^*= \frac{1}{\mathcal{F}_1(l)\nu}+ \frac{\mathcal{F}_3(l)}{\mathcal{F}_1(l)^4\nu^3}+\dots\leq \check{x}\leq \frac{1}{2}l\,,
\end{equation}
note
\begin{equation}
    \mathcal{F}_1(l)=\frac{\mathfrak{F}(\mu)\pi}{2}\,, \quad \mathcal{F}_3(l) =-\frac{\mathfrak{F}(\mu)\pi^3}{24}\,,
\end{equation}
for 
\begin{equation}
    1\leq l\leq l^*= \frac{1}{\pi}\log\lambda_D^2-\varepsilon\,.
\end{equation}

We write $\mathcal{I}_{rest}$ in terms of the closed string variables
\begin{align}
    \mathcal{I}_{rest}=&-\int_{1}^{l^*}dl\int^{\frac{1}{2}l}_{\check{x}^*}d\check{x} \frac{e^{i\vartheta}}{2\pi^3 l^2}\hat{R}\biggr[ -4\left(2\pi\check{x}-il\frac{\vartheta_{11}'(2i\check{x}|2il)}{\vartheta_{11}(2i\check{x}|2il)}\right)^2-4\pi l-4l^2\frac{\vartheta_{00}''(0|2il)}{\vartheta_{00}(0|2il)}\biggr]\delta^{(2)}\nonumber\\
    &-\int_1^{l^*}dl\frac{e^{i\vartheta}}{2\pi^3l}\hat{R} \biggr[-2il\frac{\vartheta_{11}'(2i\check{x}^*|2il)}{\vartheta_{11}(2i\check{x}^*|2il)}\biggr]\delta^{(2)}\,,
\end{align}
in the large stub limit. 

Although we will show later that the first term and the last term in $\mathcal{I}_{rest}$ are completely cancelled against the analogous terms in the annulus amplitude, for completeness, we shall compute it. We note 
\begin{equation}
    \frac{\vartheta_{11}'(2i\check{x}|2il)}{\vartheta_{11}(2i\check{x}|2il)}=-i\pi\coth(2\pi\check{x})+4\pi i \sum_{n=1}^{\infty} \frac{\check{q}^{2n}\sinh(4\pi\check{x})}{\check{q}^{4n}-2\check{q}^{2n}\cosh(4\pi \check{x})+1}\,.
\end{equation}
We find
\begin{equation}
    \int d\check{x} \check{x} \coth(2\pi \check{x})=-\frac{\check{x}^2}{2}+\frac{\check{x}}{2\pi}\log\left[1-e^{4\pi \check{x}}\right]+\frac{1}{8\pi^2}\text{Li}_2(e^{4\pi \check{x}})\,,
\end{equation}
\begin{align}
    \int d\check{x} \frac{\check{x} \check{q}^{2n}\sinh(4\pi \check{x})}{\check{q}^{4n}-2\check{q}^{2n}\cosh(4\pi \check{x})+1}=&\frac{\check{x}^2}{4}-\frac{\check{x}}{8\pi}\log\left[(1-e^{4\pi\check{x}}\check{q}^{-2n})(1-e^{4\pi\check{x}}\check{q}^{2n})\right]\nonumber\\
    &-\frac{1}{32\pi^2}\text{Li}_2(e^{4\pi \check{x}}\check{q}^{-2n})-\frac{1}{32\pi^2}\text{Li}_2(e^{4\pi\check{x}}\check{q}^{2n})\,,
\end{align}
\begin{equation}
    \int d\check{x} \coth(2\pi\check{x})^2=\check{x}-\frac{1}{2\pi}\coth(2\pi \check{x})\,,
\end{equation}
\begin{equation}
    \int d\check{x} \frac{\check{q}^{2n}\coth(2\pi\check{x})\sinh(4\pi \check{x})}{\check{q}^{4n}-2\check{q}^{2n}\cosh(4\pi\check{x})+1}=\frac{i}{4\pi}\tanh^{-1}(\coth(n\log\check{q})\tanh(2\pi\check{x}))\coth(n\log\check{q})-\frac{i}{2}\check{x}\,,
\end{equation}
\begin{align}
\int d\check{x} \left(\frac{\check{q}^{2n}\sinh(4\pi\check{x})}{\check{q}^{4n}-2\check{q}^{2n}\cosh(4\pi\check{x})+1}\right)^2=&\frac{i}{4}\check{x} -\frac{i}{16}\frac{\sinh(4\pi\check{x})}{\pi\cosh(4\pi\check{x})-\pi\cosh(2n\log\check{q})}\nonumber\\
&-\frac{i}{8\pi}\tanh^{-1}(\coth(n\log\check{q})\tanh(2\pi\check{x})\coth(2n\log\check{q})\,.
\end{align}
We use Jonquiere's inversion formula
\begin{equation}
    \text{Li}_2(z)+\text{Li}_2(z^{-1})=2\pi^2 B_2\left(\frac{1}{2\pi i}\log(z)\right)\,,
\end{equation}
for $0\leq\arg(z)<2\pi$ and $|z|\leq1,$ where $B_2$ is the Bernoulli polynomial
\begin{equation}
    B_2(x)=x^2-x+\frac{1}{6}\,,
\end{equation}
to simplify
\begin{equation}
    \int d\check{x} \check{x} \coth(2\pi \check{x})=-\frac{\check{x}^2}{2}+\frac{\check{x}}{2\pi}\log\left[1-e^{4\pi \check{x}}\right]+\frac{1}{4} B_2(2i \check{x})-\frac{1}{8\pi^2}\text{Li}_2(e^{-4\pi\check{x}})\,,
\end{equation}
and
\begin{align}
    \int d\check{x} \frac{\check{x} \check{q}^{2n}\sinh(4\pi \check{x})}{\check{q}^{4n}-2\check{q}^{2n}\cosh(4\pi \check{x})+1}=&\frac{\check{x}^2}{4}-\frac{\check{x}}{8\pi}\log\left[(1-e^{4\pi\check{x}}\check{q}^{-2n})(1-e^{4\pi\check{x}}\check{q}^{2n})\right]\nonumber\\
    &-\frac{1}{32\pi^2}\text{Li}_2(e^{-4\pi(nl-\check{x})})+\frac{1}{32\pi^2}\text{Li}_2(e^{-4\pi(nl+\check{x})})-\frac{1}{16}B_2\left(2i(nl+\check{x})\right)\,.
\end{align}

We find that the first term in $\mathcal{I}_{rest}$ therefore evaluates to
\begin{align}
    \mathcal{I}_{rest,1}\hat{R}^{-1}=\sum_{i=1}^6\mathfrak{I}_i\,,
\end{align}
where
\begin{align}
    \mathfrak{I}_1=&e^{i\vartheta}\int \frac{dl}{2\pi^3l^2}\int d\check{x} (4\pi \check{x})^2\,,\\
    =& e^{i\vartheta}\int_1^{l^*}\frac{ldl}{3\pi}\,,
\end{align}
\begin{align}
    \mathfrak{I}_2=& e^{i\vartheta}\int_1^{l^*}\frac{dl}{2\pi^3l^2}\int_{\check{x}^*}^{\frac{1}{2}l} d\check{x}(-16\pi^2l\check{x})\coth(2\pi\check{x})\,,\\
    =&e^{i\vartheta}\int_1^{l^*}\frac{dl}{2\pi^3l}\biggr[-\frac{\pi^2}{3}-2\pi^2l^2+4\pi l\text{Li}_1(\check{q})+2\text{Li}_2(\check{q})\biggr]\,,
\end{align}
\begin{align}
    \mathfrak{I}_3=&e^{i\vartheta}\int_1^{l^*}\frac{dl}{2\pi^3l^2}\int_{\check{x}^*}^{\frac{1}{2}l}d\check{x} 4\pi^2l^2\coth^2(2\pi\check{x})\,,\\
    =&e^{i\vartheta}\int_1^{l^*}\frac{dl}{2\pi^3}\biggr[2\pi^2l-2\pi+2\pi \coth(2\pi\check{x}^*)-4\pi\frac{\check{q}^2}{1-\check{q}^2}\biggr]\,,
\end{align}
\begin{align}
    \mathfrak{I}_4=&e^{i\vartheta}\int_1^{l^*}\frac{dl}{2\pi^3l^2}\int d\check{x} (64\pi^2l\check{x})\sum_{n=1}^{\infty} \frac{\check{q}^{2n}\sinh(4\pi\check{x})}{\check{q}^{4n}-2\check{q}^{2n}\cosh(4\pi \check{x})+1}\,,\\
    =&e^{i\vartheta}\int_1^{l^*}\frac{dl}{2\pi^3l}\sum_{n=1}^\infty\biggr[4\pi l\text{Li}_1(\check{q}^{2n-1})+4\pi l\text{Li}_1(\check{q}^{2n+1})-2\text{Li}_2(\check{q}^{2n-1})+2\text{Li}_2(\check{q}^{2n+1})\biggr]\,,
\end{align}
\begin{align}
    \mathfrak{I}_5=&e^{i\vartheta}\int\frac{dl}{2\pi^3l^2}\int d\check{x} (-32\pi^2l^2)\sum_{n=1}^{\infty}\left[\frac{\check{q}^{2n}\coth(2\pi\check{x})\sinh(4\pi\check{x})}{\check{q}^{4n}-2\check{q}^{2n}\cosh(4\pi\check{x})+1}\right]\,,\\
    =&e^{i\vartheta}\int_1^{l^*}\frac{2dl}{\pi^2}\sum_{n=1}^{\infty}\frac{\check{q}^{2n}}{1-\check{q}^{2n}}\biggr[ -4l\pi -(1+\check{q}^{-2n})\text{Li}_1(\check{q}^{2n-1})+(1+\check{q}^{-2n})\text{Li}_1(\check{q}^{2n+1})\biggr]\,,
\end{align}
\begin{align}
    \mathfrak{I}_{6}=&e^{i\vartheta} \int_1^{l^*}\frac{dl}{2\pi^3l^2} \int_{\check{x}^*}^{\frac{1}{2}l}d\check{x} (64\pi^2l^2) \sum_{n,m}\left[\frac{\check{q}^{2n}\sinh(4\pi \check{x})}{\check{q}^{4n}-2\check{q}^{2n}\cosh(4\pi\check{x})+1}\right]\nonumber\\
    &\times\biggr[\frac{\check{q}^{2m}\sinh(4\pi \check{x})}{\check{q}^{4m}-2\check{q}^{2m}\cosh(4\pi\check{x})+1}\biggr]\label{eqn:exp before intg}\,.
\end{align}
\subsection{Vertical Integration}
Let us recall
\begin{equation}
    \mathcal{V}_{gh}:=-\oint_{C_1} \frac{dy_1}{y_1}\oint\frac{d\bar{y}_2}{\bar{y}_2} \langle \mathcal{B}_{m_b} c\bar{c} e^{-\phi}e^{-\bar{\phi}}(2\pi x) \left[\left(\xi(y_1)-\xi(W_1) \right) \mathcal{X}(y_2)+\left(\xi(y_2)-\xi(W_2)\right)\mathcal{X}(W_1)\right] \rangle_s\,,
\end{equation}
\begin{equation}
    \mathcal{V}_{m}:=\langle \mathcal{W}(2\pi x)\rangle_s\,,
\end{equation}
with
\begin{equation}
    V^{-1,-1}=e^{-\phi}e^{-\bar{\phi}}\mathcal{W}\,.
\end{equation}

To calculate $\mathcal{V}_{gh},$ we need to work harder. Let us first recall the correlator of the $\beta-\gamma$ system on torus in the large Hilbert space \cite{Verlinde:1987sd,Lechtenfeld:1989wu,Morozov:1989ma}
\begin{align}\label{eqn:ghost one loop large H}
    \biggr\langle \prod_{i=1}^{n+1}\xi(2\pi x_i) \prod_{i=1}^n \eta(2\pi y_i)\prod_{k=1}^m e^{q_k\phi(2\pi z_k)} \biggr\rangle_s=& \frac{\prod_{j=1}^n \vartheta_s(-y_j+\sum_i x_i-\sum_i y_i+\sum_kq_k z_k|\tau)}{\prod_{j=1}^{n+1}\vartheta_s(-x_j+\sum_ix_i-\sum_i y_i+\sum_kq_kz_k|\tau)}\nonumber\\
    &\times \frac{\prod_{i<i'} E(x_i,x_{i'})\prod_{j<j'}E(y_j,y_{j'})}{\prod_{i,j}E(x_i,y_j)\prod_{k<l}E(z_k,z_l)^{q_kq_l}}\,,
\end{align}
where
\begin{equation}
    E(x,y):=2\pi \frac{\vartheta_{11}(x-y|\tau)}{\vartheta_{11}'(0|\tau)}\,.
\end{equation}
Note that $\sum_k q_k=0$ should be satisfied. And note that our conventions differ from those of \cite{Verlinde:1987sd,Lechtenfeld:1989wu,Morozov:1989ma} by the factor of $2\pi$

Of the terms in the PCOs, due to the background $\phi$ charge cancellation, only the following terms can contribute to the vertical integration
\begin{equation}
    \mathcal{X}\supset -\partial\eta b e^{2\phi} -\partial(\eta be^{2\phi})\,.
\end{equation}
Therefore, to simplify the calculation, it is useful to introduce a composite operator
\begin{equation}
    \mathcal{X}_p(2\pi y_1,2\pi y_2):=-\eta(2\pi y_1) b e^{2\phi}(2\pi y_2)\,.
\end{equation}
We shall first compute
\begin{align}
    \mathcal{V}_p(x_1,y_1,y_2):=\langle \mathcal{B}_{m_b} c\bar{c} e^{-\phi}e^{-\bar{\phi}}(2\pi x) \partial \xi(2\pi x_1) \mathcal{X}_p(2\pi y_1,2\pi y_2)\rangle_s\,,
\end{align}
and use the relation
\begin{align}
    \mathcal{V}_{gh}= &\oint_{C_1}\frac{dp_1}{p_1}\oint_{C_1}\frac{d\bar{p}_2}{\bar{p}_2} \left[\int_{p_1}^{W_1} dw (\partial_{y'}\mathcal{V}_p(w, y',p_2)|_{y'=p_2}+\partial_{p_2}\mathcal{V}_p(w,p_2,p_2))\right]\nonumber\\
    &+\oint_{C_1}\frac{dp_1}{p_1}\oint_{C_1}\frac{d\bar{p}_2}{\bar{p}_2} \left[\int_{p_2}^{W_2} dw (\partial_{y'}\mathcal{V}_p(w, y',W_1)|_{y'=W_1}+\partial_{W_1}\mathcal{V}_p(w,W_1,W_1))\right]\,,
\end{align}
to compute $\mathcal{V}_{gh}.$ To evaluate $\mathcal{V}_p$ using the formula \eqref{eqn:ghost one loop large H}, we shall insert a zero mode of $\xi$ at $x_0.$ However, the precise location of the zero mode insertion does not affect the answer. We compute
\begin{align}
    \mathcal{V}_p(x_1,y_1,y_2):=&-\langle \mathcal{B}_{m_b} c\bar{c}(2\pi x) b(2\pi y_2)\rangle\times \langle e^{-\phi}e^{-\bar{\phi}}(2\pi x) \partial \xi(2\pi x_1) \eta(2\pi y_1) e^{2\phi}(2\pi y_2) \rangle_s\,,\\
    =&-\mathcal{V}_{b,c}\times \mathcal{V}_{\beta,\gamma}\,,
\end{align}
with
\begin{align}
    \mathcal{V}_{\beta,\gamma}=&i\frac{1}{2\pi}\partial_{x_1}\biggr[\frac{\vartheta_s(-2y_1+x_0+x_1+2y_2|\tau)}{\vartheta_s(x_1-y_1+2y_2|\tau)\vartheta_s(x_0-y_1+2y_2|\tau)} \times \frac{E(x_0,x_1)E(y_2- x)^{2} E(y_2+ x)^2}{E(x_0,y_1)E(x_1,y_1) E(2 x)  } \biggr]\,,
\end{align}
\begin{align}
    \mathcal{V}_{b,c}=&2\pi\partial_{m_b} x \langle (c-\bar{c})(2\pi x)b(2\pi y_2)\rangle +\left(2\pi\partial_{m_b} \tau \oint_{C_3}dz \langle  b(z) c\bar{c}(2\pi x)b(2\pi y_2)\rangle +c.c.\right)\,.
\end{align}
To compute the b-c ghost correlator, we can ``bosonize" the b-c ghost system
\begin{equation}
    b= e^{\varphi}\,,\quad c= e^{-\varphi}\,,
\end{equation}
such that the background charge of the $\varphi$ is given as $3(g-1).$ For the one-loop amplitude to be non-trivial, we need to insert one zero mode of the b-ghost and one zero mode of the c-ghost. Contractions involving the remaining ghosts can be performed using the standard Wick contraction. We therefore find 
\begin{equation}
    \langle b(2\pi z_1)b(2\pi z_2) c(2\pi x_1)c(2\pi x_2) \rangle= -\frac{2}{ \alpha'}e^{i\vartheta}\eta(\tau)^2\left(\mathcal{G}_B'(z_2;x_1) -\mathcal{G}_B'(z_2;x_2)+\mathcal{G}_B'(z_1;x_2)-\mathcal{G}_B'(z_1;x_1)\right) \,.
\end{equation}
Note
\begin{equation}
    \mathcal{G}_B'(2\pi(y_2+x))=-\frac{\alpha'}{4\pi} \frac{\vartheta_{11}'(y_2+x|\hat{\tau})}{\vartheta_{11}(y_2+x|\hat{\tau})}\simeq -\frac{\alpha'}{4} \cot(\pi(y_2+x))+\dots\,.
\end{equation}

We can therefore rewrite $\mathcal{V}_{b,c}$ as
\begin{align}
    \mathcal{V}_{b,c}=& 4\pi\partial_{m_b}x\langle c_0 b_0\rangle -\partial_{m_b}t \int_\pi^{2\pi} d\text{Re}z \langle b(z) c(2\pi x) c(2\pi -2\pi x) b(2\pi y_2)\rangle\nonumber\\
    &-\partial_{m_b}t\int_\pi^{2\pi} d\text{Re}z \langle b(2\pi -\bar{z}) c(2\pi x) c(2\pi-2\pi x)b(2\pi y_2) \rangle\,\\
    =& -4\pi \partial_{m_b}x\langle b_0c_0\rangle+\frac{4}{\alpha'}\partial_{m_b}t \int_{\pi}^{2\pi}dz_0 \langle b_0c_0\rangle\biggr \langle\mathcal{G}_B'(2\pi y_2-2\pi x)-\mathcal{G}_B'(2\pi y_2-2\pi+2\pi x)\biggr\rangle\nonumber\\
    &+\frac{2}{\alpha'}\partial_{m_b}t\int_{\pi}^{2\pi}dz_0 \langle b_0c_0\rangle\biggr\langle\mathcal{G}_{B}'(z_0+iy_c-2\pi+2\pi x)-\mathcal{G}_B'(z_0+iy_c-2\pi x)\nonumber\\&\qquad\qquad\qquad\qquad\qquad\qquad+\mathcal{G}_B'(-z_0+iy_c+2\pi x)-\mathcal{G}_B'(2\pi-z_0+iy_c-2\pi x)\biggr\rangle\,,\\
    =&-4\pi\partial_{m_b}x\langle b_0c_0\rangle +\frac{4}{\alpha'}\partial_{m_b}t \langle b_0c_0\rangle \langle\mathcal{G}_B'(2\pi y_2-2\pi x)-\mathcal{G}_B'(2\pi y_2+2\pi x)\rangle\,.
\end{align}
Note that we used
\begin{equation}
    \bar{z}=2\pi\tau  -z\,,
\end{equation}
and
\begin{equation}
    z\equiv 2\pi +z\,.
\end{equation}

We now compute
\begin{equation}
    \mathcal{V}_m=\langle \mathcal{W}(2\pi x)\rangle_s\,,
\end{equation}
for
\begin{equation}
    \mathcal{W}=\frac{1}{4\pi }\delta g_{ab} \psi^a\bar{\psi}^b\,.
\end{equation}
We therefore find
\begin{equation}
    \mathcal{V}_m= \frac{i}{2\pi } \delta g_{ab}e^{-(k_{||}^2\mathcal{G}_{||}+k_{\perp}^2\mathcal{G}_{\perp})(4\pi x)}\eta_{||}^{ab}\mathcal{G}_{F,T^{1,2}}^{\alpha,1}(4\pi x-2\pi) \times(-1)^{\alpha+\beta+1} \frac{2^4 \vartheta_{\alpha,\beta+1}(\hat{\tau})^4\vartheta_s(\hat{\tau})}{\vartheta_{1,0}(\hat{\tau})^4\eta(\hat{\tau})^2}\delta^{(2)}(k_{||})\,.
\end{equation}

By combining all the terms, we find that the vertical integration is written as
\begin{align}
    \mathcal{V}'(x_1,y_1,y_2)=&\frac{1}{4} \sum_s \int dm_b \frac{1}{8\pi^2\alpha' t} (-4\pi)\times \frac{i}{2\pi }\times 2^4 \times \frac{i}{2\pi}\times (-1)^{\alpha+\beta+1}\nonumber\\
    &\times \frac{\vartheta_{\alpha,\beta+1}(\hat{\tau})^4\vartheta_{\alpha,\beta}(\hat{\tau})}{\vartheta_{1,0}(\hat{\tau})^4} \delta g_{ab} e^{-(k_{||}^2\mathcal{G}_{B,||}+k_{\perp}^2\mathcal{G}_{B,\perp})(4\pi x)}\eta_{||}^{ab} \mathcal{G}_{F,T^{1,2}}(4\pi x)\delta^{(2)}(k_{||})\nonumber\\
    &\times\left[\partial_{m_b} x-\frac{1}{\alpha'} \partial_{m_b}t (\mathcal{G}_{B}'(2\pi y_2-2\pi x)-\mathcal{G}_B'(2\pi y_2+2\pi x))\right] e^{i\vartheta} \nonumber\\
    &\times\partial_{x_1} \biggr[\frac{\vartheta_s(-2y_1+x_0+x_1+2y_2|\hat{\tau})}{\vartheta_s(x_1-y_1+2y_2|\hat{\tau})\vartheta_s(x_0-y_1+2y_2|\hat{\tau})} \times \frac{E(x_0,x_1)E(y_2-x)^2E(y_2+x)^2}{E(x_0,y_1)E(x_1,y_1)E(2x)}\biggr]\,,
\end{align}
with
\begin{align}\label{eqn:vert1}
    \mathcal{V}=&-\oint_{}\frac{dp_1}{p_1-x}\oint\frac{dp_2}{p_2+x} \left[\int_{p_1}^{W_1} dw (\partial_{y'}\mathcal{V}'(w, y',p_2)|_{y'=p_2}+\partial_{p_2}\mathcal{V}'(w,p_2,p_2))\right]\nonumber\\
    &-\oint\frac{dp_1}{p_1-x}\oint\frac{dp_2}{p_2+ x} \left[\int_{p_2}^{W_2} dw (\partial_{y'}\mathcal{V}'(w, y',W_1)|_{y'=W_1}+\partial_{y_1'}\mathcal{V}'(w,y_1',y_1'))|_{y_1'=W_1}\right]\,.
\end{align}
Note that we can place $x_0$ at any non-singular point, as the result does not depend on $x_0$ \cite{Verlinde:1987sd,Lechtenfeld:1989wu,Morozov:1989ma}. A comment is in order. We can change the contour integral with respect to the variable $\bar{y}_2$ to a contour integral with respect to $y_2,$ now centered at $-x_2$. 

We shall now use the identity \eqref{eqn:vertical beta gamma} to simplify the vertical integration further. Let us recall \eqref{eqn:vertical beta gamma}
\begin{align}
    \mathcal{L}_{\beta-\gamma}=&- \int_p^W d x_1 \partial_{x_1} \biggr[\frac{\vartheta_s(-2y_1+x_0+x_1+2y_2|\hat{\tau})}{\vartheta_s(x_1-y_1+2y_2|\hat{\tau})\vartheta_s(x_0-y_1+2y_2|\hat{\tau})} \times \frac{E(x_0,x_1)E(y_2-x)^2E(y_2+x)^2}{E(x_0,y_1)E(x_1,y_1)E(2x)}\biggr]\,,\\
    =&\frac{E(y_2,x)^2E(y_2,-x)^2E(W,p)\vartheta_s(-2y_1+2y_2+p+W)}{E(2x)E(p,y_1)E(W,y_1)\vartheta_s(p-y_1+2y_2)\vartheta_s(W-y_1+2y_2)}\,.
\end{align}

Let us first evaluate the vertical integration due to the jump of PCO from $x$ to $W_1.$ As we are placing the second PCO at $y_2=-x,$ $\mathcal{L}_{\beta-\gamma}$ exhibits the double zero. The only way to have a non-trivial result is to act the derivative $\partial_{y_2}$ on $\mathcal{G}_B'(2\pi y_2+2\pi x) \mathcal{L}_{\beta-\gamma}$ factor. We therefore find that the ghost factor in the first vertical integration is given by
\begin{align}
    &\lim_{y_2\rightarrow -x}\frac{1}{\alpha'}\partial_{y_2}\partial_{m_b} t \biggr[\mathcal{G}_B'(2\pi y_2+2\pi x)\frac{E(y_2,x)^2E(y_2,-x)^2E(W_1,p_1)\vartheta_s(2x+2y_2+p_1+W_1)}{E(2x)E(p_1,-x)E(W_1,-x)\vartheta_s(p_1+x+2y_2)\vartheta_s(W_1+x+2y_2)}\biggr]\,,\\
    &=\partial_{m_b}t\frac{E(W_1,x)\vartheta_s(x+W_1)}{E(W_1,-x)\vartheta_s(0)\vartheta_s(W_1-x)}\partial_{y_2} \biggr[-\frac{1}{4}\cot(\pi(y_2+x)) (2\pi)^2 (y_2+x)^2\biggr]\,,\\
    =&-\pi\partial_{m_b}t\frac{E(W_1,x)\vartheta_s(x+W_1)}{E(W_1,-x)\vartheta_s(0)\vartheta_s(W_1-x)}
\end{align}
Note that we used $p_1=x$ and
\begin{equation}
    \mathcal{G}_B'(2\pi(y_2+x))=-\frac{\alpha'}{4\pi} \frac{\vartheta_{11}'(y_2+x|\hat{\tau})}{\vartheta_{11}(y_2+x|\hat{\tau})}\simeq -\frac{\alpha'}{4} \cot(\pi(y_2+x))+\dots\,.
\end{equation}

Now let us examine the vertical integration due to the jump of PCO from $-x$ to $W_2,$ while fixing the location of the other PCO at $W_1.$ That is; we shall set $y_1$ and $y_2$ as $W_1,$ after taking the derivatives. Provided that $W_1$ is somewhat generic, $\mathcal{L}_{\beta-\gamma}$ is neither divergent nor vanishing. On the other hand, if $W_1$ is treated as a small number and at the same time $x$ is treated as a small number, we find that $\mathcal{L}_{\beta-\gamma}$ vanishes as $\mathcal{O}(x^2).$ Combined with the derivatives in $\partial_{y_1}$ and $\partial_{y_2},$ we then find that this type of vertical integration can yield diverging contributions as well, as $\mathcal{G}_{F,T^{1,2}}(4\pi x)$ and $\mathcal{G}_{B}'(2\pi y_2-2\pi x)-\mathcal{G}_B'(2\pi y_2+2\pi x)$ both scale as $\mathcal{O}(x^{-1}).$ 

$\mathcal{L}_{\beta-\gamma}$ takes the following form
\begin{align}
    \mathcal{L}_{\beta-\gamma}=&\frac{E(y_2,x)^2E(y_2,-x)^2E(W_2,-x)\vartheta_s(-2y_1+2y_2-x+W_2)}{E(2x)E(-x,y_1)E(W_2,y_1)\vartheta_s(-x-y_1+2y_2)\vartheta_s(W_2-y_1+2y_2)}\,,\\
    =&\frac{E(W_1,x)^2E(W_1,-x)^2E(W_2,-x)\vartheta_s(-x+W_2)}{E(2x)E(-x,W_1)E(W_2,W_1)\vartheta_s(-x+W_1)\vartheta_s(W_2+W_1)}\,.
\end{align}
As we are adding the derivatives in $y_1$ and $y_2,$ the relevant term is given as
\begin{equation}
    \partial\mathcal{L}_{\beta-\gamma}= -\frac{\vartheta_s(W_2-x)E(W_2+x)}{E(2x)} \partial_{W_1} \biggr[\frac{E(W_1-x)^2E(W_1+x)^2}{E(x+W_1)E(W_2-W_1) \vartheta_s(-x+W_1)\vartheta_s(W_2+W_1)}\biggr]\,. 
\end{equation}

We shall finally write down the complete form of the vertical integration
\begin{align}
    \mathcal{V}=& \mathcal{V}_{1}+\mathcal{V}_2\,,
\end{align}
with
\begin{align}
    \mathcal{V}_1=&\sum_{even~s} \int dm_b \frac{(-1)^{\alpha+\beta}}{2\pi^2\alpha' t}  \frac{\vartheta_{\alpha,\beta+1}(\hat{\tau})^4}{\vartheta_{1,0}(\hat{\tau})^4} \delta g_{ab} e^{-(k_{||}^2\mathcal{G}_{NN}+k_{\perp}^2\mathcal{G}_{DN})(4\pi x)}\eta_{||}^{ab} \mathcal{G}^{\alpha,1}_{F,T^{1,2}}(4\pi x) \delta^{(2)}(k_{||})\nonumber\\
    &\times\partial_{m_b} t\frac{E(W_1,x)\vartheta_s(x+W_1|\hat{\tau})}{E(W_1,-x)\vartheta_s(W_1-x|\hat{\tau})}e^{i\vartheta}\,,
\end{align}
\begin{align}
    \mathcal{V}_2=&\sum_{even~s} \int dm_b \frac{(-1)^{\alpha+\beta}}{2\pi^2\alpha' t}  \frac{\vartheta_{\alpha,\beta+1}(\hat{\tau})^4\vartheta_{\alpha,\beta}(\hat{\tau})}{\vartheta_{1,0}(\hat{\tau})^4} \delta g_{ab} e^{-(k_{||}^2\mathcal{G}_{NN}+k_{\perp}^2\mathcal{G}_{DN})(4\pi x)}\eta_{||}^{ab} \mathcal{G}^{\alpha,1}_{F,T^{1,2}}(4\pi x) \delta^{(2)}(k_{||})\nonumber\\
    &\times\partial_{W_1}\biggr[ \left(\partial_{m_b}x -\frac{1}{\alpha'}\partial_{m_b}t (\mathcal{G}_B'(2\pi W_1-2\pi x)-\mathcal{G}_B'(2\pi W_1+2\pi x)) \right) \mathcal{L}_{\beta-\gamma}\biggr]e^{i\vartheta}\,.
\end{align}

\subsubsection{Open string channel}
In this region, we have
\begin{equation}
    x=y_{C,o}=\frac{1}{2\pi \nu\mu}+\frac{1}{2\pi\mu^3\nu}+\dots\,.
\end{equation}
We shall average over PCO locations as follows. First, we shall move PCOs at $p_1=x$ and $p_2=-x$ to
\begin{equation}
    W_1= \left(\frac{1}{6}+\frac{\sqrt{3}}{4\pi\mu^2}-\frac{\sqrt{3}}{4\pi\mu^4}+\dots\right)\,,
\end{equation}
\begin{equation}
    W_2=\pm i\left(\frac{\sqrt{3}}{2\pi\mu\nu}+\frac{\sqrt{3}}{2\pi\mu^3\nu}+\dots\right)\,.
\end{equation}
And we shall move PCOs at $p_1=-x$ and $p_2=x$ to
\begin{equation}
    W_1= -\left(\frac{1}{6}+\frac{\sqrt{3}}{4\pi\mu^2}-\frac{\sqrt{3}}{4\pi\mu^4}+\dots\right)\,,
\end{equation}
\begin{equation}
    W_2= \pm i\left(\frac{\sqrt{3}}{2\pi\mu\nu}+\frac{\sqrt{3}}{2\pi\mu^3\nu}+\dots\right)\,,
\end{equation}
and average over two different moves of PCOs. To obtain the second choice of vertical integration, we can simply change the signs of $W_1,$ $W_2$ and $x$.

We first note
\begin{align}
    \frac{E(W_1,x)\vartheta_s(x+W_1|\hat{\tau})}{E(W_1,-x)\vartheta_s(W_1-x|\hat{\tau})}=&1-2 \partial_{W_1}\log (E(W_1) \vartheta_s(W_1|\hat{\tau}))x+\dots\,.
\end{align}
We therefore find
\begin{align}
    \mathcal{V}_1=&-e^{i\vartheta}\int dt_o \frac{1}{2\pi^2\alpha' t_o} \delta g_{ab} \eta_{||}^{ab}e^{-(k_{||}^2\mathcal{G}_{NN}+k_{\perp}^2\mathcal{G}_{DN})(4\pi x)} \mathcal{G}_{F,T^{1,2}}^{0,1}(4\pi x) \biggr[ 1-2\partial_{W_1}\log E(W_1)x\nonumber\\
    &+ \frac{2}{\vartheta_{1,0}(\hat{\tau})^4}\left(\frac{\vartheta_{0,1}(\hat{\tau})^4 \vartheta_{00}'(W_1|\hat{\tau})}{\vartheta_{0,0}(W_1|\hat{\tau})}-\frac{\vartheta_{00}(\hat{\tau})^4\vartheta_{0,1}'(W_1|\hat{\tau})}{\vartheta_{0,1}(W_1|\hat{\tau})}\right) x+\dots\biggr]\delta^{(2)}(k_{||})\,.
\end{align}
Using the fact that we need to average over two different signs, we find
\begin{equation}
    \mathcal{V}_1=-e^{i\vartheta}\int dt_o \frac{1}{2\pi^2\alpha' t_o} \delta g_{ab} \eta_{||}^{ab}e^{-(k_{||}^2\mathcal{G}_{NN}+k_{\perp}^2\mathcal{G}_{DN})(4\pi x)} \mathcal{G}_{F,T^{1,2}}^{0,1}(4\pi x)\delta^{(2)}(k_{||})\,.
\end{equation}
We shall also expand $\mathcal{L}_{\beta-\gamma}$ in small $x$ and $W_2$ limit
\begin{align}
    \mathcal{L}_{\beta-\gamma} =&\frac{E(W_1,x)^2E(W_1,-x)^2E(W_2,-x)\vartheta_s(-x+W_2)}{E(2x)E(-x,W_1)E(W_2,W_1)\vartheta_s(-x+W_1)\vartheta_s(W_2+W_1)}\,,\\
    =&\frac{E(W_1)^2\vartheta_s(0|\hat{\tau})}{\vartheta_s(W_1)^2} \frac{W_2+x}{2x}+\dots\,,\\
    =&4\pi^2\frac{1}{\vartheta_s(0|\hat{\tau})} \left(\frac{\vartheta_s''(0|\hat{\tau})}{\vartheta_s(0|\hat{\tau})}-\partial_z^2 \log \vartheta_{11}(z|\hat{\tau})|_{z=W_1}\right)^{-1}\frac{W_2+x}{2x}+\dots\,.
\end{align}
Similarly, we find
\begin{equation}
    \mathcal{G}_B'(2\pi W_1-2\pi x)-\mathcal{G}_B'(2\pi W_1+2\pi x)= -4\pi \mathcal{G}_B''(2\pi W_1)x+\dots\,.
\end{equation}
We therefore find
\begin{align}
    \mathcal{V}_2=&e^{i\vartheta}\sum_{even~s}\int d t_o \frac{4\pi(-1)^{\alpha+\beta+1}}{ \alpha't_o} \frac{\vartheta_{\alpha,\beta+1}(\hat{\tau})^4}{\vartheta_{1,0}(\hat{\tau})^4} \delta g_{ab}e^{-2k_{||}^2\mathcal{G}_{T^{1,2}}(4\pi x)}\eta_{||}^{ab} \mathcal{G}_{F,T^{1,2}}^{\alpha,1}(4\pi x)\nonumber\\
    &\times \partial_{W_1}\left[\frac{-\frac{\vartheta_s''(0|\hat{\tau})}{\vartheta_{s}(0|\hat{\tau})}}{\frac{\vartheta_s''(0|\hat{\tau})}{\vartheta_{s}(0|\hat{\tau})}-\partial_z^2\log\vartheta_{11}(z|\hat{\tau})|_{z=W_1}}\right] \frac{W_2+x}{2}+\dots\,.
\end{align}
Same as before, we need to sum over different signs of $W_1$ and $W_2,$ which lead to
\begin{equation}
    \mathcal{V}_2=\dots\,,
\end{equation}
where $\dots$ includes the terms that vanish in the large stub limit. 

As a result, we compute
\begin{equation}
    \mathcal{V}=-e^{i\vartheta}\int dt_o \frac{1}{2\pi^2\alpha' t_o} \delta g_{ab} \eta_{||}^{ab}e^{-(k_{||}^2\mathcal{G}_{NN}+k_{\perp}^2\mathcal{G}_{DN})(4\pi x)}\mathcal{G}_{F,T^{1,2}}^{0,1}(4\pi x)\delta^{(2)}(k_{||})+\dots\,.
\end{equation}
Applying the replacement rule \eqref{eqn:replace rule}, we find
\begin{align}
    \mathcal{V}=&e^{i\vartheta}\int dt_o \frac{1}{2\pi^2\alpha' t_o} R_{a\bar{b}i\bar{j}}\biggr[ 2\eta_{||}^{a\bar{b}}\mathcal{G}_{F,T^{1,2}}^{0,1}(4\pi x^*)\delta^{(2)}_{,i\bar{j}}(k_{||})-8\eta_{||}^{a\bar{b}}\eta_{||}^{i\bar{j}}\mathcal{G}_{F,T^{1,2}}^{0,1}(4\pi x^*)\mathcal{G}_{B,T^{1,2}}(4\pi x^*)\delta^{(2)}(k_{||})\biggr]\,.\label{eqn:vert mob}
\end{align}
As we claimed, the vertical integral precisely cancels $\mathcal{I}_{ex}.$

\subsubsection{Closed string channel}
Following a similar exercise, one can again show that the vertical integral precisely cancels $\mathcal{I}_{ex}$ in the closed-string channel. As this calculation is rather tedious and does not teach us anything new, we shall omit the derivation.

\section{First order: Annulus D1-D9}\label{sec:d1-d9}
\subsection{Normalization of the one-point function}
We shall treat the annulus as a square, in the open string channel, 
\begin{equation}
    [0,\pi]\times [0,2\pi t_o]
\end{equation}
with an identification
\begin{equation}
    (x,y)\equiv (\pi-x,y)\,.
\end{equation}
Note that the annulus of interest can be constructed by quotienting the torus
\begin{equation}
    [0,4\pi]\times [0,2\pi t_o]
\end{equation}
by two involutions
\begin{equation}
    \mathcal{I}_1(z)=2\pi-\bar{z}\,,\quad \mathcal{I}_2(z)=\pi-\bar{z}\,.
\end{equation}
We shall declare that the fixed locus at $\text{Re}(z)=0$ is the ED1 boundary, and the fixed locus at $\text{Re}(z)=\pi$ is the D9-brane stack.

Let us first study a few factors that affect the overall normalization. Since there are two Neumann-Neumann directions in the one-loop diagram we shall study, we need to integrate over the two-dimensional momentum space in the loop. The integral over the loop momentum induces a factor of
\begin{equation}
    \frac{1}{8\pi^2\alpha' t_o}\,.
\end{equation}
Also, we must include a factor of $1/2$ from the GSO projection and another from the orientifold projection. Note that we must include an additional factor of $2$ to take into account two possible orientations of the open string, unlike the Möbius strip amplitude. This extra factor of 2 can be understood as integrating over $0\leq x\leq1/2$ rather than $0\leq x\leq1/4$ in the case of the Möbius amplitude.

The annulus one-point function is then normalized as
\begin{equation}
    \mathcal{A}= \frac{1}{4}\sum_s \int dt_o\int_0^{1/2} dx \frac{1}{8\pi^2\alpha' t_o} \biggr\langle \mathcal{B}_t\mathcal{B}_x c\bar{c}V^{-1,-1}(2\pi x)\mathcal{X} (y_1)\mathcal{X}(y_2)\biggr\rangle_s\,,
\end{equation}
where we inserted the closed string vertex operator at $2\pi x,$ and the Beltrami differential takes the same form as that of the Möbius amplitude
\begin{equation}
    \mathcal{B}_x=-2\pi \oint_{C_1}dz b(z)+2\pi \oint_{C_1}d\bar{z}\bar{b}(\bar{z})\,,
\end{equation}
\begin{equation}
    \mathcal{B}_t=2\pi i \oint_{C_3}dzb(z)+2\pi i\oint_{C_3}d\bar{z} \bar{b}(\bar{z})+\dots\,.
\end{equation}

To simplify the calculation, as in the Möbius one-point function, we shall average over PCO locations
\begin{equation}
    \oint_{C_1}\frac{dp_1}{p_1}\mathcal{X}(p_1)\,,\quad-\oint\frac{d \bar{p}_2}{\bar{p}_2}\overline{\mathcal{X}}(\bar{p}_2)\,.
\end{equation}

As we shall see later in the calculation, the choice of PCOs we made is incompatible with the choice of PCOs for the Feynman regions. Therefore, on the boundary of the vertex region, we shall perform vertical integration explicitly to fill the gaps in PCOs. The vertical integration is then given by
\begin{align}
    \mathcal{V}=&\frac{1}{4}\sum_s \int dm_b \frac{1}{8\pi^2\alpha't}\biggr\langle \mathcal{B}_{m_b} c\bar{c} V^{-1,-1}(2\pi x)\left[\left(\xi(y_1)-\xi(W_1) \right) \mathcal{X}(y_2)+\left(\xi(y_2)-\xi(W_2)\right)\mathcal{X}(W_1)\right] \biggr\rangle_s\,.
\end{align}
\subsection{Interior contributions}
\subsubsection{CFT correlator: Even spin structure}
The structure of the CFT correlator is analogous to that of the Möbius diagram. We compute
\begin{align}
    \mathcal{I}_m=&\frac{1}{2\pi\alpha'}\exp\left(-k_{||}^2\mathcal{G}_{NN}(4\pi x)-k_{\perp}^2\mathcal{G}_{DN}(4\pi x)\right)\delta g_{ab}\biggr[\biggr( 2\eta_{||}^{ab}\mathcal{G}''_{T^{2,1}}(4\pi x)\nonumber\\
    &-(k_{||}^a\mathcal{G}'_{NN}(4\pi x)+k_{\perp}^a\mathcal{G}'_{DN}(4\pi x))(k_{||}^b\mathcal{G}'_{NN}(4\pi x)+k_{\perp}^b\mathcal{G}'_{DN}(4\pi x)\biggr)\nonumber\\
    &-\frac{\alpha'^2}{4} \left((\eta_{||}^{ab}\mathcal{G}_{F,T}^{\alpha\beta}-\eta_{\perp}^{ab}\mathcal{G}_{F,T}^{\alpha+1,\beta})(k_{||}^2\mathcal{G}_{F,T}^{\alpha,\beta}-k_{\perp}^2\mathcal{G}_{F,T}^{\alpha+1,\beta})(4\pi x)\right)\nonumber\\
    &+\frac{\alpha'^2}{4}\left(k_{||}^a\mathcal{G}_{F,T}^{\alpha,\beta}-k_{\perp}^a\mathcal{G}_{F,T}^{\alpha+1,\beta}\right)\left(k_{||}^b\mathcal{G}_{F,T}^{\alpha,\beta}-k_{\perp}^b\mathcal{G}_{F,T}^{\alpha+1,\beta}\right)(4\pi x)\biggr]\nonumber\\
    &\times (-1)^{\alpha+\beta}\times\frac{32\vartheta_{\alpha+1,\beta}^4(\tau)}{\vartheta_{0,1}^4(\tau)\eta(\tau)^2}\,.
\end{align}
Note that the $b-c$ ghost correlator gives
\begin{equation}
    2(2\pi)^2e^{i\vartheta}\eta(\tau)^2\,.
\end{equation}

Collecting every term, we find
\begin{equation}
    \mathcal{I}=\mathcal{I}_b+\mathcal{I}_f\,,
\end{equation}
\begin{align}
    \mathcal{I}_b=&\frac{1}{4}\sum_{even~s}\int dt_o\int dx \frac{2(2\pi)^2}{16\pi^3\alpha'^2t_o}(-1)^{\alpha+\beta}\frac{2^5e^{i\vartheta}\vartheta_{\alpha+1,\beta}^4(\tau)}{\vartheta_{0,1}^4(\tau)} \nonumber\\
    &\times\biggr[2\eta_{||}^{ab}\mathcal{G}_{T^{2,1}}''(4\pi x)(1-2k_{||}^2\mathcal{G}_{T^{2,1}}(4\pi x))-4k_{||}^a\mathcal{G}_{T^{2,1}}'(4\pi x)k_{||}^b\mathcal{G}_{T^{2,1}}'(4\pi x)\biggr]\delta^{(2)}(k_{||})\,.
\end{align}
By replacing $\delta g_{ab}$ with 
\begin{equation}
    -R_{a\bar{b}i\bar{j}}\frac{\p}{\p k_i}\frac{\p}{\p k_{\bar{j}}}\,,
\end{equation}
we find
\begin{align}
    \mathcal{I}_b=&\int dt_o\int dx \frac{4e^{i\vartheta}}{\pi\alpha'^2t_o} R_{a\bar{b}i\bar{j}}\biggr[4\eta_{||}^{a\bar{b}} \mathcal{G}_{T^{2,1}}''(4\pi x)\delta_{,i\bar{j}}^{(2)}-16\eta_{||}^{a\bar{b}}\eta_{||}^{i\bar{j}} \left(\mathcal{G}_{T^{2,1}}'\mathcal{G}_{T^{2,1}}\right)'\delta^{(2)}\nonumber\\
    &+8\eta_{||}^{a\bar{j}}\eta^{i\bar{b}}_{||} (\mathcal{G}_{T^{2,1}}'(4\pi x))^2\delta^{(2)}\biggr]\,.
\end{align}

Similarly, we compute
\begin{align}
    \mathcal{I}_f=&\frac{1}{4}\sum_{even~s}\int dt_o\int dx \frac{2(2\pi)^2e^{i\vartheta}}{16\pi^3\alpha'^2t_o} (-1)^{\alpha+\beta}\delta g_{ab} \frac{ 2^5\vartheta_{\alpha+1,\beta}^4(\tau)}{\vartheta_{0,1}^4(\tau)}\times \frac{\alpha'^2}{4}\nonumber\\
    &\times \left( -4\eta_{||}^{ab}k_{||}^2+4k_{||}^ak_{||}^b\right)(\mathcal{G}_{F,T^{2,1}}^{1,0})^2\delta^{(2)}(k_{||})\,.
\end{align}
Using the replacement rule \eqref{eqn:replace rule} again, and performing the spin sum, we find
\begin{align}
    \mathcal{I}_f=&-\int dt_o\int dx \frac{8e^{i\vartheta}}{\pi t_o} \hat{R}(\mathcal{G}_{F,T^{2,1}}^{1,0})^2\delta^{(2)}(k_{||})\,.
\end{align}

\subsubsection{Moduli integral: open string channel}
We shall integrate the moduli over
\begin{equation}
    \int_{1}^{t_{o}^*}dt_o \int_{x^*}^{\frac{1}{2}}dx\,,
\end{equation}
where
\begin{equation}
    t_o^*=\frac{1}{2\pi}\log\mu^2+\frac{1}{4\pi \mu^2}+\dots\,,
\end{equation}
\begin{equation}
    x^*=\frac{1}{2\pi\mu\nu}+\frac{1}{2\pi\mu^3\nu}+\dots\,.
\end{equation}
As in the Möbius amplitude, we can decompose $\mathcal{I}$ into the exact term and the rest
\begin{equation}
    \mathcal{I}=\mathcal{I}_{ex}+\mathcal{I}_{rest}\,,
\end{equation}
where $\mathcal{I}_{ex}$ is again cancelled against the vertical integration. We have
\begin{align}
    \mathcal{I}_{rest}=&\int_{1}^{t_o^*}dt_o\int_{x^*}^{\frac{1}{2}}dx \frac{32e^{i\vartheta}}{\pi\alpha'^2t_o}\hat{R} \biggr[(\mathcal{G}_{B,T^{2,1}}'(4\pi x))^2-\frac{\alpha'^2}{4} (\mathcal{G}_{F,T^{2,1}}^{1,0}(4\pi x))^2 \biggr]\delta^{(2)}(k_{||})\,,\\
    =&\int dt_o\int dx\frac{e^{i\vartheta}}{2\pi^3t_o}\hat{R} \biggr[ \left(\frac{\vartheta_{11}'(x|\tau/2)}{\vartheta_{11}(x|\tau/2)}\right)^2-\frac{\vartheta_{10}''(0|\tau/2)}{\vartheta_{10}(0|\tau/2)}+\partial_z^2\log\vartheta_{11}(z|\tau/2)|_{z=x}\biggr]\delta^{(2)}(k_{||})\,.\label{eqn:AI rest}
\end{align}
The first term in $\mathcal{I}_{rest}$ is computed as
\begin{align}
    \mathcal{I}_{rest,1}=&\int dt_o\int dx\frac{e^{i\vartheta}}{2\pi t_o}\hat{R} \biggr[\cot(\pi x)+4\sum_{n=1}^{\infty}\frac{q^{n/2}\sin(2\pi x)}{q^{n}-2q^{n/2}\cos(2\pi x)+1} \biggr]^2\delta^{(2)}(k_{||})\,,\\
    =&\int dt_o \frac{e^{i\vartheta}}{\pi t_o}\hat{R} \biggr[ -\frac{1}{4} +\frac{1}{2\pi}\cot(\pi x^*)+\sum_{n=1}^{\infty} \left(\frac{2q^{n/2}}{1-q^{n/2}}+\sum_{n=1}^{\infty} \frac{2q^{(m+n)/2}}{1-q^{(m+n)/2}}\right)\biggr]\delta^{(2)}\,,
\end{align}
the second term is
\begin{equation}
    \mathcal{I}_{rest,2}=\int dt_o \frac{e^{i\vartheta}}{\pi t_o}\hat{R}\biggr[ \frac{1}{4}+2\sum_{n=1}^{\infty} \frac{q^{n/2}}{(1+q^{n/2})^2}\biggr]\delta^{(2)}\,,
\end{equation}
and the third term is
\begin{equation}
    \mathcal{I}_{rest,3}=-\int dt_o \frac{e^{i\vartheta}}{2\pi^2t_o}\hat{R}\cot(\pi x^*)\,.
\end{equation}
We therefore conclude
\begin{equation}
    \mathcal{I}_{rest}=\int_{1}^{t_o^*}dt_o \frac{2e^{i\vartheta}}{\pi t_o} \hat{R}\sum_{n=1}^{\infty}\biggr[\frac{q^{n/2}}{(1+q^{n/2})^2}+\frac{q^{n/2}}{1-q^{n/2}}+\sum_{m=1}^{\infty}\frac{2q^{(n+m)/2}}{1-q^{(n+m)/2}} \biggr]\delta^{(2)}\,.
\end{equation}
\subsubsection{Moduli integral: closed string channel}
The closed string channel variables are obtained by applying the $S$ transformation
\begin{equation}
    \check{\tau}=il=S(\tau)=\frac{i}{t_o}\,,\quad i\check{x}=S(x)=ilx\,.
\end{equation}
Similar to the Möbius one-point function, we shall again use the following shorthand notation
\begin{equation}
    \check{q}:=\exp(-2\pi l)\,.
\end{equation}
The moduli integral must be performed over
\begin{equation}
    \int_1^{l^*}dl\int_{\check{x}^*}^{\frac{1}{2}l}d\check{x}\,.
\end{equation}
As one can check, the first and the last term in \eqref{eqn:AI rest} are identical to those of \eqref{eqn:IM rest} modulo the relative sign. As those terms, therefore, cancel out, we shall only rewrite the second term in \eqref{eqn:AI rest} in terms of the closed string variables
\begin{equation}
    \mathcal{I}_{rest,2}=-\int_1^{l^*} dl \frac{e^{i\vartheta}}{\pi^3} \hat{R} \left[\pi  +l \frac{\vartheta_{01}''(0|2il)}{\vartheta_{01}(0|2il)}\right]\delta^{(2)}\,.
\end{equation}
As a corollary, we also confirm that the closed string tadpole cancels out upon the addition of the annulus to the Möbius amplitudes. 
\subsection{Vertical integration}
As in the Möbius one-point function, the annulus with one end on the D9-branes is divergent due to an unphysical singularity $\mathcal{I}_{ex}.$ However, the vertical integration again cancels this singularity. As the calculation is an exact copy with minor modifications of the vertical integral in the Möbius amplitude, we shall not repeat the detailed analysis here.  

\begin{figure}
\centering
    \subfloat[\centering The open string channel]{{\includegraphics[width=7cm]{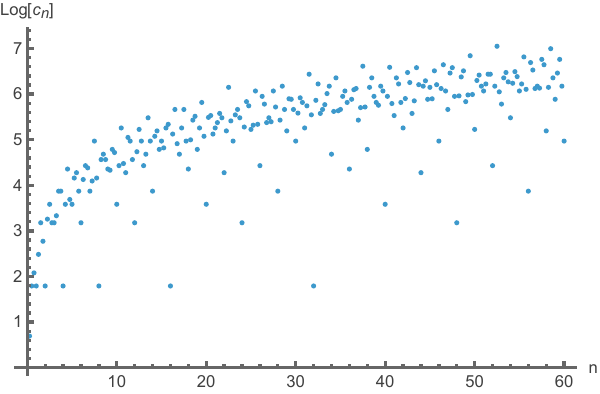} }}%
    \qquad
    \subfloat[\centering The closed string channel]{{\includegraphics[width=7cm]{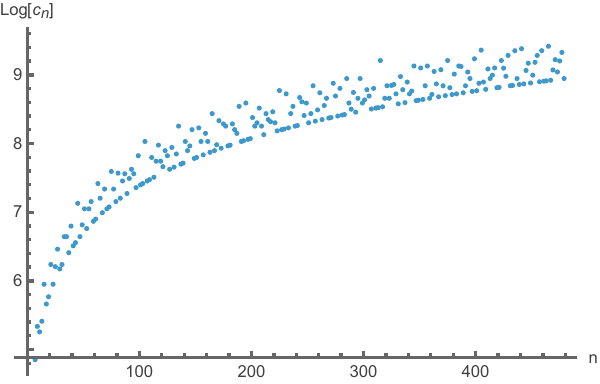} }}
    \caption{Growth of $|c_n|$}
\label{fig:growth}
\end{figure}

\section{First order: complete answer}\label{sec:first order full}
We shall now collect the results of \S\ref{sec:first d1-d1}, \S\ref{sec:first Mob}, and \S\ref{sec:d1-d9} to arrive at the full answer. As we explained in \S\ref{sec:structure}, we shall omit the open string zero modes for the numerical integral. The analysis of the overall normalization of the D-instanton amplitudes following the treatment of \cite{Alexandrov:2022mmy} will be presented in \cite{D-instanton}.

In the open string channel, we find
\begin{align}
    \frac{1}{2}(\mathcal{M}+\mathcal{A})_o=&\int_{\Sigma}d^2X\int_1^{t_o^*}dt_o\frac{e^{i\vartheta}}{4\pi^3 t_o}\hat{R} \biggr[\frac{\vartheta_{00}''(0|it_o/2)}{\vartheta_{00}(0|it_o/2)}-\frac{\vartheta_{10}''(0|it_o/2)}{\vartheta_{10}(0|it_o/2)}\biggr] +\dots\,,\\
    =&\int_{\Sigma} d^2X\int_{1}^{t_o^*}dt_o\frac{e^{i\vartheta}}{\pi t_o}\hat{R} \sum_{n=1}^{\infty}\biggr[-\frac{2q^{n/2-1/4}}{(1+q^{n/2-1/4})^2}+\frac{2q^{n/2}}{(1+q^{n/2})^2}\biggr]+\dots\,, \\
    =&\int_\Sigma d^2X \int_1^{t_o^*}dt_o \frac{e^{i\vartheta}}{\pi t_o}\hat{R} \biggr[ -2q^{1/4}+6q^{1/2}-8q^{3/4}+6q-12q^{5/4}+\dots\biggr]\,,
\end{align}
where $\dots$ out of the parentheses include terms that vanish in the large stub limit. Note that we replaced back $\delta^{(2)}$ into the spacetime integral $\int_\Sigma d^2X.$ In the closed string channel, we find
\begin{align}
    \frac{1}{2}(\mathcal{M}+\mathcal{A})_c=&-\int_\Sigma d^2X \int_1^{l^*} dl \frac{le^{i\vartheta}}{\pi^3 }\hat{R} \left[\frac{\vartheta_{00}''(0|2il)}{\vartheta_{00}(0|2il)}-\frac{\vartheta_{01}''(0|2il)}{\vartheta_{01}(0|2il)}\right]\,,\\
    =&\int_\Sigma d^2 X\int_1^{l^*} dl\frac{8le^{i\vartheta}}{\pi}\hat{R}\sum_{n=1}^{\infty}\biggr[\frac{\check{q}^{2n-1}}{(1+\check{q}^{2n-1})^2}+\frac{\check{q}^{2n-1}}{(1-\check{q}^{2n-1})^2} \biggr]\,,\\
    =&\int_\Sigma d^2X \int_1^{l^*} \frac{8l e^{i\vartheta}}{\pi} \hat{R}\biggr[2\check{q}+8\check{q}^3+12\check{q}^5+16\check{q}^7+26\check{q}^9+\dots\biggr]\,.
\end{align}
Note that 
\begin{equation}
    \int_{\Sigma}d^2X\frac{1}{4\pi}\hat{R}=\chi(\Sigma)\,,
\end{equation}
where $\chi(\Sigma)$ is the Euler characteristic of the D1-instanton on $\Sigma.$ 

By performing a numerical integral, we find
\begin{equation}
    \frac{1}{2}(\mathcal{A}+\mathcal{M})'=\chi(\Sigma)\times 4.8364 \times 10^{-1}\,.
\end{equation}
The integral converges rather quickly, as the coefficient $c_n$ of $q^n$ in the square bracket grows polynomially. We plot the growth of $c_n$ over the exponent $n$ in Fig.\ref{fig:growth}. Note that we fixed $e^{i\vartheta}=-1.$

\section{Conclusions}\label{sec:conclusions}
In this draft, we developed a practical framework for computing string loop amplitudes in Calabi–Yau orientifolds in the large-volume approximation using the patch-by-patch description of superstring field theory. The key technical point is that naive global choices of picture-changing operator (PCO) insertions can fail to reproduce the required behavior in degeneration limits, leading to spurious divergences and gauge-choice dependence. We showed how this is resolved by combining a convenient PCO prescription in the interior of moduli space with the vertical-integration boundary contributions dictated by string field theory, thereby restoring the correct factorization properties.

As a concrete application, we computed the D1-instanton one-loop partition function to first order in the large-volume expansion. In this computation, the potentially problematic exact pieces arising from PCO-mismatch are canceled by the corresponding vertical-integration terms, yielding an unambiguous answer at this order. 

The prescription we built is based on the patch-by-patch description of string field theory \cite{Frenkel:2025wko}, which provides a systematic treatment of the $\alpha'$ expansion. As treatment of Ramond-Ramond fluxes in string field theory are not very different from the treatment of non-trivial NSNS backgrounds, it is expected that the method presented in this paper can be applied to general flux compactifications in the large volume expansion.

There is much work to be done. Among the urgent problems relevant for establishing theoretical control of cosmological solutions in string theory, two natural targets for the methods developed here are: (i) loop and D-instanton corrections to the Kähler potential in flux compactifications, and (ii) the overall normalization of D-instanton amplitudes. Both require a consistent treatment of PCO choices and their behavior near the boundaries of moduli space; in this respect, algorithmic implementations of PCO-fixing and vertical integration may be valuable.
\section*{Acknowledgments}
The author thanks Ranveer Singh, Shivansh Tomar, and Raghu Mahajan for their collaboration in the early stage of this work. The author thanks Daniel Harlow for encouraging the author to work on the problem of computing string loops in the large volume limit while the author was a postdoc at MIT. The author thanks Michael Haack and Atakan Hilmi Fırat for comments on the draft. The author thanks Alex Frenkel, Andreas Schachner, Jakob Moritz, Liam McAllister, Michael Haack, Stephen Shenker, Edward Witten, and Atakan Hilmi Fırat for illuminating discussions. The author thanks the warm hospitality of KIAS, Cornell University, Peking University, and UW-Madison during the completion of this work. This work was performed in part at the Aspen Center for Physics, which is supported by a grant from the Simons Foundation (1161654, Troyer). The work of MK is supported by Simons Investigator award (MPS-SIP-00507021).

\newpage
\appendix
\section{Worldsheet conventions}\label{sec:conventions}
In this section, we shall set the convention for the worldsheet CFTs. The worldsheet CFT enjoys $(\mathcal{N},\overline{\mathcal{N}})=(1,1)$ worldsheet supersymmetry. For the matter CFT, we choose the free field CFTs with central charge $15,$ with bosonic fields $X^\mu$ and fermionic fields $\psi^\mu$ and $\bar{\psi}^\mu.$ In the ghost system, we have the usual $b,~c,~\beta,~\gamma$ fields, both holomorphic and anti-holomorphic. We shall use the bosonized $\beta,~\gamma$ fields \cite{Friedan:1985ge}
\begin{equation}
    \beta=\partial \xi e^{-\phi}\,,\quad \gamma=\eta e^\phi\,.
\end{equation}

We shall normalize the worldsheet fields such that the singular parts of the OPEs of the worldsheet fields are given as
\begin{align}
    &X^\mu(z,\bar{z}) X^\nu(0,0)\sim-\frac{\alpha'}{2}\eta^{\mu\nu}\log |z|^2\,,\quad \psi^\mu(z)\psi^\nu(0)\sim\frac{\eta^{\mu\nu}}{z}\,,\\
    &c(z)b(0)\sim\frac{1}{z}\,,\quad \xi(z)\eta(0)\sim\frac{1}{z}\,,\quad \partial\phi(z)\partial\phi(0)\sim-\frac{1}{z^2}\,,\quad e^{q_1\phi}(z)e^{q_2\phi}(0)\sim z^{-q_1q_2} e^{(q_1+q_2)\phi}(0)\,.
\end{align}
We write the matter stress-energy tensor and the superconformal current
\begin{equation}
    T_m=-\frac{1}{\alpha'}\partial X\cdot \partial X-\frac{1}{2}\psi\cdot \partial\psi\,,
\end{equation}
\begin{equation}
    G_m=i\sqrt{\frac{2}{\alpha'}}\psi\cdot\partial X\,.
\end{equation}
We normalize the BRST current as
\begin{equation}
    j_B=c\left(T_m-\frac{1}{2}(\partial\phi)^2-\partial^2\phi -\eta\partial\xi\right)+\eta e^\phi G_m+bc\partial c-\eta\partial\eta be^{2\phi}\,.
\end{equation}
The BRST charge is then, correspondingly, normalized as
\begin{equation}
    Q_B:= \oint dz j_B-\oint d\bar{z} \bar{j}_B\,.
\end{equation}
Finally, we normalize the picture changing operator (PCO) as
\begin{equation}
    \mathcal{X}:=\{Q_B,\xi\}=c\partial\xi+e^\phi G_m-\partial\eta be^{2\phi}-\partial(\eta b e^{2\phi})\,.
\end{equation}

\section{Theta function}\label{app:theta functions}
In this appendix, we summarize the definition of the theta functions and related identities. 

For $q:=e^{2\pi i\tau},$ and $y=e^{2\pi iz},$ we define
\begin{equation}
    \vartheta_{\alpha\beta}(z|\tau):=\sum_{n\in\Bbb{Z}+\frac{\alpha}{2}} e^{i\pi n\beta} q^{n^2/2}y^n\,.
\end{equation}
We also introduce a shorthand notation $\vartheta_{\alpha\beta}(\tau):=\vartheta_{\alpha\beta}(0|\tau).$ The theta functions enjoy the following product representations
\begin{align}
    \vartheta_{00}(z|\tau)=&\prod_{n=1}^\infty (1-q^n)(1+(y+y^{-1})q^{n-\frac{1}{2}}+q^{2n-1})\,,\\
    \vartheta_{01}(z|\tau)=&\prod_{n=1}^\infty (1-q^n)(1-(y+y^{-1})q^{n-\frac{1}{2}}+q^{2n-1})\,,\\
    \vartheta_{10}(z|\tau)=&q^{\frac{1}{8}}(y^{\frac{1}{2}}+y^{-\frac{1}{2}})\prod_{n=1}^\infty (1-q^n)(1+(y+y^{-1})q^n+q^{2n})\,,\\
    \vartheta_{11}(z|\tau)=&iq^{\frac{1}{8}}(y^{\frac{1}{2}}-y^{-\frac{1}{2}})\prod_{n=1}^{\infty}(1-q^n)(1-(y+y^{-1})q^n+q^{2n})\,.
\end{align}
The theta functions satisfy the following relations
\begin{equation}
    \vartheta_{\alpha\beta}(z+1/2|\tau)=\vartheta_{\alpha,\beta+1}(z|\tau)\,,
\end{equation}
\begin{equation}
    \vartheta_{\alpha\beta}(z+\tau/2|\tau)=e^{-i\pi \beta/2}q^{-\frac{1}{8}} y^{-\frac{1}{2}}\vartheta_{\alpha+1,\beta}(z|\tau)\,,
\end{equation}
with the quasi-periodicity properties
\begin{equation}
    \vartheta_{\alpha+2,\beta}=\vartheta_{\alpha\beta}\,,\quad \vartheta_{\alpha,\beta+2}=e^{i\alpha\pi}\vartheta_{\alpha\beta}\,.
\end{equation}
Note that in the literature, often, the theta functions are written as $\vartheta_i,$ for $i=1,\dots,4.$ These theta functions are related to the theta functions we defined via
\begin{equation}
    \vartheta_1=-\vartheta_{11}\,,\quad \vartheta_2=\vartheta_{10}\,,\quad \vartheta_3=\vartheta_{00}\,,\quad \vartheta_4=\vartheta_{01}\,.
\end{equation}

We also define the Dedekind eta function as
\begin{equation}
    \eta(\tau):=q^{\frac{1}{24}}\prod_{n=1}^\infty(1-q^n)\,.
\end{equation}

We now list useful identities. First, \cite{Alexandrov:2022mmy}
\begin{equation}
    \frac{\vartheta_{\alpha\beta}(z|\tau)^2\vartheta'_{11}(0|\tau)^2}{\vartheta_{11}(z|\tau)^2\vartheta_{\alpha\beta}(0|\tau)^2}=\frac{\vartheta''_{\alpha\beta}(0|\tau)}{\vartheta_{\alpha\beta}(0|\tau)}-\partial_z^2\log\vartheta_{11}(z|\tau)\,.
\end{equation}
Second, \cite{Berg:2005ja}
\begin{equation}
    \sum_{\alpha,\beta}(-1)^{\alpha+\beta} \vartheta_{\alpha,\beta}(z|\tau) \vartheta_{\alpha+h_1,\beta+g_1}(z|\tau) \prod_{i=2}^3\vartheta_{\alpha+h_i,\beta+g_i}(0|\tau)^2=\vartheta_{11}(z|\tau)\vartheta_{1+h_1,1+g_1}(z|\tau) \prod_{i=2}^3\vartheta_{1+h_i,1+g_i}(\tau)\,.
\end{equation}
A corollary of this identity is the famous abstruse identity
\begin{equation}
    \vartheta_{00}^4-\vartheta_{01}^4-\vartheta_{10}^4=0\,.
\end{equation}

\section{Green's function}\label{app:Green}
In this section, we shall summarize Green's function we need for the computation of the one-loop partition function. For convenience, we shall parametrize the torus with a coordinate $z$ that has the following identifications
\begin{equation}
    z\equiv z+1\equiv z+\tau\,.
\end{equation}
This convention is different from the main text. We shall therefore be careful in translating between different conventions.

We start by summarizing correlation functions on the worldsheet torus $T:= \Bbb{C}/(\Bbb{Z}+\tau \Bbb{Z})$ with the spin structure $(\alpha,\beta).$ The bosonic correlator is given as
\begin{equation}
    \mathcal{G}_{B,T}(z_1;z_2):=\langle X(z_1)X(z_2)\rangle_{T^2}=-\frac{\alpha'}{2} \log\left|\frac{\vartheta_{11}(z_1-z_2|\tau)}{\vartheta_{11}'(0|\tau)}\right|^2+\alpha'\pi\frac{\im(z_1-z_2)^2}{\im\tau}\,.
\end{equation}
From now on, often, we will omit the subscript $B.$ The fermionic correlator in the even spin structure is given as
\begin{equation}
    \mathcal{G}_{F,T}^{\alpha,\beta}(z_1;z_2)\eta^{\mu\nu}:=\langle\psi^\mu (z_1)\psi^\nu(z_2)\rangle_T^{\alpha\beta}=\frac{\vartheta_{\alpha\beta}(z_1-z_2|\tau)\vartheta_{11}'(0|\tau)}{\vartheta_{\alpha\beta}(0|\tau)\vartheta_{11}(z_1-z_2|\tau)}\eta^{\mu\nu}\,,
\end{equation}
and in the odd spin structure is given as
\begin{equation}
    \mathcal{G}_{F,T}^{1,1}(z_1;z_2)= \frac{\vartheta_{11}'(z_1-z_2|\tau)}{\vartheta_{11}(z_1-z_2|\tau)}+2i\pi \frac{\im (z_1-z_2)}{\tau_2}\,.
\end{equation}

We shall construct the annulus and the Möbius strip by quotienting the torus. For the annulus, we shall choose $\tau=it,$ and for the Möbius strip, we shall choose $\tau=it+1/2.$ The annulus and Möbius strip are then identified by the following involution
\begin{equation}
    I(z)=1-\bar{z}\,.
\end{equation}

Let us first summarize the CFT correlators for the annulus diagram. First, the bosonic correlator for the Neumann-Neumann directions is
\begin{align}
    \mathcal{G}_{B,NN}(z_1;z_2):=& \frac{1}{2} (\mathcal{G}_{B,T}(z_1;z_2)+\mathcal{G}_{B,T}(z_1;I(z_2))+\mathcal{G}_{B,T}(I(z_1);z_2)+\mathcal{G}_{B,T}(I(z_1);I(z_2))\,,\\
    =&\mathcal{G}_{B,T}(z_1;z_2)+\mathcal{G}_{B,T}(z_1;I(z_2))\,.
\end{align}
For the Dirichlet-Dirichlet directions, we have
\begin{align}
    \mathcal{G}_{B,DD}(z_1;z_2)=& \frac{1}{2} (\mathcal{G}_{B,T}(z_1;z_2)-\mathcal{G}_{B,T}(z_1;I(z_2))-\mathcal{G}_{B,T}(I(z_1);z_2)+\mathcal
    {G}_{B,T}(I(z_1);I(z_2))\,,\\
    =& \mathcal{G}_{B,T}(z_1;z_2)-\mathcal{G}_{B,T}(z_1;I(z_2))\,.
\end{align}

To obtain the correlation functions for the Neumann-Dirichlet and Dirichlet-Neumann, we shall work a little more. Following \cite{Epple:2004nh}, let us consider a squashed torus $T':=\Bbb{C}/(2\Bbb{Z}+\tau\Bbb{Z}).$ We shall then quotient $T'$ by two involutions
\begin{equation}
    I_1(z):=-\bar{z}\,,\quad I_2(z):=1-\bar{z}\,.
\end{equation}
The fixed points of $I_1(z)$ are $\text{Re}(z)=0,1$ and the fixed points of $I_2(z)$ are $\text{Re}(z)=1/2,~3/2.$ We shall assume that we impose the Neumann boundary condition on $\text{Re}(z)=0,$ and the Dirichlet boundary condition on $\text{Re}(z)=1/2.$ Then, the boson correlator in the annulus is found as
\begin{equation}
    \mathcal{G}_{ND}(z_1;z_2)= \mathcal{G}_{B,T'}(z_1;z_2)+\mathcal{G}_{B,T'}(z_1;I_1(z_2))-\mathcal{G}_{B,T'}(z_1;I_2(z_2))-\mathcal{G}_{B,T'}(z_1;I_1(I_2(z_2)))\,.
\end{equation}
Similarly, for the Dirichlet boundary condition on $\text{Re}(z)=0$ and the Neumann boundary condition on $\text{Re}(z)=1/2,$ the bosonic Green's function is given as
\begin{equation}
    \mathcal{G}_{DN}(z_1;z_2)=\mathcal{G}_{B,T'}(z_1;z_2)-\mathcal{G}_{B,T'}(z_1;I_1(z_2))+\mathcal{G}_{B,T'}(z_1;I_2(z_2))-\mathcal{G}_{B,T'}(z_1;I_1(I_2(z_2)))\,.
\end{equation}

We now summarize the fermion correlators. For the Neumann-Neumann directions, we have
\begin{equation}
    \langle \psi(z_1)\psi(z_2)\rangle_{A,NN}^{\alpha\beta}= \mathcal{G}_{F,T}^{\alpha\beta}(z_1;z_2)\,,\quad \langle\psi(z_1)\bar{\psi}(\bar{z}_2)\rangle_{A,NN}^{\alpha\beta}=i\mathcal{G}_{F,T}^{\alpha\beta}(z_1;I(z_2))\,,
\end{equation}
\begin{equation}
    \langle\bar{\psi}(\bar{z}_1)\psi(z_2)\rangle_{A,NN}^{\alpha\beta}=i\mathcal{G}_{F,T}^{\alpha\beta}(I(z_1);z_2)\,,\quad \langle\bar{\psi}(\bar{z}_1)\bar{\psi}(\bar{z}_2)\rangle= -\mathcal{G}_{F,T}^{\alpha\beta}(I(z_1);I(z_2))\,.
\end{equation}
We can, for example, write the fermion correlator in terms of the image charge as well
\begin{equation}
    \langle\psi(z_1)\bar{\psi}(\bar{z}_2)\rangle_{A,NN}^{\alpha\beta}=i\mathcal{G}_{F,T^{2,1}}^{1\beta}(z_1,I_1(z_2))+i(-1)^{\alpha+1}\mathcal{G}_{F,T^{2,1}}^{1\beta}(z_1,I_2(z_2))\,.
\end{equation}
For the Dirichlet-Dirichlet directions, we have
\begin{equation}
    \langle \psi(z_1)\psi(z_2)\rangle_{A,DD}^{\alpha\beta}= \mathcal{G}_{F,T}^{\alpha\beta}(z_1;z_2)\,,\quad \langle\psi(z_1)\bar{\psi}(\bar{z}_2)\rangle_{A,DD}^{\alpha\beta}=-i\mathcal{G}_{F,T}^{\alpha\beta}(z_1;I(z_2)\,,
\end{equation}
\begin{equation}
    \langle\bar{\psi}(\bar{z}_1)\psi(z_2)\rangle_{A,DD}^{\alpha\beta}=-i\mathcal{G}_{F,T}^{\alpha\beta}(I(z_1);z_2)\,,\quad \langle\bar{\psi}(\bar{z}_1)\bar{\psi}(\bar{z}_2)\rangle_{DD}^{\alpha\beta}= -\mathcal{G}_{F,T}^{\alpha\beta}(I(z_1);I(z_2))\,.
\end{equation}

To derive the Green's function of fermions for the Dirichlet-Neumann directions, we shall use the fact that the composition of the two involutions results in a constant shift
\begin{equation}
    I_1\cdot I_2(z)=1+z\,.
\end{equation}
As the fermion should change its sign under the involution $I_1,$ under the compositions of the involutions $I_1\cdot I_2,$ the Green's function must change its phase by $(-1)^\alpha$ for the spin structure $(\alpha,\beta).$ Therefore, we find
\begin{equation}
    \langle\psi(z_1)\psi(z_2)\rangle_{A,DN}^{\alpha\beta}=\mathcal{G}_{F,T}^{\alpha+1,\beta}(z_1;z_2)\,.
\end{equation}
To obtain the correlators $\langle \psi(z_1)\bar{\psi}(\bar{z}_2)\rangle$ and $\langle \bar{\psi}(\bar{z}_1)\psi(z_2)\rangle,$ we shall again go to the torus $T',$ and quotient the torus by
\begin{equation}
    I_1(z):=-\bar{z}\,,\quad I_2(z):= 1-\bar{z}\,.
\end{equation}
Then, we can write
\begin{equation}
    \langle\psi(z_1)\psi(z_2)\rangle_{A,DN}^{\alpha\beta}= \mathcal{G}_{F,T'}^{1\beta}(z_1;z_2)+(-1)^{\alpha} \mathcal{G}_{F,T'}^{1\beta}(z_1;z_2+1)\,,
\end{equation}
\begin{equation}
    \langle \psi(z_1)\bar{\psi}(\bar{z}_2)\rangle_{DN}= -i \mathcal{G}_{F,T'}^{1\beta}(z_1;-z_2) -i(-1)^\alpha\mathcal{G}_{F,T'}^{1\beta}(z_1;1-z_2)\,,
\end{equation}
\begin{equation}
    \langle\bar{\psi}(\bar{z}_1)\psi(z_2)\rangle_{A,DN}^{\alpha\beta}=-i \mathcal{G}_{F,T'}^{1\beta}(-z_1;z_2)-i(-1)^\alpha\mathcal{G}_{F,T'}^{1\beta}(-z_1;1+z_2)\,.
\end{equation}
Hence, we conclude
\begin{equation}
    \langle \psi(z_1)\bar{\psi}(z_2)\rangle_{A,DN}^{\alpha\beta}= -i\mathcal{G}_{F,T}^{\alpha+1,\beta}(z_1;-z_2)\,,
\end{equation}
and
\begin{equation}
    \langle \bar{\psi}(\bar{z}_1)\psi(z_2)\rangle_{A,DN}^{\alpha\beta}=-i\mathcal{G}_{F,T}^{\alpha+1,\beta}(-z_1;z_2)\,.
\end{equation}

We shall now summarize the relevant CFT correlation functions on the Möbius strip. The Möbius strip, as we previously mentioned, can be obtained by taking the quotient of a torus with $\tau=it+1/2,$ by an involution $I(z)=1-\bar{z},$ and treating the fundamental domain as $[1/2,1]\times [0,it].$ Most of the correlators are the same as the annulus case, provided that we appropriately change the modulus of the covering torus. We need to, however, make a modification to the fermion correlator when the fermion is along the Dirichlet direction. As the fermion goes around $z\mapsto z+\tau,$ the sign of the fermion flips additionally as the orientifold involution changes the sign of the worldsheet fermions. This, in particular, acts as changing the spin structure from $(\alpha,\beta)$ to $(\alpha,\beta+1).$ 

Let us be more explicit. A simple idea that would be helpful for us to construct the Green's function on the Möbius strip in a rather ``covariant" way is to note that the Green's function on a torus $T^{1,1}:=\Bbb{C}/(\Bbb{Z}+\tau\Bbb{Z}),$ $\mathcal{G}_{T^{1,1}},$ is related to the Green's function on a torus $T^{n,m}:= \Bbb{C}/(n \Bbb{Z}+\tau m \Bbb{Z}),$ $\mathcal{G}_{T^{n,m}},$ via the method of image charges
\begin{equation}
    \mathcal{G}_{T^{1,1}}(x;y)= \sum_{p=0}^{n-1}\sum_{q=0}^{m-1} \mathcal{G}_{T^{n,m}}(x+p+q \tau;y)\,.
\end{equation}

Now, we shall construct the Möbius strip by quotienting $T^{1,2}$ with two involutions
\begin{equation}
    I_1(z)=1-\bar{z}\,,\quad I_2(z)=z+\tau\,,
\end{equation}
such that the fundamental domain of the Möbius strip is determined to be $[1/2,1]\times [0,\tau_2].$ 

We first construct the bosonic correlators. First, let us study the bosonic correlator for the worldsheet boson that satisfies the Neumann boundary condition on the boundary and does not change the sign under the orientifolding. Following our convention of the annulus amplitudes, we shall refer to such a boundary condition as Neumann-Neumann, although $z+\tau\equiv z$ is not a boundary. The bosonic correlator is then determined as
\begin{align}
    \langle X(x) X(y)\rangle_{M,NN} =& \mathcal{G}_{T^{1,2}}(x;y) +\mathcal{G}_{T^{1,2}}(1-\bar{x};y)+\mathcal{G}_{T^{1,2}}(x+\tau;y)+\mathcal{G}_{T^{1,2}}(1+\tau-\bar{x};y)\,,\\
    =&\mathcal{G}_{T^{1,1}}(x;y)+\mathcal{G}_{T^{1,1}}(1-\bar{x};y)\,.
\end{align}
Similarly, for the Dirichlet-Dirichlet direction, we find
\begin{align}
    \langle X(x)X(y)\rangle_{M,DD}=&\mathcal{G}_{T^{1,2}}(x;y) -\mathcal{G}_{T^{1,2}}(1-\bar{x};y)+\mathcal{G}_{T^{1,2}}(x+\tau;y)-\mathcal{G}_{T^{1,2}}(1+\tau-\bar{x};y)\,,\\
    =&\mathcal{G}_{T^{1,1}}(x;y)-\mathcal{G}_{T^{1,1}}(1-\bar{x};y)\,.
\end{align}
For the Neumann-Dirichlet direction, we find
\begin{align}
    \langle X(x)X(y)\rangle_{M,ND}=&\mathcal{G}_{T^{1,2}}(x;y) +\mathcal{G}_{T^{1,2}}(1-\bar{x};y)-\mathcal{G}_{T^{1,2}}(x+\tau;y)-\mathcal{G}_{T^{1,2}}(1+\tau-\bar{x};y)\,.
\end{align}
Finally, for the Dirichlet-Neumann direction, we have
\begin{align}
    \langle X(x)X(y)\rangle_{M,DN}=&\mathcal{G}_{T^{1,2}}(x;y) -\mathcal{G}_{T^{1,2}}(1-\bar{x};y)-\mathcal{G}_{T^{1,2}}(x+\tau;y)+\mathcal{G}_{T^{1,2}}(1+\tau-\bar{x};y)\,.
\end{align}
As one can check, the Green's functions we found are compatible with T-duality.

Now, let us find the fermion correlators. For the Neumann-Neumann directions, we can write the fermion Green's function as
\begin{equation}
\langle\psi(z_1)\psi(z_2)\rangle_{M,NN}^{\alpha\beta}=\mathcal{G}_{F,T^{1,2}}^{\alpha1}(z_1;z_2)+(-1)^{\beta+1} \mathcal{G}_{F,T^{1,2}}^{\alpha1}(z_1;z_2+\tau)\,.
\end{equation}
Let us check the periodicity of the Green's function. As we shift $z_1-z_2$ to $z_1-z_2\pm 1,$ and $z_1-z_2\pm \tau,$ the Green's function must change its phase by
\begin{equation}
    \mathcal{G}\mapsto (-1)^{\alpha+1}\mathcal{G}\,,
\end{equation}
and
\begin{equation}
    \mathcal{G}\mapsto (-1)^{\beta+1}\mathcal{G}\,. 
\end{equation}
The periodicity of the Green's function under $z_1-z_2\pm1$ is automatically satisfied. Let's therefore check the periodicity under $z_1-z_2\mapsto z_1-z_2+\tau$
\begin{equation}
    \langle \psi(z_1+\tau)\psi(z_2)\rangle_{M,NN}^{\alpha,\beta}=\mathcal{G}_{F,T^{1,2}}^{\alpha,1}(z_1;z_2+\tau)+(-1)^{\beta+1} \mathcal{G}_{F,T^{1,2}}^{\alpha,1}(z_1;z_2)=(-1)^{\beta+1} \langle \psi(z_1)\psi(z_2)\rangle_{M,NN}^{\alpha,\beta}\,.
\end{equation}

We have
\begin{equation}
    \langle \psi(z_1)\psi(z_2)\rangle_{M,NN}^{\alpha\beta}= \mathcal{G}_{F,T}^{\alpha\beta}(z_1;z_2)\,,\quad \langle\psi(z_1)\bar{\psi}(\bar{z}_2)\rangle_{M,NN}^{\alpha\beta}=i\mathcal{G}_{F,T}^{\alpha\beta}(z_1;I(z_2))\,,
\end{equation}
\begin{equation}
    \langle\bar{\psi}(\bar{z}_1)\psi(z_2)\rangle_{M,NN}^{\alpha\beta}=i\mathcal{G}_{F,T}^{\alpha\beta}(I(z_1);z_2)\,,\quad \langle\bar{\psi}(\bar{z}_1)\bar{\psi}(\bar{z}_2)\rangle_{M,NN}= -\mathcal{G}_{F,T}^{\alpha\beta}(I(z_1);I(z_2))\,.
\end{equation}
For the Dirichlet-Dirichlet directions, we have
\begin{equation}
    \langle \psi(z_1)\psi(z_2)\rangle_{M,DD}^{\alpha\beta}= \mathcal{G}_{F,T}^{\alpha\beta}(z_1;z_2)\,,\quad \langle\psi(z_1)\bar{\psi}(\bar{z}_2)\rangle_{M,DD}^{\alpha\beta}=-i\mathcal{G}_{F,T}^{\alpha\beta}(z_1;I(z_2))\,,
\end{equation}
\begin{equation}
    \langle\bar{\psi}(\bar{z}_1)\psi(z_2)\rangle_{M,DD}^{\alpha\beta}=-i\mathcal{G}_{F,T}^{\alpha\beta}(I(z_1);z_2)\,,\quad \langle\bar{\psi}(\bar{z}_1)\bar{\psi}(\bar{z}_2)\rangle_{M,DD}^{\alpha\beta}= -\mathcal{G}_{F,T}^{\alpha\beta}(I(z_1);I(z_2))\,.
\end{equation}

For the Dirichlet-Neumann directions, on the other hand, we have
\begin{equation}
    \langle \psi(z_1)\psi(z_2)\rangle_{M,DN}^{\alpha\beta}=\mathcal{G}_{F,T^{1,1}}^{\alpha,\beta+1}(z_1;z_2)\,.
\end{equation}
This formula can be understood as follows. As we move along the real line, $z\mapsto z+1,$ the worldsheet fermion comes back to the original site with the phase $(-1)^\alpha.$ On the other hand, as we move along the imaginary line, $z\mapsto z+\tau_2,$ the worldsheet fermion crosses the cross cap $\Omega.$ Because $\Omega$ maps $\psi$ to $\bar{\psi},$ while the boundary condition maps $\bar{\psi}$ to $-\psi,$ the net effect of $\Omega$ is to flip the sign of $\psi,$ acting as if we inserted $(-1)^F.$ Therefore, the phase we gain in this case is $(-1)^{\beta}.$ The same fermion correlator can be found by applying the method of image charges
\begin{equation}
    \langle\psi(z_1)\psi(z_2)\rangle_{M,DN}^{\alpha\beta}= \mathcal{G}_{F,T^{1,2}}^{\alpha,1}(z_1;z_2)+(-1)^{\beta} \mathcal{G}_{F,T^{1,2}}^{\alpha,1}(z_1;z_2+\tau)\,.
\end{equation}
Similarly, we find
\begin{equation}
    \langle\psi(z_1)\bar{\psi}(\bar{z}_2)\rangle_{M,DN}^{\alpha\beta}=-i \mathcal{G}_{F,T^{1,2}}^{\alpha,1}(z_1;1-\bar{z}_2)-i(-1)^{\beta} \mathcal{G}_{F,T^{1,2}}^{\alpha,1}(z_1;1-\bar{z}_2+\tau)\,,
\end{equation}
and
\begin{equation}
    \langle \bar{\psi}(\bar{z}_1)\psi(z_2)\rangle_{M,DN}^{\alpha\beta}= -i\mathcal{G}_{F,T^{1,2}}^{\alpha,1}(1-\bar{z}_1;z_2)-i(-1)^{\beta}\mathcal{G}_{F,T^{1,2}}^{\alpha,1}(1-\bar{z}_1;z_2+\tau)\,.
\end{equation}
We can therefore rewrite the mixed fermion correlators as
\begin{equation}
    \langle \psi(z_1)\bar{\psi}(\bar{z}_2)\rangle_{M,DN}^{\alpha\beta}=-i \mathcal{G}_{F,T^{1,1}}^{\alpha,\beta+1}(z_1;1-\bar{z}_2)\,,\quad \langle \bar{\psi}(\bar{z}_1)\psi(z_2)\rangle_{M,DN}^{\alpha\beta}=-i \mathcal{G}_{F,T^{1,1}}^{\alpha,\beta+1}(1-\bar{z}_1;z_2)\,. 
\end{equation}

\section{bc ghost correlator on torus}\label{app:bc ghost}
In this section, we shall obtain the bc-ghost correlator on a torus. This result will be used for the vertical integration, in particular. 

On the torus, there is one zero mode for the b-ghost and the c-ghost. Therefore, the ghost correlator orthogonal to the zero modes satisfies the following differential equation
\begin{equation}
    \partial_{\bar{z}} \langle b(z) c(0)\rangle= 2\pi \delta^{(2)}(z)-\frac{1}{4\pi\tau_2}\,.
\end{equation}
Therefore, we find that the b-c ghost correlator orthogonal to their zero modes is given by
\begin{equation}
    \langle b(z)c(0)\rangle= -\frac{2}{\alpha'}\partial_z \mathcal{G}_{B,T}(z;0)\,.
\end{equation}

\section{Zero modes of $\xi$ in the large Hilbert space}\label{app:xi vertical}
Let us digress first on a curious observation. The total result in the vertical integration must be independent of where we insert $\xi(x_0)$ to soak up the $\xi$ zero mode in the large Hilbert space, as long as the $x_0$ does not hit vertex operators or PCOs. Now, let us do a thought experiment where we put $x_0$ at $W_1+\varepsilon,$ where $\varepsilon$ is treated as a small parameter. As long as $\varepsilon$ is not zero, the vertical integration shall not depend on $\varepsilon.$ On the other hand, as $\xi(W_1+\epsilon)$ approaches $\mathcal{X}(W_1)$ in the vertical integration, we shall find divergences that scale as
\begin{equation}\label{eqn:div xi zero}
    \xi(W_1+\epsilon) \mathcal{X}(W_1)\sim -\frac{2}{\epsilon^2} b e^{2\phi}(W_1) -\frac{1}{\epsilon}\partial (be^{2\phi})(W_1)\,.
\end{equation}
To reconcile this divergence with the independence of the results on $\epsilon,$ in light of the absence of an additional divergence that can cancel \eqref{eqn:div xi zero}, there must either be an additional vanishing contribution or the vertical integration due to $(\xi(p_2)-\xi(W_2))\mathcal{X}(W_1)$ must vanish altogether in the small $\varepsilon$ limit. However, the second option sounds surprising in that if the second option were to be true, the vertical integration would be insensitive to $W_2$ completely. 

To study if $\varepsilon$ dependence drops out, as it should, we shall explicitly evaluate the following CFT correlator in the large Hilbert space 
\begin{equation}
    \mathcal{K}_{\beta-\gamma}(x,x_0,p_2,W_1,W_2)=\langle e^{-\phi}e^{-\bar{\phi}}(2\pi x)\xi(2\pi  x_0) (\xi(2\pi  p_2)-\xi(2\pi  W_2))  \eta e^{2\phi}(2\pi  W_1)\rangle\,,
\end{equation}
with $x_0=W_1+\varepsilon.$ Note that $\mathcal{K}_{gh}$ does not directly enter the vertical integration. We find
\begin{align}
    \mathcal{K}_{\beta-\gamma}=& \frac{\vartheta_s(x_0 +p_2)}{\vartheta_s(x_0+W_1)\vartheta_s(p_2+W_1)}\frac{E(W_1-x)^2 E(W_1+x)^2 E(x_0-p_2)}{E(x_0-W_1)E(p_2-W_1)E(2x)}\nonumber\\
    &-\frac{\vartheta_{s}(x_0+W_2)}{\vartheta_s(x_0+W_1)\vartheta_s(W_2+W_1)}\frac{E(W_1-x)^2E(W_1+x)^2E(x_0-W_2)}{E(x_0-W_1)E(W_2-W_1)E(2x)}\,,\\
    =&\frac{E(W_1-x)^2E(W_1+x)^2}{E(2x)E(x_0-W_1)\vartheta_s(x_0+W_1)} \biggr(\frac{E(x_0-p_2)\vartheta_s(x_0+p_2)}{E(p_2-W_1)\vartheta_s(p_2+W_1)}\nonumber\\
    &\qquad\qquad\qquad\qquad\qquad\qquad-\frac{E(x_0-W_2)\vartheta_s(x_0+W_2)}{E(W_2-W_1)\vartheta_s(W_1+W_2)}\biggr)
\end{align}
Recall the trisecant identity \cite{fay2006theta}
\begin{align}
    \nonumber&E(x,v)E(u,y)\vartheta_s(z+x-u)\vartheta_s(z+y-v) -E(x,u)E(v,y)\vartheta_s(z+x-v)\vartheta_s(z+y-u)\\
    =&E(x,y)E(u,v)\vartheta_s(z)\vartheta_s(z+x+y-u-v)\,.
\end{align}
The goal is to use the trisecant identity to remove the $x_0$ dependence in
\begin{equation}
    \mathcal{K}'=\frac{1}{E(x_0,W_1)\vartheta_{s}(x_0+W_1)} \left(\frac{E(x_0,p_2)\vartheta_s(x_0+p_2)}{E(p_2,W_1)\vartheta_s(p_2+W_1)}-\frac{E(x_0,W_2)\vartheta_s(x_0+W_2)}{E(W_2,W_1)\vartheta_s(W_1+W_2)}\right)
\end{equation}
We shall choose 
\begin{equation}
x=x_0\,,~ v=p_2\,,~ u=W_2\,,~y=W_1
\end{equation}
and
\begin{equation}
    z=p_2+W_2\,,
\end{equation}
such that the trisecant identity reads
\begin{equation}
    \frac{E(x_0,p_2)\vartheta_s(x_0+p_2)}{E(p_2,W_1)\vartheta_s(W_1+p_2)}-\frac{E(x_0,W_2)\vartheta_s(x_0+W_2)}{E(W_2,W_1)\vartheta_s(W_1+W_2)}=\frac{E(x_0,W_1)E(W_2,p_2)\vartheta_s(p_2+W_2)\vartheta_s(x_0+W_1)}{E(p_2,W_1)E(W_2,W_1)\vartheta_s(W_1+p_2)\vartheta_s(W_1+W_2)}\,.
\end{equation}
As a result, we find
\begin{equation}
    \mathcal{K}'=\frac{E(W_2,p_2)\vartheta_s(p_2+W_2)}{E(p_2,W_1)E(W_2,W_1)\vartheta_s(W_1+p_2)\vartheta_s(W_1+W_2)}\,,
\end{equation}
and
\begin{equation}
    \mathcal{K}_{\beta-\gamma}=\mathcal{K}' \frac{E(W_1-x)^2E(W_1+x)^2}{E(2x)}\,.
\end{equation}

Great! We showed succesfully that $\mathcal{K}_{\beta-\gamma}$ is independent of $x_0.$ Let's try to do the same for 
\begin{equation}
    \mathcal{L}_{\beta-\gamma}=\langle e^{-\phi}(x)e^{-\phi}( -x)\xi(x_0) (\xi(p)-\xi(W))  \eta (y_1)e^{2\phi}(y_2 )\rangle\,.
\end{equation}
Using the Verlinde-Verlinde formula \cite{Verlinde:1987sd}, we find
\begin{align}
    \mathcal{L}_{\beta-\gamma}=&\frac{\vartheta_s(-2y_1+x_0+p+2y_2)}{\vartheta_s(x_0-y_1+2y_2)\vartheta_s(p-y_1+2y_2)}\frac{E(x_0,p) E(y_2,x)^2E(y_2,-x)^2}{E(x_0,y_1)E(p,y_1)E(2x)}\nonumber\\
    &-\frac{\vartheta_s(-2y_1+x_0+W+2y_2)}{\vartheta_s(x_0-y_1+2y_2)\vartheta_s(W-y_1+2y_2)}\frac{E(x_0,W) E(y_2,x)^2E(y_2,-x)^2}{E(x_0,y_1)E(W,y_1)E(2x)}\,.
\end{align}
The goal is to simplify the following term
\begin{align}
    \mathcal{L}'=&\frac{1}{E(x_0,y_1)\vartheta_s(x_0-y_1+2y_2)}\biggr(\frac{E(x_0,p)\vartheta_s(-2y_1+x_0+p+2y_2)}{\vartheta_s(p-y_1+2y_2)E(p,y_1)}\nonumber\\
    &\qquad\qquad\qquad\qquad\qquad\qquad\qquad-\frac{E(x_0,W)\vartheta_s(-2y_1+x_0+W+2y_2)}{\vartheta_s(W-y_1+2y_2)E(W,y_1)}\biggr)\,,
\end{align}
using the trisecant identity. We shall choose
\begin{equation}
    x=x_0\,,~v=p\,,~y=y_1\,,~u=W\,,
\end{equation}
with
\begin{equation}
    z=-2y_1+2y_2+p+W\,.
\end{equation}
Then, the trisecant identity reads
\begin{align}
    &\frac{E(x_0,p)\vartheta_s(-2y_1+2y_2+x_0+p)}{E(p,y_1)\vartheta_s(-y_1+p+2y_2)}-\frac{E(x_0,W)\vartheta_s(-2y_1+x_0+W+2y_2)}{E(W,y_1)\vartheta_s(-y_1+2y_2+W)}\nonumber\\
    =& \frac{E(x_0,y_1)E(W,p)\vartheta_s(-2y_1+2y_2+p+W)\vartheta_s(-y_1+x_0+2y_2)}{E(p,y_1)E(W,y_1)\vartheta_s(p-y_1+2y_2)\vartheta_s(W-y_1+2y_2)}\,.
\end{align}
As a result, we find
\begin{equation}
    \mathcal{L}'=\frac{E(W,p)\vartheta_s(-2y_1+2y_2+p+W)}{E(p,y_1)E(W,y_1)\vartheta_s(p-y_1+2y_2)\vartheta_s(W-y_1+2y_2)}\,,
\end{equation}
and
\begin{align}\label{eqn:vertical beta gamma}
    \mathcal{L}_{\beta-\gamma}=&\langle e^{-\phi}e^{-\bar{\phi}}( x)\xi(x_0) (\xi(p)-\xi(W))  \eta (y_1)e^{2\phi}(y_2 )\rangle\\
    =&\frac{E(y_2,x)^2E(y_2,-x)^2}{E(2x)} \mathcal{L}'\,,\\
    =&\frac{E(y_2,x)^2E(y_2,-x)^2E(W,p)\vartheta_s(-2y_1+2y_2+p+W)}{E(2x)E(p,y_1)E(W,y_1)\vartheta_s(p-y_1+2y_2)\vartheta_s(W-y_1+2y_2)}\,.
\end{align}

\section{Representations of one-loop diagrams}\label{app:one-loop reps}
In this section, we shall summarize various representations of one-loop diagrams that are related by conformal maps. The identification of such conformal maps will be useful when constructing string vertices. 

\subsection{Annulus}
\paragraph{$\Bbb{Z}_2$ quotient of a torus}
We shall embed an annulus as a $\Bbb{Z}_2$ quotient of a torus 
\begin{equation}
    z:=x+i y\equiv z+l\equiv z+it \,,
\end{equation}
with
\begin{equation}
    \tau= i\frac{t}{l}\,.
\end{equation}
We define a $\Bbb{Z}_2$ involution $\mathcal{I}$
\begin{equation}
    \mathcal{I}(z)=l-\bar{z}\,.
\end{equation}
The boundaries of the annulus in the $z$ coordinate are therefore placed at
\begin{equation}
    \re z=0\,,~ l/2\,.
\end{equation}

We can perform a modular transformation
\begin{equation}
    z'=-\frac{z}{\tau}\,,
\end{equation}
\begin{equation}
    \tau'=-\frac{1}{\tau}\,.
\end{equation}
In the $z'$ coordinate, we have
\begin{equation}
    z':=x'+iy'\equiv z'+i\frac{l^2}{t}\equiv z'+l\,.
\end{equation}
Correspondingly, the boundaries are located at
\begin{equation}
    \im z'=0\,,\quad \im z'=\frac{l^2}{2t}\,.
\end{equation}

\paragraph{Radial identification of the upper half plane}
The second representation of an annulus we will often encounter is given as a radial identification of the UHF. Let's denote the upper half coordinate by $w_o.$ And, we shall identify
\begin{equation}
    w_o\equiv \Lambda w_o\,,
\end{equation}
with $\Lambda<1.$ For the fundamental region, we shall use
\begin{equation}
    0\leq\im w_o\,,\quad \Lambda \leq|w_o|\leq1\,.
\end{equation}

Therefore, we can identify two boundaries. One at 
\begin{equation}
    \im w_o=0\,,\quad \Lambda\leq w_o\leq 1\,,
\end{equation}
and the other at
\begin{equation}
    \im w_o=0\,,\quad -1\leq w_o\leq -\Lambda\,.
\end{equation}

$w_o$ is related to $z$ via
\begin{equation}
    w_o=\exp\left(\frac{2\pi i z}{l}\right)\,,
\end{equation}
with
\begin{equation}
    \Lambda= \exp\left(-\frac{2\pi t}{l}\right)\,.
\end{equation}

\paragraph{Concentric circles}
The third representation of an annulus is given as follows. Let's denote the coordinate of $\Bbb{C}$ by $w_c.$ We shall consider a region bounded by
\begin{equation}
    \kappa \leq |w_c|\leq 1\,.
\end{equation}
$w_c$ is related to $z$ via
\begin{equation}
    w_c=\exp \left(-\frac{2\pi  z}{t}\right)\,,
\end{equation}
with
\begin{equation}
    \kappa= \exp\left(-\frac{\pi l}{t}\right)\,.
\end{equation}

Lastly, $w_0$ and $w_c$ are related by
\begin{equation}
    w_c^{-\frac{it}{l}}=w_o\,.
\end{equation}

\paragraph{Open and closed string channels}
Let us define $z_o$ as
\begin{equation}
    z_o:=\frac{z}{l}\,,
\end{equation}
such that
\begin{equation}
    z_o\equiv z_o+1 \equiv z_o+\tau_o\,,
\end{equation}
with $\tau_o:=\tau\,.$ We can perform an S transformation to obtain
\begin{equation}
    z_c= -\frac{z_o}{\tau_o}=i\frac{z}{t}\,,
\end{equation}
\begin{equation}
    \tau_c=-\frac{1}{\tau_o}=i\frac{l}{t}\,.
\end{equation}
We can then relate $w_c$ and $w_o$ to $z_c$ and $z_o$ as
\begin{equation}
    w_c=\exp \left(2\pi i z_c\right)\,,\quad w_o =\exp\left(2\pi i z_o\right)\,.
\end{equation}

\subsection{Möbius strip}
\paragraph{$\Bbb{Z}_2$ quotient of a torus}
We shall embed a Möbius strip as a $\Bbb{Z}_2$ quotient of a torus
\begin{equation}
    z:=x+iy\equiv z+l\equiv z+\frac{l}{2}+it\,,
\end{equation}
with
\begin{equation}
    \tau:=\frac{1}{2}+i\frac{t}{l}\,.
\end{equation}
We shall define a $\Bbb{Z}_2$ involution $\mathcal{I}$ as
\begin{equation}
    \mathcal{I}(z)=l-\bar{z}\,.
\end{equation}
After taking the quotient, the fundamental region of the Möbius strip is given by
\begin{equation}
    \left[\frac{l}{2},l\right]\times \left[ 0,t\right]\,.
\end{equation}
The boundary is located at 
\begin{equation}
    \re z=\frac{l}{2}\,,\quad l\,.
\end{equation}
The crosscap is located at
\begin{equation}
    \re z=\frac{3}{4}l\,.
\end{equation}

We can apply the modular transformation $TST^2S$ to find new coordinates
\begin{equation}
    z'=-\frac{z}{2\tau-1}=i\frac{l}{2t}z\,,
\end{equation}
with
\begin{equation}
    \tau'=\frac{\tau-1}{2\tau-1}=\frac{1}{2}+i\frac{l}{4t}\,.
\end{equation}
$z'$ satisfies the following equivalence relations
\begin{equation}
    z'\equiv z'+l\equiv z'+\frac{l}{2}+i\frac{l^2}{4t}\,.
\end{equation}
The fundamental region of the Möbius strip in $z'$ coordinate is given by
\begin{equation}
    [0,l]\times \left[\frac{l^2}{8t},\frac{l^2}{4t}\right]\,.
\end{equation}
Note that $\im z'=l^2/4t$ is the boundary, and $\im z'=l^2/8t$ is the crosscap. We can redefine the coordinate $z'$ such that
\begin{equation}
    \mathcal{Z}:=\frac{l^2}{4t}+\bar{z}'\,,
\end{equation}
the fundamental domain is mapped to
\begin{equation}
    [0,l]\times \left[0,\frac{l^2}{8t}\right]\,,
\end{equation}
where $\im\mathcal{Z}=0$ is the boundary and $\im\mathcal{Z}=l^2/8t$ is the crosscap.

\paragraph{Radial identification of the upper half plane}
We shall again denote the upper half coordinate by $w_o.$ We shall then identify
\begin{equation}
    w_o\equiv-\Lambda \bar{w}_o\,,
\end{equation}
where $\Lambda$ is a positive real number with $\Lambda<1.$ The fundamental region is given as
\begin{equation}
    0\leq\im w_o\,,\quad \Lambda \leq |w_o|\leq 1\,.
\end{equation}
The equivalence relation can be stated as
\begin{equation}
    r'e^{i\theta'}\equiv \Lambda r e^{i(-\theta+\pi)}\,.
\end{equation}

$w_o$ and $z$ are related by
\begin{equation}
    w_o=\exp\left(\frac{2\pi iz}{l}\right)\,.
\end{equation}
Similarly, $\Lambda$ is related to $l$ and $t$ by
\begin{equation}
    \Lambda=\exp\left(-\frac{2\pi t}{l}\right)\,.
\end{equation}

\paragraph{Concentric circles}
Let us denote the coordinate of $\Bbb{C}$ by $w_c.$ We shall put a boundary on $|w_c|=1$ and a crosscap on $|w_c|=\kappa.$ We find
\begin{equation}
    w_c= \exp\left(\frac{-\pi (z-l/2)}{t}\right)\,.
\end{equation}
We then identify $\kappa$ as
\begin{equation}
    \kappa= \exp\left(-\frac{\pi l}{4t}\right)\,.
\end{equation}

\section{String vertices and the moduli space}\label{app:vertices}
In this appendix, we shall collect details of relevant string vertices.  We shall use a simplified construction of string vertices that explicitly breaks various symmetries. This choice is justified because we are interested in computing the on-shell one-loop partition function. 

\subsection{Sphere with C-C-C}
The moduli space for a three-punctured sphere is a point. A standard choice for the local coordinates is to use the $SL(2;\Bbb{C})$ invariant one. However, for the calculations we shall perform, we just need to ensure that the local coordinates of two punctures of choice are invariant under the permutation. We shall choose $z=0$ and $z=\infty$ to be permutation invariant. This choice will simplify the construction of string vertices of various diagrams. 

To each puncture around $z_i,$ we attach a unit disk parametrized by a coordinate $w_i$ such that
\begin{equation}
w_i=\lambda_S f_i(z)\,,
\end{equation}
where $\lambda_c$ is the closed string stub parameter that will be taken to infinity at the end of the computations. We choose the local coordinates as
\begin{equation}
f_1(z) =z \,,\quad f_2(z)=(1-z) \,,\quad f_3(z)=\frac{1}{z}\,.
\end{equation}
Following \cite{Sen:2019jpm}, we shall place PCOs at symmetric locations that are invariant under the $SL(2;C)$ maps that permute over the closed string punctures
\begin{equation}
p_\pm=\frac{1}{2}\pm i\frac{\sqrt{3}}{2}\,,
\end{equation}
and average over the two choices.

\subsection{Disk with C}
The moduli space of a disk with one closed string puncture is zero-dimensional. We shall define the global coordinate on a unit disk to be $y,$ such that $|y|\leq1,$ and the global coordinate on the upper-half-plane to be $z.$ We shall place the closed string puncture at $y=0.$ The global coordinates $z$ and $y$ are related by
\begin{equation}
    z=i\frac{1-y}{1+y}\,.
\end{equation}
The closed puncture at $y=0$ is mapped to $z=i.$

We shall define the local coordinate around the closed string puncture as
\begin{equation}
    w=\lambda_D y=\lambda_D \frac{i-z}{i+z}\,.
\end{equation}

\subsection{Disk with O-O-O}
The moduli space of a disk diagram with three open-string punctures is a point. 

We shall fix the location of the open string punctures to be $0,~1,$ and $\infty.$ To each open string puncture, we shall attach a unit half disk $|w|\leq1,$ with $\im w\geq0,$ such that
\begin{equation}
    w_i=\mu g_i(z)\,,
\end{equation}
with 
\begin{equation}
    g_1(z)=\frac{2z}{2-z}\,,\quad g_2(z)=-2\frac{1-z}{1+z}\,,\quad g_3(z) =\frac{2}{1-2z}\,.
\end{equation}

If all open-string punctures are inserted with NS states, we need to insert a PCO. We shall place a PCO at the symmetric locations
\begin{equation}
    p_\pm=\frac{1}{2}\pm i\frac{\sqrt{3}}{2}\,.
\end{equation}
\subsection{Disk with C-O}
The moduli space of a disk with one closed string puncture and one open string puncture is zero-dimensional. 

We choose the local coordinates around the closed and open string punctures as
\begin{equation}
w=\lambda_D y=\lambda_D\frac{i-z}{i+z}\,,
\end{equation}
and
\begin{equation}
w=\nu z\,.
\end{equation}

If both the closed and the open string punctures are inserted with NS states, we need to insert a PCO. We shall average over PCO locations by placing them at $\pm \sqrt{3}/3.$ 

\subsection{Disk with C-C}\label{app:disk CC}
A disk with two closed string punctures has a one-dimensional moduli space. We shall place one closed string puncture at $z=i,$ and the other at $z=ix$ where $0\leq x\leq 1.$ As the calculations we will perform at this order are almost on-shell, we don't need the precise form of the local coordinates but just the range of the parameter $x$ within the string vertex. We shall determine this bound by using the plumbing fixture. 

The disk with two closed-string punctures has two degeneration channels: first, it splits into two disks with C-O, and second, into one disk with C and one sphere with C-C-C. 

\paragraph{Open string degeneration}
Let us first study the degeneration channel into two disks with C-O. We shall denote the global coordinates of disks with C-O by $x_i,$ for $i=1,~2.$ We glue those two disks with a plumbing fixture
\begin{equation}
    \nu^2 x_1x_2=-q\,,
\end{equation}
where $0\leq q\leq1.$ As one can check, $x_2=i$ is mapped to $x_1=i q\mu^{-2}.$ We shall declare therefore that the global coordinate $z$ is given by $x_1.$ Equivalently, we can rewrite the plumbing fixture using the coordinate on the unit disk
\begin{equation}
    -\nu^2 \frac{1-y_1}{1+y_1}\frac{1-y_2}{1+y_2}=q\,.
\end{equation}
Note that we used the relation between the upper half coordinate $x$ and the unit disk coordinate $y$
\begin{equation}
    y=\frac{i-x}{i+x}\,.
\end{equation}
We find that the closed string puncture at $y_2=0$ is mapped to
\begin{equation}
    \frac{1-q\nu^{-2}}{1+q\nu^{-2}}\,.
\end{equation}

The local coordinates at $y_1=0$ and $y_2=(1-q\nu^{-2})/(1+q\nu^{-2})$ are given as
\begin{equation}
    w_1= \lambda_D y_1\,,
\end{equation}
\begin{equation}
    w_2=\lambda_D \frac{y_1(1+q\nu^{-2})-(1-q\nu^{-2})}{y_1(1-q\nu^{-2})-(1+q\nu^{-2})}=\lambda_D \frac{y_1-t}{y_1 t-1}\,,
\end{equation}
with 
\begin{equation}
    t=\frac{1-q\nu^{-2}}{1+q\nu^{-2}}\,.
\end{equation}

We shall insert two PCOs. First one at
\begin{equation}
    p_{1,\pm}= \frac{1}{2} (1\pm i \sqrt{3})\,,
\end{equation}
and the second one at
\begin{align}
    p_{2,\pm}=&\pm i \frac{\nu^2+\pm i\sqrt{3}q}{\pm i \nu^2+\sqrt{3}q}\,,\\
    =&1\mp \frac{2i \sqrt{3}q}{\nu^2}-\frac{6q^2}{\nu^4}+\dots\,.
\end{align}
Note that both $p_1$ and $p_2$ are located on the boundary.

\paragraph{Closed string degeneration}
We shall now study the degeneration channel into one disk with C and one sphere with C-C-C. We shall denote the global coordinate on a disk with C by $x$ and a global coordinate on the sphere by $y.$ We shall glue the closed string puncture at $x=0$ to a closed string puncture at $y=0$ via the plumbing fixture
\begin{equation}
    \lambda_S\lambda_D x y=q\,,
\end{equation}
where $|q|\leq1.$ We find
\begin{equation}
    x= q\lambda_D^{-1}\lambda_{S}^{-1}y^{-1}\,,
\end{equation}
and therefore punctures at $y=1\,,~\infty$ are mapped to
\begin{equation}
    x=q\lambda_D^{-1}\lambda_S^{-1}\,,
\end{equation}
and
\begin{equation}
    x=0\,,
\end{equation}
respectively. Using the $SL(2;R),$ we can fix $q=e^{-s}.$ 

The local coordinates are given as
\begin{equation}
    w_1=q^{-1}\lambda_D\lambda_S^{2}x  \,, 
\end{equation}
and
\begin{equation}
        w_2= \lambda_S \frac{x-q\lambda_D^{-1}\lambda_S^{-1}}{x}\,. 
\end{equation}

The PCOs are located at
\begin{equation}
    p=\frac{q}{2\lambda_D\lambda_S}(1\pm i\sqrt{3})\,.
\end{equation}

\paragraph{Vertex region}
We shall now construct the vertex regions. We shall denote the location of the movable closed string puncture by the modulus $t,$ whose range is given as
\begin{equation}
   \lambda_D^{-1}\lambda_S^{-1}\leq t \leq \frac{1-\nu^{-2}}{1+\nu^{-2}}\,.
\end{equation}
We shall now construct the local coordinates
\begin{equation}
    w_1= f(t) x\,,\quad w_2=\frac{g_1(t) (x-t)}{g_2(t) x+g_3(t)}\,,
\end{equation}
with the boundary conditions, modulo the phase,
\begin{equation}
   f\left(\frac{1-\nu^{-2}}{1+\nu^{-2}}\right)=\lambda_D\,,\quad f(\lambda_D^{-1}\lambda_S^{-1})=\lambda_D\lambda_S^2\,,
\end{equation}
\begin{equation}
    g_1\left(\frac{1-\nu^{-2}}{1+\nu^{-2}}\right)=\lambda_D\,,\quad g_1\left(\lambda_D^{-1}\lambda_S^{-1}\right)=\lambda_S \,,
\end{equation}
\begin{equation}
    g_2\left(\frac{1-\nu^{-2}}{1+\nu^{-2}}\right)=\frac{1-\nu^{-2}}{1+\nu^{-2}}\,,\quad g_2\left(\lambda_D^{-1}\lambda_S^{-1}\right)=1\,,
\end{equation}
\begin{equation}
    g_3\left(\frac{1-\nu^{-2}}{1+\nu^{-2}}\right)=-1\,,\quad g_3\left(\lambda_D^{-1}\lambda_S^{-1}\right)=0\,.
\end{equation}
We shall let $f,$ and $g_i$ to be linear functions in $t$ that interpolate the boundary conditions. 

We shall also let the PCO locations interpolate the boundary conditions. Let us recall the boundary conditions for the PCO locations. When $t_{D,o}^*=(1-\nu^{-2})/(1+\nu^{-2}),$ PCOs are located at
\begin{equation}
    p_{1,\pm}= e^{\pm\frac{i\pi}{3}}\,,
\end{equation}
\begin{equation}
p_{2,\pm}=  \frac{\nu^2\pm i\sqrt{3}}{\nu^2 \mp i\sqrt{3}}= e^{i\vartheta_{DCC,\pm}}\,,
\end{equation}
where $\vartheta_{DCC,\pm}$ is defined by
\begin{equation}
    e^{i\vartheta_{DCC,\pm}/2}= \frac{\nu^2\pm i\sqrt{3}}{\sqrt{3+\nu^4}}\,.
\end{equation}
When $t_{D,c}^*=\lambda_D^{-1}\lambda_S^{-1},$ the PCOs are located at
\begin{equation}
    p_{\pm}= \frac{1}{\lambda_D\lambda_S} e^{\pm \frac{i\pi}{3}}=e^{\pm\frac{i\pi}{3} -\log\lambda_D\lambda_S}\,.
\end{equation}

We shall construct PCOs
\begin{equation}
    p_1= \exp\left(\pm \frac{i\pi}{3}-\log \lambda_D \lambda_S \frac{t-t_{D,o}^*}{t_{D,c}^*-t_{D,o}^*}\right)\,,
\end{equation}
\begin{equation}
    p_2=\exp\left( i\vartheta_{DCC,\mp}+ \mp\left(\frac{i\pi}{3}-\log\lambda_D\lambda_S\right)\frac{t-t_{D,o}^*}{t_{D,c}^*-t_{D,o}^*}\right)\,.
\end{equation}

\subsection{$\Bbb{RP}^2$ with C}
A crosscap with a closed string puncture has no non-trivial moduli space. Hence, we shall declare that the entire $\Bbb{RP}^2$ with C belongs to the string vertex. 

The cross-cap can be represented by a unit disk $|z|\leq1,$ and the extension 
\begin{equation}
    z=-\frac{1}{\bar{z}}
\end{equation}
for $|z|\geq1.$ Note that, unlike the disk, there cannot be a fixed point of
\begin{equation}
    z=-\frac{1}{\bar{z}}\,.
\end{equation}

We shall place the closed string puncture at $z=0,$ and attach the local coordinate
\begin{equation}
    w=\lambda_R z\,.
\end{equation}
\subsection{$\Bbb{RP}^2$ with C-C}
$\Bbb{RP}^2$ with two closed string punctures has a one-dimensional moduli space. The Feynman region is obtained by gluing $\Bbb{RP}^2$ with C to a sphere with C-C-C. Let us denote the coordinates of the $\Bbb{RP}^2$ and the sphere by $x_1$ and $x_2.$ We shall glue the puncture at $x_1=0$ to $x_2=\infty$ such that
\begin{equation}
    \lambda_R \lambda_S x_1 \frac{1}{x_2}=q\,.
\end{equation}
The closed string punctures at $x_2=0$ and $x_2=1$ are mapped to
\begin{equation}
    x_1= 0\,,
\end{equation}
and
\begin{equation}
    x_1= \frac{q}{\lambda_R\lambda_S}\,,
\end{equation}
respectively. Note that we can fix the phase of $q,$ such that $q=e^{-s}.$

The local coordinates at the closed string punctures are then determined as
\begin{equation}
    w_1= \lambda_R\lambda_S^2q^{-1} x_1\,,
\end{equation}
and
\begin{equation}
    w_2= -\lambda_S \frac{x_1-q\lambda_R^{-1}\lambda_S^{-1}}{q\lambda_R^{-1}\lambda_S^{-1}}\,.
\end{equation}
We denote the modulus $t$ by
\begin{equation}
t=\frac{q}{\lambda_R\lambda_S}\,.
\end{equation}

We shall now construct the vertex region. We shall use the following ansatz
\begin{equation}
    w_1=f(t) x_1\,,\quad w_2= g(t) (x-t)\,.
\end{equation}
We shall impose the following boundary conditions
\begin{equation}
    f(1)=\lambda_R\lambda_S^2\,,\quad f(\lambda_R^{-1}\lambda_S^{-1})=\lambda_R\lambda_S^2\,,
\end{equation}
\begin{equation}
    g(1)=-\lambda_R\lambda_S^2\,,\quad g(\lambda_R^{-1}\lambda_S^{-1})=-\lambda_R\lambda_S^2\,.
\end{equation}
We shall place PCOs at 
\begin{equation}
    \frac{t}{2}(1\pm i \sqrt{3})\,.
\end{equation}
\subsection{Disk with C-O-O}
The moduli space of a disk diagram with one closed string puncture and two open string punctures is one dimension. We shall place the closed string at the center of the disk, $z=i,$ and place the open string punctures at $z=\pm t,$ while $t$ is the modulus. 

The Feynmann region of the disk amplitude with C-O-O is obtained by joining open string punctures of one disk with C-O and the other disk with O-O-O. Let us denote the global coordinate in the upper half plane of the first disk by $x_1,$ the global coordinate of the second disk by $x_2,$ and the coordinate of the disk with C-O-O by $z.$ We shall glue the open string punctures at $x_1=x_2=0$ via
\begin{equation}
    \mu \nu x_1 g_1(x_2)=-q\,.
\end{equation}
As a result, we find that the open string punctures at $x_2=1$ and $x_2=\infty$ are mapped to
\begin{equation}
    x_1=-\frac{1}{2}q\mu^{-1}\nu^{-1}\,,
\end{equation}
and
\begin{equation}
    x_1=\frac{1}{2}q\mu^{-1}\nu^{-1}\,,
\end{equation}
respectively. Similarly, at $q=1,$ PCOs at $x_2=p_\pm$ are mapped to
\begin{equation}
    x_1=\pm i\frac{\sqrt{3}}{2}\mu^{-1}\nu^{-1}\,.
\end{equation}

We shall identify $z$ with $x_1,$ and $t$ with $q\mu^{-1}\nu^{-1}/2.$

We determine the local coordinates of the open string punctures as
\begin{equation}
    w_1=\mu g_2(x_2)=\mu \frac{2 (q\mu^{-1}\nu^{-1}+2x_1) }{3q\mu^{-1}\nu^{-1}-2x_1}\,,\quad w_2=\mu g_3(x_2)= -\mu \frac{2(q\mu^{-1}\nu^{-1}-2x_1)}{3q\mu^{-1}\nu^{-1}+2x_1}\,.
\end{equation}
The local coordinate of the closed string puncture is
\begin{equation}
    w=\lambda_D \frac{i-z}{i+z}\,.
\end{equation}

The other Feynmann region around $t\sim\infty$ can be determined by replacing $x_1$ with $-1/x_1.$

We shall determine the local coordinates and the location of PCO for $\mu^{-1}\nu^{-1}/2\leq t\leq1.$ Following \cite{Sen:2020eck}, we choose the local coordinates of the open string punctures as
\begin{equation}\label{eqn: O of COO}
    w_i=\mu^2\nu\frac{4\mu^2\nu^2+1}{4\mu^2\nu^2} \frac{z-z_i}{(1+z_iz)+\mu\nu h(z_i) (z-z_i)}\,,
\end{equation}
where $z_1=-t,$ and $z_2=t.$ We choose $h(z)$ to be a holomorphic function that interpolates 
\begin{equation}
    h\left(\pm\frac{1}{2\mu\nu}\right)= \pm\frac{4\mu^2\nu^2-3}{8\mu^2\nu^2}\,,
\end{equation}
and 
\begin{equation}
    h(\pm 1)=0\,.
\end{equation}
We shall therefore choose
\begin{equation}
    h(z)=-\mu^{-1}\nu^{-1}\frac{3-4\mu^2\nu^2}{4-16\mu^2\nu^2}\left(z-\frac{1}{z}\right)\,.
\end{equation}
Note that the range of $h$ for the above representation is $(2\mu\nu)^{-1}\leq |z|\leq 1.$ For the rest of the region for $z,$ we shall use the following relation to extend the range of the validity of $h(z)$
\begin{equation}
    h(z)=h(-1/z)=-h(1/z)\,.
\end{equation}

We shall now determine the locations of the PCOs. As the closed string puncture is located at $z=i,$ it is not a good idea to interpolate the PCO locations
\begin{equation}
 p_\pm (\mu^{-1}\nu^{-1}/2)=\pm i\frac{\sqrt{3}}{2}\mu^{-1}\nu^{-1}\,,
\end{equation}
and
\begin{equation}
    p_{\pm}(2\mu\nu)=\pm i\frac{2\sqrt{3}}{3}\mu\nu\,.
\end{equation}
To interpolate the PCO locations without hitting the closed string puncture, we shall proceed as follows. Let us use the conformal map from the upper half plane to a unit disk
\begin{equation}
    w=\frac{i-z}{i+z}\,,
\end{equation}
or equivalently
\begin{equation}
    z=-i\frac{w-1}{w+1}\,,
\end{equation}
such that the closed string puncture at $z=i$ is mapped to $w=0$ with $|w|\leq1.$ Then the boundary conditions for the PCOs are given as
\begin{equation}
    w_{\pm}\left(\frac{1}{2\mu\nu}\right)=\left(\frac{2\mu\nu-\sqrt{3}}{2\mu\nu+\sqrt{3}}\right)^{\pm}
\end{equation}
\begin{equation}
    w_{\pm}(2\mu\nu)=-\left(\frac{2\mu\nu-\sqrt{3}}{2\mu\nu+\sqrt{3}}\right)^{\pm}\,.
\end{equation}
We shall connect the two PCOs by
\begin{equation}
    y_{\pm}(t) = r_{\pm}(t) e^{i \theta_{\pm}(t)}\,,
\end{equation}
with the boundary conditions
\begin{equation}
    r_\pm(1/2\mu\nu)= \left(\frac{2\mu\nu-\sqrt{3}}{2\mu\nu+\sqrt{3}}\right)^\pm\,,\quad r_\pm(2\mu\nu)= r_\pm(1/2\mu\nu)\,,
\end{equation}
\begin{equation}
    \theta_{\pm}(1/2\mu\nu)=0\,,\quad \theta_\pm(2\mu\nu)=1\,.
\end{equation}

The other PCOs are located at $\pm\sqrt{3}/3$ when $t_D=1/2\mu\nu,$ and $\pm \sqrt{3}$ when $t_D=2\mu\nu. $ As the open string punctures are located at $z=\pm t_D,$ where $(2\mu\nu)^{-1}\leq t_D\leq 2\mu\nu,$ we cannot simply interpolate the locations of PCOs. Let $z_{pco}'(t_D)$ to be the PCO location interpolating $\pm\sqrt{3}^{-1}$ and $\pm\sqrt{3}.$ The goal is to choose simple $z_{pco}'(t_D)$ such that
\begin{equation}
    \frac{i}{2\pi}\log \left(\frac{y_{pco}'(t_D)+t_D}{y_{pco}'(t_D)-t_D}\right)
\end{equation}
takes a simple form. We shall require
\begin{equation}
    y_{pco}'(t_D)
\end{equation}
never hits the vertex operators. 

\subsection{Empty annulus}
Here we shall construct the string vertex for the empty annulus. This will serve as a stepping stone towards constructing string vertices for the Möbius strip and the annulus diagram with more complicated insertions. 

The only Feynman region of the empty annulus is given by sewing two disks with C. Let us denote the global coordinate of the disks by $z_i,$ for $i=1,2$ with $|z_i|\leq1.$ Note that $z_i=1/\bar{z}_i.$ We glue those two Riemann surfaces by the plumbing fixture
\begin{equation}
    \lambda_D^2 z_1z_2=-q\,,
\end{equation}
with $|q|\leq1.$ We shall declare that the global coordinate of the annulus diagram is $z:=z_1,$ with two boundaries given by
\begin{equation}
    |z|=1\,,
\end{equation}
and
\begin{equation}
    |z|=\frac{|q|}{\lambda_D^2}\,.
\end{equation}
By flipping $z\mapsto 1/z,$ we can also bound the strip by the boundary $|z|=|q|/\lambda_D^2$ and $|z| =\lambda_D^2/|q|.$ This determines that the annulus can be obtained by the $z\mapsto 1/z$ qotient of the toroidal strip where
\begin{equation}
    \lambda_D^2/q\equiv \bar{q}/\lambda_D^2\,.
\end{equation}

We shall map the concentric disks to a strip parametrized by the flat coordinate
\begin{equation}
    y_c=y_{c,1}+iy_{c,2}\,,
\end{equation}
via
\begin{equation}
    z= \exp \left(2\pi i  y_c\right)\,,
\end{equation}
where the range of $y_c$ is given by
\begin{equation}
    [0,1]\times [0,t_c]\,.
\end{equation}
This describes the annulus diagram in the closed string channel with $\tau_c=i\frac{1}{\pi}\log\lambda_D^2/|q|$. 

We now define $y_o$ such that
\begin{equation}
    y_o=\frac{y_c}{\tau_c}=-\tau_o y_C\,,
\end{equation}
with $\tau_o=-1/\tau_c.$ Then, $y_{o,1}$ ranges over $[0,1/2]$ and $w_2$ ranges over $[0,\pi/\log (\lambda_D^2/|q|)].$ We therefore declare that the fundamental region of the empty annulus is given by
\begin{equation}
    \frac{\pi}{\log\lambda^2}\leq t \leq \infty\,.
\end{equation}

\subsection{Empty Möbius}
The construction of the string vertex for the empty Möbius is quite analogous to the construction of the empty annulus. The Feynman region of the empty Möbius is given by gluing a disk with C to an $\Bbb{RP}^2$ with C. We shall glue them using the plumbing fixture
\begin{equation}
    \lambda_D\lambda_R z_1z_2=q\,.
\end{equation}
Note that we can also glue two disks with the plumbing fixture with the orientation reversal
\begin{equation}
    \lambda_D^2 z_1\bar{z}_2=q\,.
\end{equation}
This shows that we should fix $\lambda_D=\lambda_R.$

Similar to the annulus, if we use $z_1$ as the global coordinate, the boundary is located at $|z_1|=1,$ and the cross-cap is located at 
\begin{equation}
    z_1=\frac{|q|e^{i\theta}}{\lambda_D\lambda_R}\,. 
\end{equation}
By using the relation
\begin{equation}
    z=-\frac{1}{\bar{z}}\,,
\end{equation}
we can map the cross-cap to 
\begin{equation}
    z_1= -\frac{\lambda_D\lambda_R}{|q|} e^{i\theta}\,.
\end{equation}
As a result, we can view the Möbius strip as a quotient of a torus with 
\begin{equation}
    \tau_C=\frac{1}{2}+\frac{i}{\pi} \log \lambda_D\lambda_R q^{-1}\,,
\end{equation} 
if we relate $z_1$ to a new torus coordinate $y_c$ 
\begin{equation}
    z_1= \exp \left(- 2\pi i y_c\right)\,,
\end{equation}
so that $y_c$ ranges over 
\begin{equation}
    \left[0,1\right] \times \left[0,\frac{1}{2\pi} \log q\lambda_D^{-1}\lambda_R^{-1}\right]\,.
\end{equation}
We can embed this Möbius strip in the closed string channel into a torus parametrized by $y_c$ with
\begin{equation}
    y_c\equiv y_c+1\equiv y_c+\tau_C\,,
\end{equation}
with
\begin{equation}
    \tau_C=\frac{1}{2}+\frac{i}{\pi}\log \lambda_D\lambda_Rq^{-1}\,,
\end{equation}
by imposing the following quotient
\begin{equation}
    I(y)=\tau_C-\bar{y}\,.
\end{equation}
Note that by applying the modular transformation,
\begin{align}
    T S T^2 S:& \tau\mapsto-\frac{1}{\tau}\mapsto \frac{2\tau-1}{\tau}\mapsto -\frac{\tau}{2\tau-1}\mapsto\frac{\tau-1}{2\tau-1}\,,\\
    & y\mapsto -\frac{y}{\tau}\mapsto -\frac{y}{\tau}\mapsto -\frac{y}{2\tau-1}\mapsto \frac{y}{2\tau-1}\,,
\end{align}
this worldsheet is mapped to the Möbius strip in the open string channel with $y_o$ and $\tau_o$
\begin{equation}
    y_o= \frac{y_c}{2\tau_c-1}\,,
\end{equation}
\begin{equation}
    \tau_o= \frac{1}{2}+ i \frac{\pi}{4\log\frac{\lambda_D\lambda_R}{q}}\,.
\end{equation}
Note that $y_0$ ranges over
\begin{equation}
    \left[0,\frac{1}{2}\right]\times \left[0,\frac{\pi}{4\log\lambda_D\lambda_Rq^{-1}}\right]\,.
\end{equation}
Upon identifying $\lambda_R=\lambda_D,$ we find that the lower bound for the Möbius modulus in the vertex region is given by
\begin{equation}
   \frac{\pi}{4\log\lambda_D^2} \leq\tau_2\,,
\end{equation}
whereas the lower bound for the annulus modulus is
\begin{equation}
    \frac{\pi}{\log\lambda_D^2}\leq t\,.
\end{equation}
Note that this agrees with the analysis of \cite{Gimon:1996rq}.

\subsection{Möbius with O}
Möbius with one open string puncture has two Feynman regions. First, a disk with C-O glued to an $\Bbb{RP}^2$ with C. Second, a disk with O-O-O, where two open string punctures are identified with the orientation reversal. 

For the construction of the vertex region, we shall divide the moduli space into $t\geq1$ and $t<1$ for the open-string channel modulus
\begin{equation}
    \tau_o=\frac{1}{2}+it\,.
\end{equation}

\paragraph{Closed string degeneration}
Let us first construct the Feynman region of the first case, where a disk with C-O is glued to an $\Bbb{RP}^2$ with C. We shall denote the global coordinate of the disk by $z_1$ and the global coordinate of the $\Bbb{RP}^2$ by $z_2.$ In the disk, we shall place a closed string puncture at $z=0,$ and an open string puncture at $z=1.$ This Feynman region is identical to that of a disk with C glued to an $\Bbb{RP}^2$ with C. 

The local coordinate around the open string puncture is given by
\begin{equation}
    w=\nu z_{U}=-i\nu\frac{z-1}{z+1} =-i \nu \frac{\exp(-2\pi i y_c)-1}{\exp(-2\pi i y_c)+1}\,,
\end{equation}
where $y_c$ ranges over
\begin{equation}
    [0,1]\times \left[0,\frac{1}{2\pi} \log q\lambda_D^{-1}\lambda_R^{-1}\right]\,.
\end{equation}
We shall use the open string channel coordinate $y_o$
\begin{equation}
    y_o:= \frac{y_c}{2\tau_c-1}=(1-2\tau_o) y_c\,,
\end{equation}
with 
\begin{equation}
    \tau_o=\frac{\tau_c-1}{2\tau_c-1}=\frac{1}{2} +i\frac{\pi}{4\log \lambda_D\lambda_R/q}
\end{equation}
Then, the local coordinate is given as
\begin{equation}
    w=-i\nu \frac{\exp\left(\frac{2\pi i y_o}{2\tau_o-1}\right)-1}{\exp\left(\frac{2\pi i y_o}{2\tau_o-1}\right)+1}\,.
\end{equation}

Note that we placed PCO at $z_U=\pm 1/\sqrt{3}$ in the upper-half-plane. This PCO location translates to
\begin{equation}
    z=\frac{1}{2}\pm i\frac{\sqrt{3}}{2}\,,
\end{equation}
that is equivalent to
\begin{equation}
   2\pi i y_c=\pm i\frac{\pi}{3}\,,
\end{equation}
and equivalently
\begin{equation}
    2\pi i y_o=\pm i\frac{\pi (2\tau_o-1)}{3}\,.
\end{equation}

\paragraph{Open string degeneration}
For the construction of the second Feynman region, we shall glue to the open string punctures at $z=1$ and $z=\infty,$ such that \cite{FarooghMoosavian:2019yke}
\begin{equation}
    -4\mu^2 \frac{1-z_1}{1+z_1} \frac{1}{1-2\bar{z}_\infty}\equiv q\,.  
\end{equation}
Compared to the plumbing fixture of the open string punctures with the same orientation, there is an additional negative sign. This is because $\Omega$ maps $z$ to $-\bar{z}$ for the open string punctures. For clarity, we denoted the local coordinate around $1$ by $z_1$, and similarly, the local coordinate around $\infty$ by $ z_\infty$. It is important to note that the coordinates $z_1$ and $z_\infty$ are coordinates in the upper half plane, which extends to the lower half plane via 
\begin{equation}
    z'=\bar{z}\,.
\end{equation}

It is instructive to understand the worldsheet topology that results from the plumbing fixture. First, there is a boundary at 
\begin{equation}
    z=\bar{z}\,.
\end{equation}
We also cut out a small half-disk around $1$ and $\infty$, which are identified with the orientation reversal. 

As it stands, despite the correct topology for the Möbius strip, it is not very convenient to work with the current form of the worldsheet. Therefore, we shall find a new coordinate $y$ that admits the following identifications
\begin{equation}
    y\equiv -\bar{y}+\tau\,,
\end{equation}
with $\tau=\frac{1}{2}+it.$

We can rewrite the identification imposed by the plumbing fixture as
\begin{equation}\label{eqn:mob id1}
    \bar{z}_1= \frac{ q^{1/2}\mu^{-1} z_\infty-q^{1/2}\mu^{-1}/2-2 q^{-1/2}\mu}{-q^{1/2}\mu^{-1}z_\infty+ q^{1/2}\mu^{-1}/2-2q^{-1/2}\mu}\,.
\end{equation}
Following \cite{Sen:2020eck}, we shall first find the SL(2;C) transformation that brings the coordinate $z$ into a form $w$ in which the identification corresponds to the conformal rescaling. Then, we shall map this new coordinate $w$ into the standard Möbius strip coordinate $y.$\footnote{For a different strategy, see \cite{Erler:2017pgf}.}

We shall define $w$ such that
\begin{equation}
    w=M_1(z):=\frac{az+b}{cz+d}\,,
\end{equation}
where
\begin{align}
    \left(\begin{array}{cc}
      a   &  b\\
        c & d
    \end{array}\right)=&\left(
\begin{array}{cc}
 \frac{2 q}{\mu ^2 \sqrt{\frac{9 q^2}{\mu ^4}+\frac{40 q}{\mu ^2}+16}} & \frac{\sqrt{\frac{9 q^2}{\mu ^4}+\frac{40 q}{\mu ^2}+16}+\frac{q}{\mu ^2}+4}{2 \sqrt{\frac{9 q^2}{\mu ^4}+\frac{40 q}{\mu ^2}+16}} \\
 -\frac{2 q}{\mu ^2 \sqrt{\frac{9 q^2}{\mu ^4}+\frac{40 q}{\mu ^2}+16}} & -\frac{-\sqrt{\frac{9 q^2}{\mu ^4}+\frac{40 q}{\mu ^2}+16}+\frac{q}{\mu ^2}+4}{2 \sqrt{\frac{9 q^2}{\mu ^4}+\frac{40 q}{\mu ^2}+16}} \\
\end{array}
\right)\,,\\
=&\left(
\begin{array}{cc}
 \frac{q}{2 \mu ^2}+O\left( \mu^{-3}\right) & 1-\frac{q}{2 \mu ^2}+O\left(\mu^{-3}\right) \\
 -\frac{q}{2 \mu ^2}+O\left(\mu^{-3}\right) & \frac{q}{2 \mu ^2}+O\left(\mu^{-3}\right) \\
\end{array}
\right)\,.
\end{align}
Then, the identification \eqref{eqn:mob id1} can be rewritten as
\begin{equation}\label{eqn:mob id2}
    \bar{w}=-A w\,,
\end{equation}
with
\begin{align}
    A=&q^{-1}\mu^2+\frac{1}{2}- \frac{7q}{16\mu^2}+\dots\,.
\end{align}

A few comments on the new coordinate $w$ are in order. What we need to ensure is that a half-circle starting on the right-hand side of $1$ and ending on the left-hand side of $1$ represents a line connecting the boundaries of the fundamental cell of the Möbius strip. In the final coordinate $y$, we shall find that this corresponds to $y\mapsto y+1/2.$ We can get some idea about how $y$ and $w$ ought to be related. When $z$ is on the real line, $w$ is also on the real line. When $z=x$ is larger than $1,$ $w$ is on the negative real line. On the other hand, if $z=x$ is less than $ 1$, then $w$ lies on the positive real line. The identification \eqref{eqn:mob id2} introduces the $\Omega$ operator in the radial direction. Therefore, we conclude that the worldsheet takes the following form.

We shall identify $w$ with 
\begin{equation}
    w=\exp(2\pi i(y_o-y_0))\,,
\end{equation}
such that $y$ takes the value in
\begin{equation}
    \left[0,\frac{1}{2}\right]\times \left[0, it\right]\,, 
\end{equation}
such that 
\begin{equation}
    \exp \left(-2\pi t\right)= A^{-1}\,.
\end{equation}
Note that $y_0$ is chosen such that $y=0$ corresponds to $z=0.$

The local coordinate around the open string puncture is then determined as
\begin{equation}
   w= \mu g_1(z)= \mu g_1( M_1^{-1}(\exp(2\pi i (y-y_0)))=2 \mu\frac{4+q\mu^{-2}}{\sqrt{9q^2\mu^{-4}+40q\mu^{-2}+16}} \frac{\exp(2\pi iy)-1}{\exp(2\pi iy)+1} \,. 
\end{equation}
The PCO locations are given as
\begin{align}
   e^{2\pi iy_{pco}}=& -\frac{4 \left(\sqrt{3}\pm3 i\right) \mu ^2+\left(\sqrt{3}\mp i\right) \sqrt{16 \mu ^4+9 q^2+40 \mu ^2 q}+\left(\sqrt{3}\pm3 i\right) q}{-4 \left(\sqrt{3}\pm3 i\right) \mu ^2+\left(\sqrt{3}\mp i\right) \sqrt{16 \mu ^4+9 q^2+40 \mu ^2 q}-\left(\sqrt{3}\pm3 i\right) q}\,,\\
   =&\frac{1}{2}(1\pm i\sqrt{3}) \pm \frac{i(\pm 3i +\sqrt{3})q}{4\mu^2}+\dots\,.
\end{align}

Collecting the Feynman regions, we determine that the vertex region is given by
\begin{equation}
    \frac{\pi}{4\log \lambda_D^2}\leq t\leq \frac{1}{2\pi} \log A|_{q=1}=\frac{1}{2\pi}\log\mu^2+\frac{1}{4\pi\mu^2}+\dots\,.
\end{equation}

\paragraph{Vertex region}
We shall determine the local coordinate of the vertex region by interpolating the local coordinate of the Feynman regions. Note that, as we are working with the level-matched string field theory, the phase of the local coordinate data does not affect the calculation. 

Let us recall the local coordinates around the closed string degeneration and the open string degeneration channels, with $q=1$, respectively
\begin{align}
    w=&i\nu \frac{\exp(-2\pi iy_c)-1}{\exp(-2\pi iy_c)+1}=i\nu \frac{\exp\left(\frac{2\pi iy_o}{2\tau_o-1}\right)-1}{\exp \left(\frac{2\pi i y_o}{2\tau_o-1}\right)+1}\,,\\
    =&\pi \nu y_c+\frac{1}{3} \pi^3\nu y^3_c+\frac{2}{15} \pi^5 \nu y_c^5+\dots\,,\\
    =&i\frac{\pi \nu}{2t_o} y_o-i\frac{\pi^3 \nu}{24t_o^3} y_o^3+i\frac{\pi^5\nu}{2^4\times 3\times 5\times t_o^5}y_o^5+\dots\,,
\end{align}
and
\begin{align}
    w=&\mathfrak{F}(\mu) \frac{\exp(2\pi iy_o)-1}{\exp(2\pi i y_o)+1}
    =\mathfrak{F}(\mu) \frac{\exp\left(\frac{2\pi iy_c}{2\tau_c-1}\right)-1}{\exp(\frac{2\pi iy_c}{2\tau_c-1})+1}\,,\\
    =&i\mathfrak{F}(\mu)\left(\pi y_o+\frac{1}{3} \pi^3y_o^3+\frac{2}{15}\pi^5 y_o^5+\dots\right)\,,\\
    =&\mathfrak{F}(\mu)\left(\frac{\pi}{2t_c}y_c-\frac{\pi^3}{24 t_c^3}y_c^3+\frac{\pi^5}{2^4\times 3\times 5\times t_c^5}y_c^5\right)\,,
\end{align}
where
\begin{equation}
    \mathfrak{F}(\mu)=2\mu \frac{4+\mu^{-2}}{\sqrt{9\mu^{-4}+40\mu^{-2}+16}}=2 \mu -\frac{2}{\mu }+\frac{7}{2 \mu ^3}-\frac{53}{8 \mu ^5}+\dots\,.
\end{equation}

We shall now fill up the vertex region. As we previously mentioned, we shall split the vertex region into $t_o>1/4\equiv t_c<1,$ and $t_o\leq 1/4\equiv t_c\geq1.$ For simplicity, we shall set $\mu=\nu.$ In the region $t_o>1,$ we shall declare that the local coordinate is given by
\begin{equation}
    w= \mathfrak{F}(\mu) \frac{\exp(2\pi i y_o)-1}{\exp(2\pi i y_o)+1}\,,
\end{equation}
which is equal to the local coordinate in the open string degeneration channel with $q=1.$ 

On the other hand, for $t_o\leq 1/4\equiv t_c\geq 1,$ we shall use the polynomial approximation for the local coordinate such that
\begin{equation}
    w=\mathcal{F}_1(t_c) y_c+\mathcal{F}_3(t_c)y_c^3+\mathcal{F}_5(t_c)y_c^5+\dots\,,
\end{equation}
where
\begin{equation}
    \mathcal{F}_1(1)=\frac{\mathfrak{F}(\mu)\pi}{2t_c}\,,\quad \mathcal{F}_1(t_c^*)=\pi \nu\,,
\end{equation}
\begin{equation}
    \mathcal{F}_3(1)=-\frac{\mathfrak{F}(\mu)\pi^3}{24t_c^3}\,,\quad \mathcal{F}_3(t_c^*)=\frac{1}{3}\pi^3\nu\,,
\end{equation}
\begin{equation}
    \mathcal{F}_5(1)=\frac{\mathfrak{F}(\mu) \pi^5}{2^4\times 3\times 5 \times t_c^5}\,,\quad \mathcal{F}_5(t_c^*)= \frac{2}{15}\pi^5\nu\,,
\end{equation}
with
\begin{equation}
    t_c^*=\frac{1}{\pi}\log\lambda_D\lambda_R\,.
\end{equation}
We shall let $\mathcal{F}_i(t_c)$ be a function that interpolates the boundary conditions.

To simplify the calculation, we shall choose $\mathcal{F}_i$ such that
\begin{equation}
    \mathcal{F}_i(t_c)=\mathcal{F}_i(1)\,,
\end{equation}
for
\begin{equation}
    1\leq t_c\leq t_c^*-\varepsilon=\frac{1}{\pi}\log\lambda^2_D-\varepsilon\,,
\end{equation}
and $\mathcal{F}_i(t_c)$ is a linear function in moduli for 
\begin{equation}
    t_c^*-\varepsilon\leq t_c\leq t_c^*\,,
\end{equation}
where $\varepsilon$ is treated as a perturbatively small number. In terms of the open string channel modulus, the region where we shall keep the local coordinates fixed will be given by
\begin{equation}
  \frac{1}{4(\log\lambda_D^2/\pi-\varepsilon)}\simeq \frac{\pi}{4\log\lambda_D^2}+\frac{\pi^2\varepsilon}{4(\log\lambda_D^2)^2}+\dots \leq t_o \leq \frac{1}{2\pi}\log\mu^2+\frac{1}{4\pi\mu^2}+\dots\,,
\end{equation}

Let us now determine the PCO location. For $t_o>1/4,$ we shall declare that the PCO is placed at
\begin{align}
    y_{o,pco,\pm}=&\frac{1}{2\pi i}\log \left(\frac{4 \mu ^2\pm i \sqrt{3} \sqrt{16 \mu ^4+40 \mu ^2+9}-3}{8 \mu ^2+6}\right)\,,\\
    =&\pm\left( \frac{1}{6}+\frac{\sqrt{3}}{4\pi\mu^2}-\frac{\sqrt{3}}{4\pi\mu^4}+\dots\right)\,.
\end{align}

For $t_o\leq1/4\equiv t_c\geq1,$ we shall continuously change the PCO location. Let us recall the PCO location at the closed string degeneration channel at $t_c=\frac{1}{\pi} \log\lambda_D\lambda_R,$
\begin{equation}
    y_{o,pco,\pm}=\pm i\frac{1}{3} t_o\equiv y_{c,pco,\pm}=\pm \frac{1}{6}\,.
\end{equation}
On the other hand, at $t_o=1/4\equiv t_c=1,$ the pco location is given as
\begin{align}
    y_{c,pco,\pm}=&i\frac{1}{2t_o} y_{o,pco,\pm}=-\frac{i \log \left(\frac{4 \mu ^2\pm i \sqrt{3} \sqrt{16 \mu ^4+40 \mu ^2+9}-3}{8 \mu ^2+6}\right)}{2 \log \left(\frac{4 \left(4 \mu ^2+\sqrt{16 \mu ^4+40 \mu ^2+9}+2\right) \mu ^2-3 \sqrt{16 \mu ^4+40 \mu ^2+9}+9}{32 \mu ^2}\right)}\,,\\
    =&\pm\left(\frac{\pi }{12 \log (\mu )}+\frac{6 \sqrt{3} \log (\mu )-\pi }{48 \mu ^2 \log ^2(\mu )}+\frac{-48 \sqrt{3} \log ^2(\mu )-12 \sqrt{3} \log (\mu )+9 \pi  \log (\mu )+2 \pi }{384 \mu ^4 \log ^3(\mu )}+\dots\right)\,.
\end{align}
We shall linearly interpolate the PCO location from $t_c=1$ to $t_c= \frac{1}{\pi}\log\lambda_D\lambda_R.$

\subsection{Möbius with C}
\begin{figure}[]
    \centering
    \begin{tikzpicture}
        \draw[thick,->] (0,0) -- (8,0) node[anchor=north west] {$e^{-t}$};
        \draw[thick,->] (0,0) -- (0,4) node[anchor=south east] {$x$};
        \draw[dashed] (8,3) --(0,3) node[anchor=east] {$x=1/4$};
        \draw[thick] (1,0)--(1,3);
        \draw[thick] (0,1)--(8,1);
        \draw[thick] (6,0)--(6,3);
        \draw[thick] (6,2)--(8,2.8);
        \draw[thick] (6,2)--(8,1.2);
        \node at (0.5,0.5) {(b)};
        \node at (0.5,2) {(a)};
        \node at (3.5,0.5) {(c)};
        \node at (7.5,0.5) {(d)};
        \node at (6.5,1.35) {(e)};
        \node at (7.5,2) {(f)};
        \node at (6.5,2.65) {(g)};
        \node at (3.5,2) {(h)};
    \end{tikzpicture}
    \caption{Schematic illustration of the moduli space of the Möbius strip with one closed string puncture.}
    \label{fig:Mobius with C}
\end{figure}
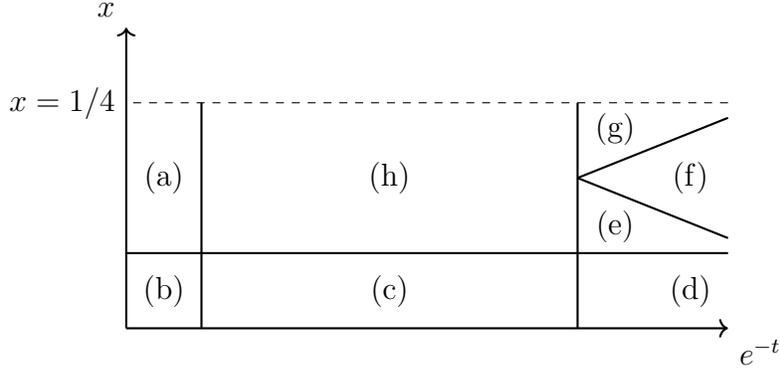

Möbius with C has a total of 7 Feynman regions, including one fundamental vertex region.
\begin{enumerate}[(a)]
    \item $D^2$ with $C-O_1-O_2.$ We sew $O_1$ to $O_2$ with the orientation reversal.
    \item $D_1^2$ with $C-O,$ $D_2^2$ with $O_1-O_2-O_3.$ We sew $O$ of $D_1^2$ to $O_1$ of $D_2^2$ and sew $O_2$ and $O_3$ of $D_2^2$ with the orientation reversal.
    \item $D^2$ with $C-O,$ $\mathcal{M}$ with $O.$ We glue $O$ of $D^2$ to $O$ of $\mathcal{M}.$
    \item $D^2_1$ with $C-O,$ $D_2^2$ with $C-O,$ and $\Bbb{RP}^2$ with C. We glue $O$ of $D^2_1$ to $O$ of $D_2^2$ and glue $C$ of $D_2^2$ to $C$ of $\Bbb{RP}^2.$
    \item $D^2$ with $C_1-C_2$, $\Bbb{RP}^2$ with $C.$ We glue $C_2$ of $D^2$ to $C$ of $\Bbb{RP}^2.$
    \item $D_1^2$ with $C$, $S^2$ with $C_1-C_2-C_3$, $\Bbb{RP}^2$ with $C$. We glue $C$ of $D_1^2$ to $C_1$ of $S^2,$ and glue $C_3$ of $S^2$ to $C$ of $\Bbb{RP}^2.$ 
    \item $D^2$ with $C$, $\Bbb{RP}^2$ with $C_1-C_2.$ We glue $C$ of $D^2$ to $C_1$ of $\Bbb{RP}^2.$   
    \item The fundamental vertex for Möbius with one closed string puncture.
\end{enumerate}
We sketched the stratification of the moduli space in Fig. \ref{fig:Mobius with C}.

\subsubsection{The region (a)}
We shall start by recalling the string vertex for $D^2$ with C-O-O. The open string punctures are located at $z_1=-t_D$ and $z_2=t_D,$ and to each of the open string punctures, we append the data of the local coordinates
\begin{equation}
    w_i=\frac{f_i(t_D)(z-z_i)}{g_i(t_D) z+l_i(t_D)}=\mu^2\nu \frac{4\mu^2\nu^2+1}{4\mu^2\nu^2} \frac{z-z_i}{(1+ z_iz)+\mu\nu h(z_i) (z-z_i)}\,,
\end{equation}
with
\begin{equation}
    h(z)=-\mu^{-1}\nu^{-1}\frac{3-4\mu^2\nu^2}{4-16\mu^2\nu^2} \left(z-\frac{1}{z}\right)\,,
\end{equation}
and
\begin{equation}
    f_i(t_D)= \mu^2\nu \frac{4\mu^2\nu^2+1}{4\mu^2\nu^2}\,,\quad g_i(t_D)=\mu\nu  h(z_i)+z_i\,,\quad l_i(t_D)=1-\mu\nu  h(z_i)z_i\,.
\end{equation}

We shall sew the open string punctures with the orientation reversal, via the plumbing fixture,
\begin{equation}
    w_1 \bar{w}_2= q\,.
\end{equation}
The plumbing fixture leads to the identification
\begin{equation}\label{eqn:MobC id1}
    \bar{z}_1\equiv \frac{\alpha z_2+\beta}{\gamma z_2+\delta}\,,
\end{equation}
with
\begin{equation}
    \alpha= f_1(t_D)f_2(t_D) t_D-l_1(t_D)g_2(t_D)q\,, \quad \beta= - f_1(t_D)f_2(t_D) t_D^2-l_1(t_D)l_2(t_D)q\,,
\end{equation}
\begin{equation}
    \gamma= -f_1(t_D)f_2(t_D) +g_1(t_D)g_2(t_D)q\,,\quad \delta=f_1(t_D)f_2(t_D)t_D+g_1(t_D)l_2(t_D) q\,.
\end{equation}

We shall redefine the coordinate 
\begin{equation}
    w:= \frac{A z+B}{Cz+D}\,,
\end{equation}
where
\begin{align}
    A=&\frac{16 \nu ^4 q \left(\mu  \nu  h\left(t_D\right)+t_D\right){}^2+\left(4 \mu ^2 \nu ^3+\nu \right)^2}{2 \sqrt{\nu ^4 \left(16 \nu ^2 q \left(\mu  \nu  h\left(t_D\right)+t_D\right){}^2+\left(4 \mu ^2 \nu ^2+1\right)^2\right) }}\nonumber\\
    &\times\frac{1}{\sqrt{\left(t_D \left(8 \mu  \nu ^2 \left(2 \nu  \left(\mu  \nu  t_D \left(q h\left(t_D\right){}^2+\mu ^2\right)-2 q h\left(t_D\right)\right)+\mu  t_D\right)+t_D\right)+16 \nu ^2 q\right)}}\,,
\end{align}

\begin{align}
    B=\frac{1}{2}\,,
\end{align}
\begin{align}
    C=&-\frac{\nu ^2 \left(16 \nu ^2 q \left(\mu  \nu  h\left(t_D\right)+t_D\right){}^2+\left(4 \mu ^2 \nu ^2+1\right)^2\right)}{2 \sqrt{\nu ^4 \left(16 \nu ^2 q \left(\mu  \nu  h\left(t_D\right)+t_D\right){}^2+\left(4 \mu ^2 \nu ^2+1\right)^2\right) }}\nonumber\\
    &\times \frac{1}{\sqrt{\left(t_D \left(8 \mu  \nu ^2 \left(2 \nu  \left(\mu  \nu  t_D \left(q h\left(t_D\right){}^2+\mu ^2\right)-2 q h\left(t_D\right)\right)+\mu  t_D\right)+t_D\right)+16 \nu ^2 q\right)}}\,,
\end{align}
\begin{align}
    D=\frac{1}{2}\,,
\end{align}
such that the identification \eqref{eqn:MobC id1} turns into
\begin{equation}
    \bar{x}\equiv -\mathcal{A} x\,,
\end{equation}
where
\begin{align}
    \mathcal{A}=&\frac{\mu ^2 \nu ^2 (-u) t_D h\left(t_D\right){}^2-\mu  \nu  u \left(t_D^2-1\right) h\left(t_D\right)+t_D \left(u-\mu ^2 \nu ^2\right)+\mathcal{K}}{\mu ^2 \nu ^2 t_D \left(u h\left(t_D\right){}^2+1\right)+\mu  \nu  u \left(t_D^2-1\right) h\left(t_D\right)-u t_D+\mathcal{K}}\,,\\
    =&\frac{u \left(t_D^2+1\right){}^2}{4 \mu ^2 \nu ^2 t_D^2}-\frac{u^2 \left(\left(t_D^2+1\right){}^2 \left(2 \mu  \nu  t_D h\left(t_D\right)+t_D^2-1\right){}^2\right)}{8 \left(\mu ^4 \nu ^4 t_D^4\right)}+\dots\,,
\end{align}
where
\begin{equation}
    u=\frac{q\mu^{-2}}{\left(1+\frac{1}{4\mu^2\nu^2}\right)^2}\,,
\end{equation}
and
\begin{equation}
    \mathcal{K}=\sqrt{\left(u \left(\mu  \nu  h\left(t_D\right)+t_D\right){}^2+\mu ^2 \nu ^2\right) \left(\mu  \nu  t_D \left(\mu  \nu  t_D \left(u h\left(t_D\right){}^2+1\right)-2 u h\left(t_D\right)\right)+u\right)}
\end{equation}

We define
\begin{equation}
    w:=\exp (-2\pi i (y_{o}-1/2))\,,
\end{equation}
such that $y$ takes value in
\begin{equation}
    [0,1/2]\times [0,i\tau_2]\,,
\end{equation}
with
\begin{equation}
    \exp(-2\pi \tau_2)=A^{-1}\,.
\end{equation}
The coordinates $z$ and $y_o$ are related via
\begin{equation}
    y_o=\frac{i}{\pi} \text{arctanh}(t_D z^{-1}) -\frac{i u z \left(t_D^2+1\right) \left(2 \mu  \nu  t_D h\left(t_D\right)+t_D^2-1\right)}{2 \pi  \mu ^2 \nu ^2 t_D \left(z-t_D\right) \left(t_D+z\right)}+\dots\,.
\end{equation}
Note 
\begin{equation}
    \text{arctanh}(x)=\frac{1}{2}\log\left(\frac{1+x}{1-x}\right)=-i\arctan(ix)\,.
\end{equation}

The closed string puncture at $i$ is mapped to
\begin{align}
    y_{C,o}=&\frac{1}{2}-\frac{1}{2\pi i} \log\left(\frac{i+t_D}{-i+t_D}\right)-\frac{u \left(2 \mu  \nu  t_D h\left(t_D\right)+t_D^2-1\right)}{2 \pi  \mu ^2 \nu ^2 t_D}+\dots\,,\\
    =&\frac{1}{2}-\frac{1}{\pi}\tan^{-1}(t_D^{-1})-\frac{u \left(t_D^2-1\right) \left(4 \mu ^2 \nu ^2+1\right)}{4 \pi  \mu ^2 \nu ^2 t_D \left(4 \mu ^2 \nu ^2-1\right)}+\dots\,,
\end{align}
where
\begin{equation}
    \frac{1}{\pi }\tan^{-1}(t_D^{-1})=\frac{1}{2\pi i}\log \left(\frac{i+t_D}{-i+t_D}\right)\,.
\end{equation}
For convenience, we also record $\sin(2\pi y_{C,o})$ 
\begin{equation}
    \sin(2\pi y_{C,o})=\frac{2 t_D}{t_D^2+1}+\frac{u \left(t_D^2-1\right){}^2 \left(4 \mu ^2 \nu ^2+1\right)}{2 \mu ^2 \nu ^2 t_D \left(t_D^2+1\right) \left(4 \mu ^2 \nu ^2-1\right)}+\dots\,.
\end{equation}
We also record the open string modulus
\begin{align}
    t_o:=&-\frac{1}{2\pi}\log\mathcal{A}\,,\\
    =&\frac{1}{2\pi}\log \left(\frac{4t_D^2 \mu^2\nu^2}{(1+t_D^2)^2u}\right)+\frac{u \left(2 \mu  \nu  t_D h\left(t_D\right)+t_D^2-1\right){}^2}{4 \pi  \mu ^2 \nu ^2 t_D^2}+\dots\,, \\
    =&\frac{1}{2\pi}\log \left(\frac{4t_D^2 \mu^2\nu^2}{(1+t_D^2)^2u}\right)+\frac{u \left(t_D^2-1\right){}^2 \left(4 \mu ^2 \nu ^2+1\right)^2}{16 \pi  \mu ^2 \nu ^2 t_D^2 \left(4 \mu ^2 \nu ^2-1\right)^2}+\dots\,.
\end{align}

Let us now study the region of the moduli space covered by the Feynman diagram we have been studying. First, let us note
\begin{equation}
    \frac{1}{2\mu\nu}\leq t_D\leq1\,, \qquad 0\leq u\leq \frac{\mu^{-2}}{\left(1+\frac{1}{4\mu^2\nu^2}\right)^2}\,.
\end{equation}

The first boundary of the Feynman region is obtained by fixing $t_D=1,$ which yields
\begin{equation}
    y_{C,o}=\frac{1}{4}\,,
\end{equation}
and
\begin{equation}
    t_o=-\frac{1}{2\pi} \log\left(\frac{u}{\mu^2\nu^2}\right)\,.
\end{equation}
The interface between the Feynman region and the vertex region is obtained by fixing $u=\mu^{-2}/(1+1/(4\mu^2\nu^2))^2,$
\begin{equation}
    y_{C,o}=\frac{1}{2}-\frac{1}{\pi}\tan^{-1}t_D^{-1}-\frac{4 \nu ^2 \left(t_D^2-1\right)}{\pi  t_D \left(16 \mu ^4 \nu ^4-1\right)}+\dots\,,
\end{equation}
\begin{align}\label{eqn:y to A}
    \mathcal{A}=&\frac{\left(t_D^2+1\right){}^2}{4 \mu ^4 \nu ^2 t_D^2}-\frac{ \left(t_D^2+1\right){}^2}{8 \left(\mu ^6 \nu ^4 t_D^2\right)}+\dots\,,\\
    =&\frac{1}{\mu^4\nu^2\sin^2(2\pi y_{C,o})}\left(1+\frac{1}{4\mu^2\nu^2}\right)^{-2}\biggr(1+2(\cot^2(2\pi y_{C,o})+\mu^2\nu^2 h(t_D)^2) \mu^{-4}\nu^{-2} \left(1+\frac{1}{4\mu^2\nu^2}\right)^{-4}\biggr)\,,
\end{align}
or equivalently
\begin{equation}
    t_o=-\frac{1}{2\pi}\log\left[\frac{(1+t_D)^2}{4\mu^4\nu^2t_D^2}\right]+\frac{1}{4\pi\mu^2\nu^2}+\frac{ \left(t_D^2-1\right){}^2}{16 \pi  \mu ^4 \nu ^2 t_D^2}+\dots\,.
\end{equation}
We shall also express $y_{C,o}$ in terms of $\mathcal{A}$ for future purposes. We shall invert \eqref{eqn:y to A} perturbatively
\begin{equation}
    y_{C,o}=y_{C,o}^{(0)}+y_{C,o}^{(1)}+\dots\,,
\end{equation}
where
\begin{equation}
    y_{C,o}^{(0)}=\frac{1}{2\pi}\sin^{-1}\left[ \mathcal{A}^{-1/2} \mu^{-2}\nu^{-1} \left(1+\frac{1}{4\mu^2\nu^2}\right)\right]=\frac{1}{2\pi} \sin^{-1}\biggr[ e^{\pi t_o}\mu^{-2}\nu^{-1} \left(1+\frac{1}{4\mu^2\nu^2}\right)\biggr]\,,
\end{equation}
and
\begin{equation}
    y_{C,o}^{(1)}=\frac{\tan(2\pi y_{C,o}^{(0)})}{2\pi}\left(\cot^2(2\pi y_{C,o}^{(0)})-\mu^2\nu^2 h(\tan(\pi y_{C,o}^{(0)}) )\right)\mu^{-4}\nu^{-2}\left(1+\frac{1}{4\mu^2\nu^2}\right)^{-4}\,.
\end{equation}
Note
\begin{equation}
  \frac{1}{\pi}\log\mu+\frac{1}{4\pi\mu^2}+\dots\leq  t_o\leq \frac{1}{\pi} \log \left(\mu^2\nu \left(1+\frac{1}{4\mu^2\nu^2}\right)^{-1}\right)+\dots\,.
\end{equation}

On the other boundary obtained by fixing $t_D=1/(2\mu\nu),$ we obtain
\begin{equation}
    y_{C,c}=\frac{1}{2\pi\mu\nu}+\frac{u}{2\pi\mu\nu}-\frac{(1+9u^2\mu^2\nu^2)}{24\pi\mu^3\nu^3}+\dots\,,
\end{equation}
\begin{equation}
    \mathcal{A}=u-\frac{u^2}{2}+\frac{8u+11u^3\mu^2\nu^2}{16\mu^2\nu^2}+\dots\,,
\end{equation}
or equivalently
\begin{equation}
    t_o=-\frac{1}{2\pi}\log u+\frac{u}{4\pi}-\frac{8+9u^2\mu^2\nu^2}{32\pi\mu^2\nu^2}+\dots\,.
\end{equation}

The PCO locations are determined as
\begin{align}
    w_{p_1}=&\pm\frac{i \log \left(\frac{3 t+\sqrt{3}}{\sqrt{3}-3 t}\right)}{2 \pi }+\frac{i \sqrt{3}  \left(t^4-1\right)}{4 \pi  \mu ^3 \nu  t \left(1-3 t^2\right)}+\dots\,,\\
    =&\pm\frac{i \log \left(\frac{3 \sin \left(\pi  w_c\right)+\sqrt{3} \cos \left(\pi  w_c\right)}{\sqrt{3} \cos \left(\pi  w_c\right)-3 \sin \left(\pi  w_c\right)}\right)}{2 \pi} \,,
\end{align}
\begin{equation}
    w_{p_2}=\frac{i}{2\pi} \log \left(\frac{i \left(-1+e^{\pm i \theta }\right)-\left(1+e^{\pm i \theta }\right) \tan \left(\pi  w_c\right)}{\left(1+e^{\pm i \theta }\right) \tan \left(\pi  w_c\right)+i \left(-1+e^{\pm i \theta }\right)}\right)-\frac{1}{2\pi}\frac{2 \sqrt{3} e^{\pm i \theta } \sin \left(2 \pi  w_c\right)}{\mu  \nu  \left(-2 e^{\pm i \theta } \cos \left(2 \pi  w_c\right)+e^{\pm 2 i \theta }+1\right)}+\dots\,.
\end{equation}
\subsubsection{The region (b)}
Let us denote the global coordinate of $D^2$ with C-O by $x_1$ and the global coordinate of $D^2$ with O-O-O by $x_2.$ We then shall glue the open string puncture at $x_1=0$ to the open string puncture at $x_2=0.$ Similarly, we shall sew open string punctures at $x_2=1$ and $x_2=\infty$ with an orientation reversion 
\begin{equation}
    \mu\nu x_1 g_1(x_2)=-q_1\,,
\end{equation}
\begin{equation}
    \mu^2 g_2(x_{2,1})g_3(\bar{x}_{2,\infty})= q_2\,.
\end{equation}

\subsubsection{Region (c)}
As in the case of the open-string one-point function on the Möbius strip, we shall divide the moduli space into $t_o> 1/4$ and $t_o\leq 1/4.$ We shall first construct the Feynman region for 
\begin{equation}
    1/4<t_o\leq \frac{1}{2\pi}\log\mu^2+\frac{1}{4\pi\mu^2}+\dots\,.
\end{equation}

Let us denote the coordinate of $D^2$ with respect to C-O by $x_1,$ and the Möbius coordinate by $ y_o$. We shall glue the open string puncture at $x_1=1$ to the open string puncture at $y_o=0$ via the plumbing fixture
\begin{equation}
    \left(-i\nu \frac{x_1-1}{x_1+1}\right) \left(\mathfrak{F}(\mu) \frac{\exp(2\pi iy_o)-1}{\exp(2\pi iy_o)+1}\right)=-q_1\,,
\end{equation}
which yields
\begin{equation}
    \exp\left(2\pi iy_o\right)= \frac{-q_1-q_1x_1+i\mathfrak{F}(\mu)\nu-i\mathfrak{F}(\mu) \nu x_1}{q_1+q_1x_1+i\mathfrak{F}(\mu)\nu-i\mathfrak{F}(\mu) \nu x_1}\,.
\end{equation}

Therefore, the closed string puncture at $x_1=0$ is mapped to
\begin{equation}
    \exp\left(2\pi iy_{C,o}\right)= \frac{-q_1+i\mathfrak{F}(\mu)\nu}{q_1+i\mathfrak{F}(\mu)\nu}\,,
\end{equation}
equivalently
\begin{equation}
    y_{C,o}=\frac{q_1}{2\pi\nu\mu}+\frac{q_1}{2\pi\mu^3\nu}+\dots\,.
\end{equation}
The PCOs are located at 
\begin{equation}
    y_{o,pco1,\pm}=\pm\left(\frac{1}{6}+\frac{\sqrt{3}}{4\pi\mu^2}-\frac{\sqrt{3}}{4\pi\mu^4}+\dots\right)\,,
\end{equation}
and
\begin{equation}
    y_{o,pco2,\pm}=\pm i\left(\frac{\sqrt{3}q_1}{2\pi\nu\mu}+\frac{\sqrt{3}q_1}{2\pi\mu^3\nu}+\dots\right)\,.
\end{equation}

\subsubsection{Region (c)'}
We shall now study the region
\begin{equation}
    \frac{\pi}{4\log\lambda_D^2}\leq t_o\leq \frac{1}{4}\quad\equiv\quad 1\leq t_c\leq\frac{\log\lambda_D^2}{\pi} \,.
\end{equation}
Let us denote the coordinate of $D^2$ with C-O by $x_1$ and the Möbius coordinate by $y_c.$ We shall glue the open string puncture at $x_1=1$ to the open string puncture at $y_c=0$ via the plumbing fixture
\begin{equation}
    \left(-i\nu \frac{x_1-1}{x_1+1}\right)\left(\mathcal{F}_1(t_c) y_c+\mathcal{F}_3(t_c) y_c^3+\mathcal{F}_5(t_c)y_c^5+\dots\right)=-q_1\,.
\end{equation}

We shall perturbatively solve $y_c,$ treating $y_c=\mathcal{O}(\mu^{-1}\nu^{-1}),$ with $\mathcal{O}(\mu)=\mathcal{O}(\nu)$
\begin{equation}
    y_c=\sum_n y_c^{(2n+1)}\,,
\end{equation}
\begin{equation}
    y_c^{(1)}=-i\frac{q_1}{\mathcal{F}_1(t_c)\nu}\frac{x_1+1}{x_1-1}\,,\quad y_c^{(3)}=-\frac{\mathcal{F}_3(t_c) (y_c^{(1)})^3}{\mathcal{F}_1(t_c)}\,,
\end{equation}
and
\begin{equation}
    y_c^{(5)}= \frac{(3\mathcal{F}_3(t_c)^2-\mathcal{F}_1(t_c)\mathcal{F}_5(t_c))(y_c^{(1)})^5}{\mathcal{F}_1(t_c)^2}\,,\quad \dots\,.
\end{equation}

The closed string puncture at $x_1=0$ is mapped to
\begin{equation}
    y_{C,c}^{(1)}=i\frac{q_1}{\mathcal{F}_1(t_c)\nu}\,,
\end{equation}
and the PCO at $x_1=e^{\pm i\pi/3}$ is mapped to
\begin{equation}
    y_{c,pco2,\pm}^{(1)}=\pm \frac{\sqrt{3}q_1}{\mathcal{F}_1(t_c)\nu}\,.
\end{equation}

\subsubsection{Region (d)}
Let us denote the coordinate of the first $D^2$ with C-O by $x_1,$ the coordinate of the second $D^2$ with C-O by $x_2,$ and the coordinate of $\Bbb{RP}^2$ with C by $x_3.$ We shall glue open string punctures and the closed string puncture at $x_1=0$ and $x_3=0$ via plumbing fixtures
\begin{equation}
    -\nu^2 \frac{x_1-1}{x_1+1}\frac{x_2-1}{x_2+1} =-q_1\,,
\end{equation}
\begin{equation}
    \lambda_D^2 x_1x_3=q_2\,. 
\end{equation}

The closed string puncture at $x_2=0$ is mapped to
\begin{equation}
    x_1=\frac{1-q_1\nu^{-2}}{1+q_1\nu^{-2}}\,,
\end{equation}
and the crosscap at $x_3=e^{i\theta}$ is mapped to
\begin{equation}
    x_1= q_2\lambda_D^{-2} e^{-i\theta}\,.
\end{equation}
The PCOs are located at
\begin{equation}
    x_{pco,1,\pm}= e^{\pm\frac{i\pi}{3}}\,,
\end{equation}
and
\begin{equation}
    x_{pco,2,\pm}=\frac{\nu^2\pm i\sqrt{3}q_1}{\nu^2\mp i\sqrt{3}q_1}=e^{i\vartheta_{DCC,\pm}}\,.
\end{equation}

We shall define the Möbius coordinate $y_c$ such that
\begin{equation}
    x_1=\exp(-2\pi i y_c)\,,
\end{equation}
where
\begin{equation}
    \exp(-\pi t_c)=q_2\lambda_D^{-2}\,,
\end{equation}
or equivalently
\begin{equation}
    t_c=\frac{1}{\pi} \log \lambda_D^2q_2^{-1}\,.
\end{equation}
The closed string puncture is located at 
\begin{equation}
    y_{C,c}=\frac{1}{2\pi i} \log\left(\frac{1-q_1\nu^{-2}}{1+q_1\nu^{-2}}\right)\,,
\end{equation}
and the PCOs are located at 
\begin{equation}
    y_{c,pco,1}=\pm\frac{1}{6}\,,
\end{equation}
and
\begin{equation}
    y_{c,pco,2}= \frac{1}{2\pi} \vartheta_{DCC,\pm}\,.
\end{equation}
\subsubsection{Region (e)}
Let us denote the global coordinate of $D^2$ with C-C by $x_1,$ and the global coordinate of $\Bbb{RP}^2$ with C by $x_2.$ We shall use the unit disk coordinates. 

We shall glue the puncture at $x_1=0$ to the puncture at $x_2=0$ via the plumbing fixture
\begin{equation}
    \lambda_Df_D(t_D)x_1 x_2=q_2\,,
\end{equation}
where $\lambda_D^{-1}\lambda_S^{-1}\leq t_D\leq (1-\nu^{-2})/(1+\nu^{-2}).$ The modulus $t_D$ denotes the location of the closed string puncture at $x_1=t_D.$ The crosscap at $x_2=e^{i\theta}$ is mapped to
\begin{equation}
    x_1= q_2\lambda_D^{-1}f_D(t_D)^{-1} e^{-i\theta}\,.
\end{equation}
The PCOs are located at
\begin{equation}
    x_{pco,1}=\exp\left(\pm \frac{i\pi}{3} -\log\lambda_D\lambda_S \frac{t_D-t_{D,o}^*}{t_{D,c}^*-t_{D,o}^*}\right)\,,
\end{equation}
and
\begin{equation}
    x_{pco,2}=\exp\left(i\vartheta_{DCC,\mp}\mp \left(\frac{i\pi}{3} -\log\lambda_D\lambda_S \right) \frac{t_D-t_{D,o}^*}{t_{D,c}^*-t_{D,o}^*}\right)\,.
\end{equation}

Let us define the Möbius coordinate $y_c$ such that
\begin{equation}
    x_1=\exp(2\pi iy_c)\,,
\end{equation}
where
\begin{equation}
    q_2\lambda_D^{-1} f_D(t_D)^{-1}=\exp(-\pi t_c)\,,
\end{equation}
or equivalently
\begin{equation}
    t_c=\frac{1}{\pi}\log (\lambda_Df_D(t_D) q_2^{-1})\,.
\end{equation}
In the $y_c$ coordinate, the closed puncture is located at
\begin{equation}
    y_c=\frac{1}{2\pi i}\log t_D\,,
\end{equation}
and the PCOs are located at
\begin{equation}
    y_{c,pco,1}= \pm \frac{1}{6}+\frac{i}{2\pi}\log\lambda_D\lambda_S \frac{t_D-t_{D,o}^*}{t_{D,c}^*-t_{D,o}^*}\,,
\end{equation}
and
\begin{equation}
    y_{c,pco,2}=\frac{1}{2\pi}\vartheta_{DCC,\mp}\pm \frac{i}{2\pi} \left(\frac{i\pi}{3} -\log\lambda_D\lambda_S \right) \frac{t_D-t_{D,o}^*}{t_{D,c}^*-t_{D,o}^*}\,.
\end{equation}

\subsubsection{Region (f)}
We shall denote the global coordinate of the disk by $x_1,$ the coordinate of $S^2$ by $x_2,$ and the coordinate of $\Bbb{RP}^2$ by $x_3.$ We shall glue the closed string puncture at $x_1=0$ to the closed puncture at $x_2=0,$ and the closed puncture at $x_2=\infty$ to the closed string puncture at $x_3=0\,$
\begin{equation}
    \lambda_D \lambda_S x_1 x_2=q_1\,,\quad \lambda_D \lambda_S x_3 \frac{1}{x_2}=q_2\,. 
\end{equation}

Let us map the locations of the crosscap and the closed-string puncture to the disk parametrized by $ x_1$. The closed string puncture at $x_2=1$ is mapped to
\begin{equation}
    x_{c,1}=\frac{q_1}{\lambda_D\lambda_S}\,,
\end{equation}
while the crosscap at $x_3=e^{i\theta}$ is mapped to
\begin{equation}
    x_{C,1}=\frac{q_1q_2}{\lambda_D^2\lambda_S^2} e^{-i\theta}\,.
\end{equation}

We define
\begin{equation}
    x=\exp(2\pi i y_c)\,,
\end{equation}
where
\begin{equation}
    q_1q_2\lambda_D^{-2}\lambda_S^{-2}=\exp\left(-\pi t_c\right)\,,
\end{equation}
equivalently
\begin{equation}
    t_c= \frac{1}{\pi}\log \lambda_D^2\lambda_S^2q_1^{-1}q_2^{-1}\,.
\end{equation}
The closed string puncture is therefore mapped to
\begin{equation}
    y_c=\frac{i}{2\pi } \log q_1^{-1}\lambda_D\lambda_S\,.
\end{equation}
The PCOs are located at
\begin{equation}
    y_{c,pco}=\frac{1}{2\pi i} \log e^{\pm \pi i/3} q_1 \lambda_D^{-1}\lambda_S^{-1}=y_c\pm \frac{1}{6}\,.
\end{equation}

\subsubsection{Region (g)}
Let us denote the global coordinate of $D^2$ with $C$ by $x_1,$ and the global coordinate of $\Bbb{RP}^2$ with $C-C$ by $x_2.$ We shall use the unit disk coordinates. 

We shall glue the puncture at $x_1=0$ to the puncture at $x_2=0$ via a plumbing fixture, which is given by
\begin{equation}
    \lambda_D x_1 f(t) x_2=q_1\,,
\end{equation}
where $\lambda_D^{-1}\lambda_S^{-1}\leq t\leq1.$ Then the closed string puncture at $x_2=t$ is mapped to
\begin{equation}
    x_1= q_1\lambda_D^{-1}t^{-1}f(t)^{-1}\,,
\end{equation}
the crosscap is mapped to
\begin{equation}
    x_1= q_1 \lambda_D^{-1} f(t)^{-1} e^{-i\theta}\,,
\end{equation}
and the PCOs are mapped to
\begin{equation}
    x_1=q_1 \lambda_D^{-1} e^{\pm i\pi/3}t^{-1}f(t)^{-1}\,.
\end{equation}

We shall define the Möbius coordinate $y_c$ such that
\begin{equation}
    x_1=\exp(2\pi iy_c)\,,
\end{equation}
where
\begin{equation}
    q_1\lambda_D^{-1}f(t)^{-1}=\exp(-\pi t_c)\,,
\end{equation}
equivalently,
\begin{equation}
    t_c=\frac{1}{\pi}\log \left(\lambda_D f(t)q_1^{-1}\right)\,.
\end{equation}
In the $y_c$ coordinate, the closed string puncture and the PCOs are given as
\begin{equation}
    y_{c,C}=\frac{i}{2\pi} \log\left(\lambda_Dt f(t)q_1^{-1}\right)\,,
\end{equation}
\begin{equation}
    y_{c,PCO,\pm}=\frac{i}{2\pi} \log \left(\lambda_Dtf(t)q_1^{-1} \right)\pm\frac{1}{6}\,.
\end{equation}

\subsubsection{Vertex region}
Since the closed-string vertex operator we insert into the Möbius strip is on-shell, the only off-shell data we need to construct is the boundary of the vertex region. The boundary of the vertex region can be obtained by setting the sewing parameters $q=1$ of the Feynman regions.  

\paragraph{Boundary of region (a)}
We have
\begin{equation}
    y_{C,o}=\frac{1}{2} -\frac{1}{\pi} \tan^{-1}t_D^{-1}-\frac{4\nu^2(t_D^2-1)}{\pi t_D (16\mu^4\nu^4-1)}+\dots\,,
\end{equation}
\begin{equation}
    t_o=-\frac{1}{2\pi} \log\left[\frac{(1+t_D)^2}{4\mu^4\nu^2t_D^2}\right]+\frac{1}{4\pi\mu^2\nu^2}+\frac{(t_D^2-1)^2}{16\pi\mu^4\nu^2t_D^2}+\dots\,,
\end{equation}
for
\begin{equation}
    \frac{1}{2\mu\nu}\leq t_D\leq1\,.
\end{equation}

\paragraph{Boundary of region (c)}
We have
\begin{equation}
    y_{C,o}=\frac{1}{2\pi\mu\nu}+\frac{1}{2\pi\mu^3\nu}+\dots\,,
\end{equation}
and
\begin{equation}
   \frac{1}{4}\leq t_o\leq \frac{1}{2\pi}\log\mu^2+\frac{1}{4\pi\mu^2}+\dots\,.
\end{equation}

\paragraph{Boundary of region (c)'}
We have
\begin{equation}
    \frac{\pi}{4\log\lambda_D^2}\leq t_o\leq \frac{1}{4}\quad\equiv\quad 1\leq t_c\leq t_c^*= \frac{\log\lambda_D^2}{\pi}\,,
\end{equation}
\begin{equation}
    y_{C,c}=i\frac{1}{\mathcal{F}_1(t_c)\nu}+i\frac{\mathcal{F}_3(t_c)}{\mathcal{F}_1(t_c)^4\nu^3}+i\frac{(3\mathcal{F}_3(t_c)^2-\mathcal{F}_1(t_c)\mathcal{F}_5(t_c))}{\mathcal{F}_1(t_c)^7\nu^5}+\dots\,,
\end{equation}
where
\begin{equation}
    \mathcal{F}_1(t_c)=\frac{\mathfrak{F}(\mu)\pi}{2} +\frac{t_c-1}{t_c^*-1}\left(\pi \nu-\frac{\mathfrak{F}(\mu)\pi}{2}\right)\,,~ \mathcal{F}_3(t_c)=-\frac{\mathfrak{F}(\mu)\pi^3}{24}+\frac{t_c-1}{t_c^*-1}\left(\frac{1}{3}\pi^3\nu+\frac{\mathfrak{F}(\mu)\pi^3}{24}\right)\,,
\end{equation}
\begin{equation}
    \mathcal{F}_5(t_c)=\frac{\mathfrak{F}(\mu)\pi^5}{2^4\times 3\times 5}+\frac{t_c-1}{t_c^*-1}\left(\frac{2}{15}\pi^5\nu-\frac{\mathfrak{F}(\mu)\pi^5}{2^4\times 3\times 5}\right)\,,
\end{equation}
with
\begin{equation}
    \mathfrak{F}(\mu)=2\mu \frac{4+\mu^{-2}}{\sqrt{9\mu^{-4}+40\mu^{-2}+16}}\,.
\end{equation}
\paragraph{Boundary of region (e)}
We have
\begin{equation}
    y_{C,c}=\frac{i}{2\pi}\log t_D^{-1}
\end{equation}
and
\begin{equation}
    t_c=\frac{1}{\pi}\log(\lambda_Df_D(t_D))\,,
\end{equation}
where
\begin{equation}
    t_{D,l}=\lambda_D^{-1}\lambda_S^{-1}\leq t_D\leq t_{D,u}=\frac{1-\nu^{-2}}{1+\nu^{-2}}\,,
\end{equation}
and
\begin{equation}
    f(t_D)=\lambda_D \lambda_S^2+\frac{t_D-t_{D,l}}{t_{D,u}-t_{D,l}}(\lambda_D-\lambda_D\lambda_S^2)\,.
\end{equation}
\paragraph{Boundary of region (g)}
We find
\begin{equation}
    y_{C,c}=\frac{i}{2\pi}\log (\lambda_D^2\lambda_S^2 t_{\Bbb{RP}^2} )\,,
\end{equation}
\begin{equation}
    t_c=\frac{1}{\pi}\log(\lambda_D^2\lambda_S^2)\,,
\end{equation}
with
\begin{equation}
t_{\Bbb{RP}^2,l}=\lambda_D^{-1}\lambda_S^{-1}\leq     t_{\Bbb{RP}^2}\leq t_{\Bbb{RP}^2,u}=1\,.
\end{equation}
Note that
\begin{equation}
    y_{C,o}=\frac{1}{4}
\end{equation}
corresponds to
\begin{equation}
    y_{C,c}=\frac{it_c}{2}\,.
\end{equation}
\subsection{Annulus with O}
Annulus with one open string puncture has 2 Feynman regions: a disk with C-O glued to a disk with C, a disk with O-O-O with two open string punctures identified. As in the case of the vertex region of the Möbius strip with one open string puncture, we shall divide the moduli space of the vertex region into two regions $t_o\geq1$ and $t<1$ for the open-string channel modulus
\begin{equation}
    \tau_o=it\,.
\end{equation}

\paragraph{Closed string degeneration}
We shall first construct the Feynman region representing the closed-string degeneration. We shall denote the global coordinate of the disk with C-O by $z_1$ with $|z_1|\leq1$ and the global coordinate of the other disk by $z_2$ with $|z_2|\leq1.$

The plumbing fixture gives
\begin{equation}
    \lambda_D^2z_1z_2=-q\,,
\end{equation}
with $|q|\leq1.$ We shall declare that the global coordinate is $z:=z_1,$ where two boundaries of the annulus are located at
\begin{equation}
    |z|=1\,,
\end{equation}
and
\begin{equation}
    |z|=\frac{|q|}{\lambda_D^2}\,.
\end{equation}
The local coordinate around the open string puncture is given by
\begin{equation}
    w=i\nu \frac{1-z}{1+z}\,.
\end{equation}

We shall define a new coordinate $y_c$ such that
\begin{equation}
    z=\exp(2\pi iy_c)\,,
\end{equation}
where the range of $y_c$ is given by
\begin{equation}
    [0,1]\times [0,t_c/2]\,,
\end{equation}
with $t_c=\frac{1}{\pi}\log\lambda_D^2/|q|.$

We define the open string channel variables
\begin{equation}
    \tau_o:=-\frac{1}{\tau_c}=\frac{i}{t_c}\,,
\end{equation}
\begin{equation}
    y_o=\frac{y_c}{\tau_C}=-\frac{iy_c}{t_c}=-it_o y_c \,.
\end{equation}
such that $y_o$ ranges
\begin{equation}
    [0,1/2]\times [0,t_o]\,.
\end{equation}

The local coordinate around the open string puncture is then given as
\begin{equation}
    w=i\nu \frac{1-z}{1+z}=i\nu \frac{1-\exp(2\pi i y_c)}{1+\exp(2\pi i y_c)}=i\nu \frac{1-\exp\left(-\frac{2\pi y_o}{t_o}\right)}{1+\exp\left(-\frac{2\pi y_o}{t_o}\right)}\,.
\end{equation}
The PCO is placed at
\begin{equation}
    z=\frac{1}{2}\pm i \frac{\sqrt{3}}{2}\quad\equiv\quad y_c=\pm\frac{1}{6}\,.
\end{equation}

\paragraph{Open string degeneration}
The local coordinate of the open string degeneration channel can be simply obtained by replacing $q$ with $-q$ in the analogous diagram in the Möbius strip
\begin{equation}
    w=2\mu \frac{4-q\mu^{-2}}{\sqrt{9q^2\mu^{-4}-40q\mu^{-2}+16}} \frac{\exp(2\pi iy_o)-1}{\exp(2\pi iy_o)+1}\,,
\end{equation}
with the PCO locations
\begin{align}
   e^{2\pi iy_{pco}}=&\frac{1}{2}(1\pm i\sqrt{3}) \pm \frac{i(\pm 3i +\sqrt{3})q}{4\mu^2}+\dots\,.
\end{align}

\paragraph{Vertex region}
To construct the vertex region, we shall interpolate two Feynman regions. As we noted earlier, we shall divide the vertex region into $t_o\geq1/2$ and $t_o<1/2.$

The local coordinate around the open string puncture at the closed string degeneration is given by
\begin{align}
    w=&i\nu \frac{1-\exp(2\pi iy_c)}{1+\exp(2\pi iy_c)}=i\nu \frac{1-\exp\left( -\frac{2\pi y_o}{t_o}\right)}{1+\exp\left(-\frac{2\pi y_o}{t_o}\right)}\,,\\
    =&\pi \nu y_c+\frac{1}{3}\pi^3\nu y_c^3+\frac{2}{15}\pi^5\nu y_c^5+\dots\,,\\
    =&i\frac{\pi \nu y_o}{t_o}-i\frac{\pi^3\nu y_o^3}{3t_o^3}+\frac{2 i \pi^5\nu y_o^5}{15t_o^5}+\dots\,,
\end{align}
with the PCO locations $y_{PCO,c}=\pm\frac{1}{6}\,.$

Similarly, the local coordinate around the open string degeneration is given by
\begin{align}
    w=& \mathcal{F}'(\mu) \frac{\exp(2\pi i y_o)-1}{\exp(2\pi i y_o)+1}\,,\\
    =&i \mathcal{F}'(\mu)\left(\pi y_o +\frac{1}{3} \pi^3 y_o^3+\frac{2}{15} \pi^5 y_o^5+\dots\right)\,,\\
    =&\mathcal{F}'(\mu)\left(\frac{\pi y_c}{t_c}-\frac{\pi^3 y_c^3}{3t_c^3}+\frac{2 \pi^5 y_c^5}{15t_c^5}+\dots\right)\,,
\end{align}
with the PCO locations $y_{PCO,o}=\pm\frac{1}{6}+\dots\,,$
and
\begin{equation}
    \mathcal{F}'(\mu)=2\mu \frac{4-\mu^{-2}}{\sqrt{16-40\mu^{-2}+9\mu^{-4}}}\,.
\end{equation}

We shall declare that the local coordinates and the PCO locations for $t_o\geq1/2$ are given as
\begin{equation}
    w=i\mathcal{F}'(\mu)\frac{\exp(2\pi iy_o)-1}{\exp(2\pi i y_o)+1}\,,
\end{equation}
and
\begin{equation}
    y_{PCO}=\pm\frac{1}{6}+\dots\,.
\end{equation}
For $t_o<1/2,$ we shall again use the polynomial approximation as in the case of the Möbius strip, such that
\begin{equation}
    w= \mathcal{F}_1'(t_c)y_c+\mathcal{F}_3'(t_c)y_c^3+\mathcal{F}_5(t_c)y_c^5+\dots\,,
\end{equation}
where
\begin{equation}
    \mathcal{F}_1'(2)=2\mathcal{F}'(\mu)\pi \,,\quad \mathcal{F}_1'(t_c^*)= \pi \nu\,,
\end{equation}
\begin{equation}
    \mathcal{F}_3'(2)=-\frac{8\pi^3}{3} \mathcal{F}'(\mu)\,,\quad \mathcal{F}_3'(t_c^*)=\frac{1}{3}\pi^3\nu\,,
\end{equation}
where
\begin{equation}
    t_c^*=\frac{1}{\pi} \log\lambda_D^2\,.
\end{equation}
We shall let $\mathcal{F}_i'(t_c)$ to be a function that interpolates the aforementioned two boundary values. To simplify our construction, we shall again treat $\mathcal{F}'_i$ as a constant for 
\begin{equation}
   2\leq t_c\leq \frac{1}{\pi}\log\lambda_D^2-\varepsilon\,,
\end{equation}
and let $\mathcal{F}_i'$ to be a linear function for
\begin{equation}
    \frac{1}{\pi}\log\lambda_D^2 -\varepsilon\leq t_c\leq \frac{1}{\pi}\log\lambda_D^2\,.
\end{equation}
Similarly, we shall fix the PCO locations to be
\begin{equation}
    y_{PCO,c}= \pm \frac{1}{3}\,,
\end{equation}
for
\begin{equation}
    2\leq t_c \leq\frac{1}{\pi}\log\lambda_D^2-\varepsilon\,,
\end{equation}
and let the PCO locations to interpolate the boundary conditions linearly 
\begin{equation}
    \frac{1}{\pi}\log\lambda_D^2 -\varepsilon\leq t_c\leq \frac{1}{\pi}\log\lambda_D^2\,.
\end{equation}

\paragraph{Open string puncture on the other disk}
The annulus diagrams we shall consider in this draft have boundaries that end on two different types of D-branes: D1-instanton and D9-branes. So, unlike the annulus diagrams studied in \cite{Sen:2020eck}, for example, the annulus amplitude is not symmetric under
\begin{equation}
    y\mapsto \frac{1}{2}-y\,,
\end{equation}
and we shall explicitly sum over the annulus diagram with the open-string puncture on the other boundary. To construct the vertex region for the open string on the other boundary, we can use the map
\begin{equation}
    y\mapsto \frac{1}{2}-y\,.
\end{equation}
\subsection{Annulus with C}
Annulus with C has in total 5 Feynamnn regions
\begin{itemize}
    \item Disk with C, sphere with C-C-C, disk with C. 
    \item Disk with C-C, disk with C
    \item Disk with C-O-O, with O-O sewed.
    \item Disk with C-O, annulus with O.
    \item Disk with C-O, disk with O-O-O, with O-O sewed.
\end{itemize}
and one fundamental vertex.

We shall only need to construct the Feynman region covered by the disk diagram with C-O glued to an annulus with O. 

\paragraph{Feynman region}
Let us first construct the Feynman region for which the closed string puncture is closed to the boundary at $y_o=0.$

Let's first consider $t>1/2.$ The plumbing fixture gives
\begin{equation}
    \nu z \mathcal{F}'(\mu) \frac{\exp(2\pi i y_o)-1}{\exp(2\pi i y_o)+1}=-q\,,
\end{equation}
where $0\leq q\leq1.$ The closed string puncture at $z=i$ is mapped to
\begin{equation}
    y_o=\frac{1}{\mathcal{F}'(\mu)\pi \nu}-\frac{q^3}{3(\mathcal{F}'(\mu))^3 \pi \nu^3}+\dots\,,
\end{equation}
and the PCOs at $z=\pm1/\sqrt{3}$ are mapped to
\begin{equation}
    y_{PCO,o}= \pm i\left(\frac{\sqrt{3}q}{\mathcal{F}'(\mu)\pi \nu}+\frac{\sqrt{3}q}{\mathcal{F}'(\mu)^3\pi\nu^3}+\dots\right)\,.
\end{equation}
Note 
\begin{equation}
    \mathcal{F}'(\mu)=2\mu+\dots\,.
\end{equation}

Let us now construct the Feynman region for $ 2\leq t_c\leq t_c^*-\varepsilon.$ We find that the closed string puncture is mapped to
\begin{equation}
    y_c=\frac{2i}{\mathcal{F}'(\mu)\pi\nu}-\frac{2iq^3}{3(\mathcal{F}'(\mu))^3\pi\nu^3}+\dots\,,
\end{equation}
and the PCOs at $z=\pm 1/\sqrt{3}$ are mapped to
\begin{equation}
    y_{PCO,c}=\pm\left(\frac{2\sqrt{3}q}{\mathcal{F}'(\mu)\pi \nu}+\frac{2\sqrt{3}q}{\mathcal{F}'(\mu)^3\pi\nu^3}+\dots\right)\,.
\end{equation}

To construct the Feynman region when the closed string puncture is located near the boundary at $y_o=1/2,$ we can simply use the map 
\begin{equation}
    y_o\mapsto \frac{1}{2}-y_o\,,
\end{equation}
or equivalently
\begin{equation}
    y_c\mapsto \frac{t_c}{2}-y_c\,.
\end{equation}

\newpage
\bibliographystyle{JHEP}
\bibliography{refs}

\end{document}